\documentclass[11pt]{article}
\usepackage{amssymb,amsmath}
\usepackage{bm} %math bold symbols
\usepackage{booktabs} %some alignment stuff
\usepackage{array}
\usepackage{latexsym}
\usepackage{graphicx}
\usepackage{color}
\usepackage{datetime}
\usepackage[nosort]{cite}
\usepackage{verbatim}
\usepackage{enumerate}
\usepackage{enumitem}
\usepackage{chngpage} % allows for temporary adjustment of side margins
\usepackage{mathrsfs}
\usepackage{euscript}
\usepackage{psfrag}

\usepackage{mciteplus}

\usepackage[colorlinks=true,      linkcolor=blue,      urlcolor=blue,
            filecolor=blue,      citecolor=blue,       pdfstartview=FitH,
                        pdfpagemode=UseNone,      bookmarksopen=true]{hyperref}  % LINKS
\usepackage[all]{hypcap}     %should be loaded AFTER hyperref, which should otherwise be loaded last.

\usepackage{tensor}
\usepackage{physics}
\usepackage{tikz}
\usepackage{mathtools}

\renewcommand{\comm}[1]{} %for commenting out blocks of text

\def\({\left(}
\def\){\right)}
\def\[{\left[}
\def\]{\right]}

\def\coeff#1#2{{\textstyle \frac{#1}{#2}}}

\def\One{{\hbox{ 1\kern-.8mm l}}}

\def\barray{\begin{array}}
\def\earray{\end{array}}
\def\be{\begin{equation}}
\def\ee{\end{equation}}
\def\bea{\begin{eqnarray}}
\def\eea{\end{eqnarray}}
\def\bal{\begin{align}}
\def\eal{\end{align}}

\def\nBPS#1{$\frac{1}{#1}$-BPS}

\numberwithin{equation}{section} % replaces the hack below

\definecolor{cardinal}{rgb}{0.6,0,0}
\definecolor{darkgreen}{rgb}{0,0.4,0}
\definecolor{golden}{rgb}{0.92, 0.7, 0}
\definecolor{midnight}{rgb}{0, 0, 0.5}
\definecolor{darkblue}{rgb}{0, 0, 0.7}
\definecolor{darkred}{rgb}{0.6, 0, 0}
\definecolor{purple}{rgb}{0.5, 0, 0.5}

\def\oneone{\rlap 1\mkern4mu{\rm l}}

\def\IR{\mathbb{R}}

\def\ZZ{\mathbb{Z}}

\def\cB{{\cal B}}

\def\cD{{\cal D}}
\def\cF{{\cal F}}

\def\cI{{\cal I}}

\def\cL{{\cal L}}

\def\cO{{\cal O}}
\def\cP{{\cal P}}

\def\cR{{\cal R}}

\def\nBPS#1{$\frac{1}{#1}$-BPS}

\def\RAdS{{R_{AdS}}}
\def\RAdSsq{{R^2_{AdS}}}

\newcommand{\stP}{\mathcal{P}}

\begin{document}

\phantom{AAA}
\vspace{-10mm}

\begin{flushright}
%
%IPHT-T19/021\\
%
\end{flushright}

\vspace{1.9cm}

\begin{center}

{\huge {\bf Q-Balls Meet Fuzzballs:}}\\
{\huge {\bf \vspace*{.25cm} Non-BPS Microstate Geometries }}

\vspace{1cm}

{\large{\bf {Bogdan Ganchev$^1$,~Anthony Houppe$^{1}$   and  Nicholas P. Warner$^{1,2,3}$}}}

\vspace{1cm}

$^1$Institut de Physique Th\'eorique, \\
Universit\'e Paris Saclay, CEA, CNRS,\\
Orme des Merisiers, Gif sur Yvette, 91191 CEDEX, France \\[12 pt]
\centerline{$^2$Department of Physics and Astronomy}
\centerline{and $^3$Department of Mathematics,}
\centerline{University of Southern California,} 
\centerline{Los Angeles, CA 90089, USA}

\vspace{10mm} 
{\footnotesize\upshape\ttfamily bogdan.ganchev @ ipht.fr,  anthony.houppe @ ipht.fr, warner @ usc.edu} \\

\vspace{2.2cm}
 
\textsc{Abstract}

\end{center}

\begin{adjustwidth}{3mm}{3mm} % to adjust the L and R margins
 
%\begin{abstract}
\vspace{-1.2mm}
\noindent
We construct a three-parameter family of non-extremal microstate geometries, or ``microstrata,'' that are dual to states and deformations of the D1-D5 CFT.   These families are non-extremal analogues of superstrata. We find these microstrata by using a Q-ball-inspired Ansatz that reduces the equations of motion to solving for eleven functions of one variable.   We then solve this system both perturbatively and numerically and the results match extremely well.  We find that the solutions have normal mode frequencies that depend upon the amplitudes of the excitations.  We also show that, at higher order in perturbations, some of the solutions, having started with normalizable modes,  develop a ``non-normalizable''  part, suggesting that the microstrata represent states in a  perturbed form of the D1-D5 CFT.  This paper is intended as a ``Proof of Concept'' for the Q-ball-inspired approach, and we will describe how it opens the way to many interesting follow-up calculations both in supergravity and in the dual holographic field theory.

%While this result suggests that an observer will not fall unharmed into the structure replacing the black hole horizon, the strong tidal disruption almost certainly reflects the high  coherence of the states represented by particular microstate geometries.
% Iosif:the last sentence was giving a detail that is not very useful the abstract
%
%
%\end{abstract}
\end{adjustwidth}

\bigskip 

\centerline{\it Dedicated to the memory of Sidney Coleman}

%\end{titlepage}
\thispagestyle{empty}
\newpage

%%%%%%%%%%%%%%%%%%%%%%%%%%%%%%%%%%%%%

\baselineskip=17pt
\parskip=5pt

\setcounter{tocdepth}{2}
\tableofcontents
%\newpage

\baselineskip=15pt
\parskip=3pt

\newpage

%%%%%%%%%%%%%%%%%%%%%%%%%%%%%%%%%%%%%
\section{Introduction}
\label{sec:Intro}
%%%%%%%%%%%%%%%%%%%%%%%%%%%%%%%%%%%%%

Microstate geometries  have already yielded remarkable results in the face of seemingly impossible odds\footnote{For a more detailed review of this, see the introduction to \cite{Gibbons:2013tqa}.} that ranged from ``No-Go'' theorems and apparently insuperable non-linearities in the geometry, to the Horowitz-Polchinski correspondence  principle \cite{Horowitz:1996nw,Damour:1999aw}  that suggested that  microstructure must collapse to Planck-scale decoration of a singularity.   In retrospect, the needle that has been threaded by  microstate geometries highlights the fact that they represent the primary mechanism\footnote{It is the unique mechanism if all the fields are  time-independent.}  \cite{Gibbons:2013tqa} through which one can describe the smooth gravitational expression of coherent microstate structure  at the horizon scale.  Indeed, the last few years have seen extensive holographic  confirmation that the class of microstate geometries known as ``superstrata''  describe families of coherent states of the D1-D5 CFT that underlie  the three-charge black hole in five dimensions  \cite{Kanitscheider:2006zf,Kanitscheider:2007wq,Taylor:2007hs,Giusto:2015dfa,Bombini:2017sge,Giusto:2019qig, Tormo:2019yus, Rawash:2021pik}.  In the last twenty years, microstate geometries have gone from being a chimera to becoming a standard laboratory for testing holographic CFT and supporting horizon-scale microstructure. 

The current challenge for microstate geometries is to get beyond  supersymmetry and extremality.   Supersymmetry and the BPS equations have provided  immense technical simplifications that have brought vast families of microstate geometries within range of analytic construction and exploration. Supersymmetry and extremality also impose a huge physical simplification: classical stability through positive mass theorems and  quantum stability because of the vanishing Hawking temperature.  The information problem is thus simplified to an ``information storage'' problem at zero temperature.

Families of supersymmetric microstate geometries are necessarily time independent, however they have to depend on at least two (spatial) coordinates.  Indeed, the fully generic superstratum depends on five spatial coordinates, and the most important families of analytically known, ``deep, scaling'' superstrata must depend, non-trivially, on three spatial coordinates \cite{Bena:2015bea,Bena:2016ypk,Bena:2017geu,Bena:2017upb,Bena:2017xbt,Ceplak:2018pws,Heidmann:2019zws,Heidmann:2019xrd,Shigemori:2020yuo,Mayerson:2020tcl}. The construction of non-extremal, time-dependent,  Hawking radiating generalizations of these superstrata would seem to be an impossible  ambition, even by the remarkable standards of the Microstate Geometry Program.   This paper provides an important step in turning this fantasy into a reality, through the construction of new families of non-extremal microstate geometries in a setting in which the holographic dictionary is known.

There are  many well-known, analytically-constructed examples of non-BPS microstate geometries. Perhaps the most well known is the  JMaRT solution \cite{Jejjala:2005yu}. There are generalizations of this to multi-centered solutions, and there are new bubbled non-BPS solutions  \cite{Bobev:2011kk, Vasilakis:2011ki,Bena:2015drs,Bena:2016dbw,Bossard:2017vii,Bah:2020pdz,Bah:2021owp}.  All  these solutions are interesting but they represent very atypical  states of the underlying black hole: they tend to have very high angular momentum, and often lie in the ``over-spinning'' sector of the theory.  The current status of non-BPS microstate geometries is somewhat reminiscent of the early results on  their supersymmetric counterparts, where the solutions were also highly specialized.    

There were two breakthroughs that followed the early work on supersymmetric microstate geometries:  the construction of scaling geometries that accessed the typical sector of the dual CFT, and then the construction of superstrata, whose geometries have a precise holographic correspondence with states of the D1-D5 CFT \cite{Kanitscheider:2006zf,Kanitscheider:2007wq,Taylor:2007hs,Giusto:2015dfa,Bombini:2017sge,Giusto:2019qig, Tormo:2019yus, Rawash:2021pik}.

Our goal here is to build upon the success of superstrata by finding non-extremal analogues of superstrata, which we will refer to as ``microstrata.''  These will be non-BPS geometries  in the six-dimensional supergravity with a  holographic correspondence to the D1-D5 CFT.  This means that the  CFT dual states can ultimately be determined.  Moreover, we will also construct classes of microstrata as excitations of superstrata and argue that some members of these classes, even if we do not construct them explicitly, exhibit a scaling behavior similar to superstrata, which suggests they can also access the typical sector of the CFT.

There are three components of our strategy for surmounting the array of technical obstacles to constructing such non-BPS solutions.  First, we will find solutions that are asymptotic to AdS$_3$ $\times S^3$.  This means that we will put the microstate geometry ``in a box'' that prevents its thermal decay.   Second, we use the recent discovery that a significant family of superstrata are encoded in a consistent truncation of six-dimensional supergravity on $S^3$  down to a gauged supergravity in three dimensions \cite{Mayerson:2020tcl,Houppe:2020oqp}.  The  inherent power of this consistent truncation is that the complicated dependence of solutions on the  $S^3$ directions is handled entirely by the machinery of the consistent truncation.  We therefore only have to work with dependence on the three coordinates, $(t,r,v)$, of the three-dimensional supergravity.

The hallmark of the superstratum is that it is dual to a CFT state that involves purely left-moving states, while the right moving sector remains in the Ramond ground state, preserving the right-moving supersymmetries.  This means that the configuration is \nBPS{8}.  There is similarly the \nBPS{8} anti-superstratum, with purely right-moving excitations and a left-moving Ramond ground state. In this paper we construct solutions that have non-trivial dependence on both $t$  and $v$, and so the excitations travel {\it inside} the light cone of the CFT, and are thus a superposition of both left-moving and right-moving excitations.  This breaks all the supersymmetries and leads to families of genuinely non-extremal, non-BPS solutions.

The third part of our strategy is to use  Sidney Coleman's Q-ball trick (and the related ``coiffuring trick''  in microstate geometries \cite{Bena:2013ora,Bena:2014rea,Bena:2015bea,Bena:2017xbt}) to reduce the core of the problem to functions of one variable, $r$.  That is, we find configurations within the three-dimensional supergravity in which the scalar fields depend on all the variables, but their energy-momentum tensor and electromagnetic currents only depend on the radial variable, $r$.  The end result is a family of coupled, non-linear differential equations for eleven functions of $r$.

Much of the effort in this paper focusses on finding interesting families of perturbative and numerical solutions to this still rather daunting system.   Indeed, we will  show that this system of equations is  extremely well-adapted to perturbative analysis and we are able to construct the solution to fourth  order, and sometimes to much higher orders.  These perturbative results  provide powerful confirmation of the accuracy of, and new features discovered in, our numerical analysis. 

It should be emphasized at the outset that, while we are making use of the ``Q-ball trick,''  Sidney Coleman's Q-ball construction led to a significant change in the effective potential that produced, for an appropriate range of frequencies, new families of classically (and quantum mechanically) stable solitons that minimize an energy functional.  Given the complexity of our solutions, and the gravitational back-reaction, it would be technically  difficult to see if our solutions could also  be quantum mechanically stable.  We are simply going to use  the Q-ball construction as a means to break supersymmetry while keeping the equations of motion in a simple form,  thereby enabling the construction of new supergravity solutions.   Ultimately, some of  the solutions we create here may have a more prosaic interpretation as gravitational bound states of more ordinary scalar excitations.

The reductions and simplifications that we make in constructing our new solutions mean that they are necessarily extremely specialized.  However,  as we will discuss in the final section, this paper  provides a powerful  leverage point for the construction of far more general non-BPS microstate geometries.  Amongst the many possible threads for  future work, we note that the experience with superstrata suggests that, once one has asymptotically-AdS solutions, one should be able to couple them to flat space and obtain asymptotically-flat microstrata.  While this will involve some additional technical challenges, it is evident from the work of \cite{Bena:2019azk,Bena:2020yii} that one can analyze such geometries as a tunneling problem using  WKB methods. In this way,  the  Hawking radiation emitted by microstrata into flat space can probably be analyzed  as a tunneling process from microstrata constructed in the box of AdS$_3$. 

A major motivation for this work is to provide a very important ``proof of concept.''  Despite the immense successes of supersymmetric microstate geometries, some suggested  that non-BPS microstate geometries may well prove unstable to collapse to a black hole.  It is still an open question as to  whether small perturbations can destabilize the supersymmetric geometries, especially given the seeming non-linear instability of AdS$_4$ \cite{Bizon:2011gg,Choptuik:2017cyd,Moschidis:2018ruk,Moschidis:2018kcf} (and references therein).   Thus there was a real concern that any fully back-reacted perturbation away from BPS would simply fold the whole microstate geometry up into a black hole, or some other singularity.  The fact that we have now constructed explicit examples of non-extremal microstata finally puts this issue to rest. The results presented here tells us that there are microstrata with ``large\footnote{Here ``large'' means a finite, as opposed to an infinitessimal, fraction of the object's mass is involved in the non-BPS deormation.}'' non-BPS deformations.    It is still possible to find singular limits in some corners of microstrata moduli space and we will  discuss this further in Section \ref{sec:Conclusions}.

One should also remember that some classes of instabilities of microstate geometries represent a ``feature'' rather than a ``bug,'' because such instabilities will prove essential in the scrambling of matter and in the generation of Hawking radiation.  Indeed, this seems to be precisely the correct interpretation of the instabilities discussed in \cite{Eperon:2016cdd,Marolf:2016nwu,Bena:2020yii}.   As we have already noted, we expect these physical instabilities to become important when we couple microstrata to flat space.

In addition to providing a ``proof of concept,''  our results also have significant implications for the microstate geometry and fuzzball programs more broadly.  

On a technical holographic level, there is the obvious question of looking at the precision holography of the new microstrata \cite{Kanitscheider:2006zf,Kanitscheider:2007wq,Taylor:2007hs,Giusto:2015dfa,Bombini:2017sge,Giusto:2019qig, Tormo:2019yus, Rawash:2021pik}.   This might seem to be something of a challenge because precision holography often involves correlators that are protected by supersymmetry.  On the other hand, it is possible that the microstrata constructed here are sufficiently specialized, coherent states that their holographic dual might be sufficiently protected by large-$N$ coherence.  At a minimum, the effectiveness of the perturbation theory we find on the gravity side should have some computational counterpart within the CFT.

On a fundamentally more physical level are the frequency shifts of the normal modes for microstrata. 

One of the tensions between supersymmetric microstate geometries and the microstructure of black holes is that while both have $e^{S}$ microstates, the former have rationally spaced energy levels with high occupation numbers while the latter have energy levels spaced by $e^{-S}$ with occupation numbers of $\cO(1)$.  Thus supersymmetric objects, with vanishing Hawking temperature, have very rigid structures with apparently very sharp resonances (see, for example, \cite{Bena:2019azk,Chakrabarty:2019ujg,Bena:2020yii}), spaced out by an energy gap  $\cO(\frac{1}{N_1 N_5})$.  The issue for microstate geometries is how non-BPS microstate geometries can give rise to a ``transition to chaos'' with the energy gap and occupation numbers that are characteristic of a black hole.   

In this paper we find that the interactions and gravitational back-reaction of the excitations can lead to  shifts in the normal mode frequencies of the microstate geometries, and that these shifts depend non-linearly on the amplitudes of the excitations.  As a result, the back-reaction of excitations creates a complicated set of resonances whose frequencies shift with the interactions and excitations of new modes.  Thus, from these first examples of non-BPS microstate geometries, we  see that the energy levels of the fully back-reacted microstate geometry are highly non-trivial functions of all the modes and their interactions: the generic non-BPS excitations may therefore be expected to have a rich, and far more chaotic spectrum.

In Section \ref{Sect:3Dsugr}  we summarize the relevant details   of the underlying three-dimensional gauged supergravity, and in Section \ref{sec:Qball} we discuss further truncations of this theory motivated by the $Q$-ball/coiffuring trick.   In Section  \ref{sec:ActionBCGC},  we restrict ourselves to one of the simplest possible families of solutions based on the observations in Section \ref{sec:Qball}.  This family contains the $(1,0,n)$ superstratum and microstrata generalizations.  We also discuss all the details of gauge fixing, coordinate choices and boundary conditions for microstrata.  In Section  \ref{sec:Charges} we discuss how to read off the mass and charges for our three-dimensional solutions and characterize extremality from the three-dimensional perspective.   Section \ref{sec:Perturbative} contains an extensive discussion of the perturbative solutions and how we find them.  This section shows that there are two distinct families of microstrata that fall within the Ansatz of Section  \ref{sec:ActionBCGC}.   Section \ref{sec:Perturbative}  also contains the perturbative results for the frequency shifts.  The numerical algorithms are described in Section \ref{sec:numerics} and the results of both the perturbation theory and the numerical solution are shown and compared in Section \ref{sec:results}.  The numerics and the perturbative analysis are in excellent agreement and the combined picture gives compelling evidence for the existence of two distinct families of microstrata, providing a precise description of all the underlying fields and  their frequency shifts.  Our final comments appear in   Section \ref{sec:Conclusions}.

%%%%%%%%%%%%%%%%%%%%%%%%%%%%%%%%%%%%%
\section{Three-Dimensional Gauged Supergravity}
\label{Sect:3Dsugr}
%%%%%%%%%%%%%%%%%%%%%%%%%%%%%%%%%%%%%

The supergravity theory of interest is the  $SO(4,5)$ theory described in  \cite{Mayerson:2020tcl, Houppe:2020oqp}.  (Here we will use the notation and conventions  of \cite{ Houppe:2020oqp}.)   This theory has  eight supersymmetries: four two-component spinors in three dimensions, transforming under an  $SO(4)$ $\cR$-symmetry.   The theory has a graviton, four gravitini, 20 ``spin-$\frac{1}{2}$'' fermions, 6 gauge fields and 14 scalars.   Since we are going to focus on non-supersymmetric  solutions to the equations of motion, we restrict our attention to the bosons, and their action.

%%%%%%%%%%%%%%%%%%%%%%%%
\subsection{The supergravity action}
\label{ss:sugra-action}
%%%%%%%%%%%%%%%%%%%%%%%%

The gauge group is $SO(4)$, and $I,J,K,\dots$ will denote indices transforming in the vector representation of $SO(4)$.  The gauge fields live in the adjoint representation and will be denoted by:  $A_\mu^{IJ} = -A_\mu^{JI}$.  The scalars live in the $9 +4 +1$ representations of $SO(4)$, and are most conveniently  represented as a vector,  $\chi_I$, and a symmetric matrix, $m_{IJ} = m_{JI}$, with non-vanishing determinant.  The determinant is the $SO(4)$ singlet.  This matrix may be thought of as being parametrized by the non-compact generators of $GL(4\,,\IR)$ and the inverse of $m_{IJ}$ will be denoted as $m^{IJ}$.  

The minimal couplings involve the   $SO(4)$ duals of gauge fields: 

\begin{equation}
\begin{aligned}
\cD_\mu \chi_I   ~\equiv~ &  \partial_\mu \chi_I   ~-~   2\, g_0\,\widetilde A_\mu{}^{IK} \chi_K   \,   \,. \\
\cD_\mu m_{IJ}   ~\equiv~   & \partial_\mu m_{IJ}  ~-~   2\, g_0\,\widetilde A_\mu{}^{IK} m_{KJ }~-~   2\, g_0\,\widetilde A_\mu{}^{JK} m_{IK }   \,   \,.
\end{aligned}
\label{Dscalars}
\end{equation}
where 
\begin{equation}
{\widetilde A_\mu}{}^{IJ}  ~\equiv~ \coeff{1}{2} \,\epsilon_{IJKL}\,{A_\mu}^{KL} \,. \label{dualGFs}
\end{equation}
This means that the field strengths, and their $SO(4)$ duals, are given by:
\begin{equation}
F_{\mu \nu}{}^{IJ}  ~=~ \coeff{1}{2}\, \epsilon_{IJKL} \,  \widetilde F_{\mu \nu}{}^{KL}    ~=~  \partial_{\mu}   A_{\nu}{}^{IJ}   ~-~  \partial_{\nu}   A_{\mu}{}^{IJ}   ~-~ 2 \,  g_0 \, \big(   A_{\mu} {}^{IL} \,\widetilde  A_{\nu} {}^{LJ}  ~-~ A_{\mu} {}^{JL} \,\widetilde  A_{\nu} {}^{LI}\big)  \,.
\label{fieldstrength2}
\end{equation}

It is also convenient to define the currents:
\begin{equation}
\begin{aligned}
Y_{\mu \, IJ}  ~\equiv~&  \chi_J \,\cD_\mu \chi_I ~-~  \chi_I \,\cD_\mu\chi_J \,, \\
 \end{aligned}
\label{Ydefinition}
\end{equation}

The bosonic action is then  \cite{Mayerson:2020tcl, Houppe:2020oqp}:
\begin{equation}
\begin{aligned}
\cL ~=~ & -\coeff{1}{4} \,e\,R 
 ~+~ \coeff{1}{8}\,e \, g^{\mu \nu} \, m^{IJ}   \, (\cD_\mu\, \chi_{I})  \, (\cD_\nu\, \chi_{J})    
~+~  \coeff{1}{16}\,e \, g^{\mu \nu} \,  \big( m^{IK} \, \cD_\mu\, m_{KJ}  \big)   \big( m^{JL} \, \cD_\nu\, m_{LI}  \big) \\
& ~-~e\, V  ~-~   \coeff{1}{8}\, e \, g^{\mu \rho}  \, g^{\nu \sigma} \, m_{IK} \,m_{JL}\,  F_{\mu \nu }^{IJ}  \, F_{\rho \sigma }^{KL}   \\
& ~+~  \coeff{1}{2}\,e  \, \varepsilon^{\mu \nu \rho} \, \Big[  g_0 \,\big(A_\mu{}^{IJ}\, \partial_\nu  \widetilde A_\rho{}^{IJ}  ~+~\coeff{4}{3}\,  g_0 \, A_\mu{}^{IJ} \,  A_\nu{}^{JK}\, A_\rho{}^{KI} \,\big) ~+~  \coeff{1}{8}\,  {Y_\mu}{}^{IJ}  \, F_{\nu \rho}^{IJ} \Big]
\end{aligned}
\label{eq:3Daction}
\end{equation}
where  $e= \sqrt{g}$ and $V$ is the scalar potential:
\begin{equation}
V~=~   \coeff{1}{4}\, g_0^2   \,  \det\big(m^{IJ}\big) \, \Big [\, 2 \,\big(1- \coeff{1}{4} \,  (\chi_I \chi_I)\big)^2    ~+~ m_{IJ} m_{IJ}  ~+~\coeff{1}{2} \,  m_{IJ} \chi_I \chi_J  ~-~\coeff{1}{2} \,  m_{II}  \,  m_{JJ}\, \Big]  \,.
\label{potential3}
\end{equation}
Note that because this theory has an $SO(4)$ gauge symmetry, we can fix the gauge by choosing $m_{IJ}$ to be diagonal.  

Solving the equations of motion in this action automatically leads to solutions of the six-dimensional supergravity that is dual to a sector of the D1-D5 CFT.  The details of how to uplift such solutions to six-dimensions can be found in \cite{Mayerson:2020tcl} and details of the further uplift to the IIB supergravity, and the holographic duality can be found in \cite{Giusto:2013rxa, Bena:2015bea}.  Here we will simply focus on the intrinsically three-dimensional description of the geometries and the microstates. 

We also note that the potential has a supersymmetric critical point when $ \chi_I = 0$ and $m_{IJ} = \delta_{IJ}$, at which point $V$ takes the value
\begin{equation}
V_0 ~=~    -  \coeff{1}{2}\, g_0^2   \,.
\label{susypt}
\end{equation}
Setting all the other fields to zero, the Einstein equations give: 
\begin{equation}
R_{\mu \nu} ~=~ - 4 \, V_0 \, g_{\mu \nu} ~=~    2 \, g_0^2 \, g_{\mu \nu}  \,.
\label{susyvac}
\end{equation}
and the supersymmetric vacuum\footnote{One should note that because we are using a metric signature $(+ - - )$ the cosmological constant of AdS is positive, contrary to the more standard and rational choice of signature.} is an AdS$_3$ of radius, $g_0^{-1}$. 

It will also be important to note that this supersymmetric critical point is part of a family of flat directions for $V$.  Specifically, if $ \chi_I = 0$  then there is a line of critical points with $V = V_0$ when $m_{IJ}$ has eigenvalues $\lambda, \lambda, \lambda^{-1},  \lambda^{-1}$ for any $\ \lambda >0$.   If the background only involves non-trivial scalar vevs  then the critical point only leads to a  supersymmetric background for $\lambda =1$  \cite{Houppe:2020oqp}.  We will refer to this as the {\it standard, supersymmetric critical point}.  The other critical points will then break supersymmetry.  As we will discuss in  \cite{GHW1}, this conclusion is not valid if there are also non-trivial gauge configurations: supersymmetry can be broken by the gauge fields, or supersymmetry can be restored for   $\lambda \ne 1$ by appropriately tuned gauge potentials.

%%%%%%%%%%%%%%%%%%%%%%%%
\subsection{The three-dimensional metric}
\label{ss:3metric}
%%%%%%%%%%%%%%%%%%%%%%%%

Following  \cite{Houppe:2020oqp} we will use  a metric signature of $(+ - -)$ and, as noted in  \cite{Houppe:2020oqp}, one can use coordinate choices to recast the three-dimensional metric in the following form:
\begin{equation}
ds_{3}^{2}  ~=~  \RAdSsq \, \bigg[ \,\Omega_1^{2} \, \bigg(d \tau +   \frac{k}{(1- \xi^{2})} \, d\psi \bigg)^2~-~\,\frac{\Omega_0^{2}}{(1-\xi^{2} )^{2}} \, \big( d \xi^2 ~+~ \xi^2 \, d \psi^2 \big) \, \bigg] \,,
\label{genmet1}
\end{equation}
for three arbitrary functions $\Omega_0$,  $\Omega_1$ and $k$ of the three coordinates, $(\tau ,\xi,\psi)$ with 
\begin{equation}
0 ~\le~ \xi ~ < 1 \,, \qquad \psi  ~\equiv~ \psi ~+~ 2 \, \pi \,.
\label{psi-period}
\end{equation}
We have also introduced an overall scale, $ \RAdS$, so that the metric functions and coordinates can be chosen to be dimensionless.  This scale will, of course, become the radius of the AdS metric at infinity.

To see that one can reduce a general metric to this form, one first uses the spatial coordinates to make the spatial base conformally flat.   Then one uses the freedom to shift $\tau$ by an arbitrary function to reduce the angular momentum vector, $k$, to a single component.   We have introduced the additional factors of $(1- \xi^{2})$ to anticipate and simplify the global AdS$_3$   limit of this metric.

To relate this to the standard superstratum form of the metric, one uses the change of variables:
\begin{equation}
\xi ~=~\frac{r}{\sqrt{r^2+ a^2}} \,,  \qquad  \tau~=~ \frac{t}{R_y}\,  \,,  \qquad  \psi~=~ \frac{\sqrt{2}\, v }{R_y}\,, 
\label{xidef}
\end{equation}
and
\begin{equation}
u ~\equiv~\frac{1}{\sqrt{2}} \, \big( t ~-~y  \big) \,, \qquad  v ~\equiv~\frac{1}{\sqrt{2}} \, \big(t ~+~y )\,, 
\label{uvtyreln}
\end{equation}
where $y$ is periodically identified as 
\begin{equation}
y ~\equiv~ y ~+~ 2 \pi \,R_y\,,
\end{equation}
One then obtains the metric:
\begin{equation}
ds_{3}^{2}  ~=~   \RAdSsq \, \bigg[ \, \frac{\Omega_1^{2}}{R_y^2} \, \bigg(dt  + \frac{\sqrt{2}}{a^2} \, (r^2  + a^2 ) \, k  \, dv  \bigg)^2~-~  \Omega_0^{2}\,\bigg(\frac{dr^2}{r^2 + a^2} ~+~\frac{2}{R_y^2 \, a^4 } \,r^2\,(r^2 + a^2) \, dv^2 \bigg)  \, \bigg]\,,
\label{genmet2}
\end{equation}
If one further sets:
\begin{equation}
\Omega_0 ~=~ \Omega_1 ~=~ 1\,, \qquad  k ~=~ \xi^2    ~=~  \frac{  r^2 }{ (r^2+ a^2) }  \,,
\label{AdSvals}
\end{equation}
then (\ref{genmet2}) becomes the standard metric\footnote{Up to rescaling the coordinates as $r\rightarrow\frac{a}{R_{AdS}}\,\tilde{r}$, $t\rightarrow\frac{R_y}{R_{AdS}}\,\tilde{\tau}$ and $y\rightarrow R_y\,\tilde{y}$} of global AdS$_3$:
\begin{equation}
ds_{3}^{2}  ~=~ \RAdSsq \, \bigg[ \,   \bigg(1 + \frac{r^2}{a^2} \bigg)\,  \bigg(\frac{dt}{R_y}\bigg)^2~-~  \frac{dr^2}{r^2 + a^2}  ~-~ \frac{r^2}{a^2}\,  \bigg(\frac{dy}{R_y}\bigg)^2 \, \bigg] \,.
\label{AdSmet}
\end{equation}

As noted after (\ref{susyvac}),  the supersymmetric AdS vacuum of the gauged supergravity has a radius given by:
\begin{equation}
 \RAdS ~=~ \frac{1}{g_0} \,,
\label{RAdSscale}
\end{equation}
and so we will henceforth use this to set the overall scale of the general metric (\ref{genmet1}).

We will also set the orientation as in  \cite{Houppe:2020oqp}.  We take the coordinates to be $(x^0, x^1, x^2)  = (\tau, \xi, \psi)$  or $(\hat x^0, \hat  x^1,\hat  x^2)  = (t,r,v)$ and  set:
\begin{equation}
\epsilon^{012}   ~=~ \epsilon_{012} ~=~ +1 \,.
\label{epsdefn}
\end{equation}
The covariant $\varepsilon$-symbol is then 
\begin{equation}
\varepsilon_{\mu \nu \rho}   ~=~ e \, \epsilon_{\mu \nu \rho}  \,, \qquad \varepsilon^{\mu \nu \rho}   ~=~ e^{-1} \, \epsilon^{\mu \nu \rho}   \,,
\label{varepsdefn}
\end{equation}
where $e = \sqrt{| g|}$ is the frame determinant.   The volume form is then:
\begin{equation}
vol_3   ~=~\coeff{1}{6}\, \varepsilon_{\mu \nu \rho} \, dx^\mu \wedge dx^\nu   \wedge dx^\rho ~=~   e\, d \tau \wedge d \xi \wedge d\psi~=~   \hat e\, dt \wedge dr \wedge dv   \,. 
\label{volorient}
\end{equation}
%

%%%%%%%%%%%%%%%%%%%%%%%%
\subsection{A further truncation: A \texorpdfstring{$U(1)$}{U(1)}-invariant sector}
\label{ss:truncation}
%%%%%%%%%%%%%%%%%%%%%%%%

The three-dimensional supergravity can capture what are known as the $(1,0,n) + (1,1,n)$ families of superstrata, which are encoded by two independent holomorphic functions of one variable.   However, for simplicity it is convenient to reduce the theory to a sub-sector that contains only the $(1,0,n)$ family of superstrata.  

This truncation is defined by requiring the configuration to be invariant under the $U(1)$ rotation in the $(3,4)$ internal directions.  This reduces the gauge symmetry to $U(1) \times U(1)$ with gauge connections, $A_\mu^{12}$ and $A_\mu^{34}$.  One also must set $\chi_3 = \chi_4 =0$ and take $m_{IJ}$ to have the form:
\begin{equation}
m_{IJ}~=~ \begin{pmatrix}
e^{2\,\mu_1} \, S_{2 \times 2}  &0 _{2 \times 2}  \\
0 _{2 \times 2}    &  e^{2\,\mu_2} \, \oneone_{2 \times 2}    
\end{pmatrix}\,,
 \label{mmatformq}
\end{equation}
where
\begin{equation}
S ~=~ {\cal O}^T  \, \begin{pmatrix}
e^{2\,\mu_0}  &0 \\
0   &e^{-2\,\mu_0}   
\end{pmatrix}\,  {\cal O} \,, 
\qquad 
 {\cal O}  ~=~  \begin{pmatrix}
\cos  \sigma  & \sin  \sigma \\
- \sin  \sigma  &\cos  \sigma   
\end{pmatrix}\,,
  \label{SOmatform}
\end{equation}
for some scalar fields, $\mu_0$, $\mu_1$, $\mu_2$ and $\sigma$.  As we noted above, the gauge invariance can be used to diagonalize $m_{IJ}$, and here this reduces to  the freedom to use one of the  $U(1)$ gauge invariances to set $\sigma$ to zero.

%%%%%%%%%%%%%%%%%%%%%%%%
\subsection{The \texorpdfstring{$(1,0,n)$}{(1,0,n)} family of superstrata and the AdS\texorpdfstring{$_3$}{3} vacuum}
\label{ss:superstrata}
%%%%%%%%%%%%%%%%%%%%%%%%

The $(1,0,n)$ family of superstrata is then given by introducing a holomorphic coordinate:
\begin{equation}
\zeta ~\equiv~ \xi \, e^{i \psi} ~\equiv~\frac{r}{\sqrt{r^2+ a^2}} \, e^{i \frac{\sqrt{2} v}{R_y} }\,. 
\label{zetadef}
\end{equation}
and taking 
\begin{equation}
 \chi_1 + i \chi_2 ~=~  \frac{a}{\sqrt{r^2 + a^2}}\ \,F(\zeta)    \label{chi_superstrata}
\end{equation}
for a holomorphic function, $F$:
\begin{equation}
F=\sum_{n=1}^{\infty}b_n\zeta^n \,.
\end{equation}
Regularity  of the solutions requires that the coefficients, $b_n$, satisfy:
\begin{equation}
\frac{2\,Q_1\,Q_5}{R_y^2}~=~ 2\,a^2+\sum_{n=1}^{\infty}b_n^2,
\label{ss-smoothness}
\end{equation}
where $Q_1$ and $Q_5$ are D1  and D5 supergravity charges.  In the three-dimensional formulation, these charges set the scale of the AdS$_3$ and thus determine the coupling constant of the gauged supergravity:
\begin{equation}
  g_0 ~=~ \big(Q_1\, Q_5\big)^{-\frac{1}{4}}  \,.
    \label{g0reln}
\end{equation}
The regularity condition then becomes:
\begin{equation}
\frac{1} { a^2\, g_0^ 4 \,R_y^2}=  1 ~+~ \frac{1}{2} \,\sum_{n=1}^{\infty} \,\frac{b_n ^2}{a^2} \,.
\label{ss-smoothness2}
\end{equation}

The complete solution is then given by choosing $\RAdS$ as in (\ref{RAdSscale}) and setting
\begin{align}
 \mu_1 ~=~ & \coeff{1}{2}\,  \log \Big[ \, 1 - \coeff{1}{4}\, \big(\chi_1^2 +\chi_2^2\big) \, \Big]   \,, \qquad \mu_0 ~=~ \mu_2 ~=~ \sigma ~=~ 0\,,
  \label{mu_superstrata} \\ 
 \Omega_0 ~=~ &\sqrt{\, 1 - \coeff{1}{4}\, \big(\chi_1^2 +\chi_2^2\big)}    \,, \qquad  \Omega_1 ~=~ 1 \,, \qquad  k  ~=~ \xi^2
  \label{Omk_superstrata} \\ 
\tilde A^{12} ~=~ & \frac{1}{2\, g_0} \, d\tau  \,, \\
 \tilde A^{34} ~=~  &- \frac{1}{2\, g_0} \,\Big[ \, 1 - \coeff{1}{4}\, \big(\chi_1^2 +\chi_2^2\big) \, \Big]^{-1} \bigg[  \,  d\tau~+~ \frac{\xi^2}{4\,(1 - \xi^2)} \,\big(\chi_1^2 +\chi_2^2\big) \, d\psi   \, \bigg] \,. 
  \label{gauge_superstrata}
\end{align}
Note that the gauge connections contain explicit factors of $g_0^{-1}$ that cancel the explicit factors of $g_0$ in the minimal couplings, thereby preserving the scale invariance of functions in the solution.

We impose the boundary conditions that $F(\zeta)$ is bounded as $|\zeta| \to 1$ so that $\chi_{1,2}$ vanish as  $r \to \infty$  ($\xi \to 1$).  The metric then limits to that of AdS$_3$ of radius $g_0^{-1}$, as defined by (\ref{AdSmet}) and (\ref{RAdSscale}). Similarly, the ``vacuum'' solution, with $\chi_{1,2} \equiv 0$, is simply the global AdS$_3$ of radius $g_0^{-1}$ described in Section \ref{ss:sugra-action}.

%%%%%%%%%%%%%%%%%%%%%%%%%%%%%%%%%%%%%
\section{``Q-ball'' truncations}
\label{sec:Qball}
%%%%%%%%%%%%%%%%%%%%%%%%%%%%%%%%%%%%%

The core of the ``Q-ball trick''  is to isolate a complex scalar field and give it a phase dependence of the form $e^{i \omega t}$ while arranging that these phases cancel in the currents and in the energy-momentum tensor.  The  result is to produce a background in which some of the scalars oscillate in time while the gauge fields and the metric are completely independent of $t$.  The important effect of such time-dependent scalars  is that they produce an effective shift in the scalar potential, changing the energetics.  As we stated in the introduction, we are simply using this technique as a way to break supersymmetry and access new families of solutions, and are not making broader claims about quantum stability of the resulting solitons. 

%%%%%%%%%%%%%%%%%%%%%%%%
\subsection{The ``simplest'' microstratum Ansatz}
\label{ss:simpAnsatz}
%%%%%%%%%%%%%%%%%%%%%%%%

There are several sectors of the three-dimensional supergravity in which this can be implemented.  The first, and most obvious lies in the truncation defined in Section \ref{ss:truncation}. Indeed, the obvious step to making a ``microstratum'' is to replace (\ref{chi_superstrata}) by 
\begin{equation}
 \chi_1 + i \chi_2 ~=~  \frac{a}{\sqrt{r^2 + a^2}}\ \,F(\zeta, \bar \zeta) \,   e^{i \omega t}\,, \label{chi_microstratum}
\end{equation}
where we allow for the fact that a general non-supersymmetric solution will not necessarily lead to holomorphy.   The potential only depends on $|\chi|^2$ and so the time-dependence cancels there.  However, the time-dependence does not cancel in the equations for $m_{IJ}$ and some of these scalars must also be made time dependent.  Indeed, consistency requires that one also takes  
\begin{equation}
\sigma  ~=~ \omega \, t \label{sigma_microstratum}
\end{equation}
in (\ref{SOmatform}).  Having made these changes, all the equations of motion remain consistent with the assumption that all the functions only depend on $(r,\psi)$.  

In retrospect, this is obvious.  This introduction of time dependence through these phases is  gauge equivalent to making a constant shift in the Coulomb potential of $\tilde A^{12}$.   This does not mean that these phases are trivial, but simply that they are gauge equivalent to applying a voltage to the background.   We can also think of the fields within this Ansatz as being precisely those that preserve the global U(1) rotation by an angle $\alpha$ in the $(1,2)$-direction, combined with a time translation $t  \to t - \alpha/\omega$.  

This Ansatz still involves arbitrary functions of $\xi$ and $\psi$, and while there might be rich families of such solutions, finding them is still too much of a challenge at this point.  Instead we simplify further by electing to generalize the single-mode superstratum \cite{Bena:2016ypk,Bena:2017upb,Heidmann:2019zws}.  That is, we make an Ansatz based on (\ref{chi_superstrata}) and (\ref{chi_microstratum}), in which 
\begin{equation}
 \chi_1 + i \chi_2 ~=~  \frac{a}{\sqrt{r^2 + a^2}}\,   \nu(\xi)\, e^{i (n \psi + \omega t)}  ~=~  \sqrt{1- \xi^2}\,   \nu(\xi)\, e^{i (n \psi + \omega t)}   \,, \label{singlemode1}
\end{equation}
for some integer, $n$, and some function, $\nu(\xi) $.  Note that we are retaining the explicit factor of $\sqrt{1- \xi^2}$ in our Ansatz as this is somewhat more convenient for the numerical analysis.

The other scalars $\mu_j(\xi)$, $j=0,1,2$ are also taken to be only functions of $\xi$, with the phase in   (\ref{SOmatform}) now having the form:
\begin{equation}
 \sigma  ~=~ n \, \psi ~+~\omega \, t \,. \label{singlemode2}
\end{equation}
We make an Ansatz for the gauge fields:
\begin{equation}
\tilde A^{12} ~=~ \frac{1}{g_0} \,\big[\,  \Phi_1(\xi)  \, d\tau ~+~  \Psi_1(\xi)  \, d\psi \, \big]\,, \qquad  \tilde A^{34} ~=~ \frac{1}{g_0} \,\big[\,\Phi_2(\xi)  \, d\tau ~+~  \Psi_2(\xi)  \, d\psi    \, \big] \,,
  \label{gauge_ansatz}
\end{equation}
where we have, once again, introduced explicit factors of $g_0^{-1}$ so as to cancel the $g_0$'s in the minimal coupling and thus render the fields and interactions scale independent. We have also fixed the gauges in $\tilde A^{12}$ and $\tilde A^{34}$ by removing the components proportional to $d\xi$.  Finally, we assume that all the metric functions, $\Omega_0(\xi)$,  $\Omega_1(\xi)$  and  $k(\xi)$, only depend on $\xi$.  

The Ansatz involves eleven arbitrary functions of  one variable, $\xi$, which we assemble into a list:
\begin{equation}
{\cal F} ~\equiv~ \big\{\, \nu  \,, \ \  \mu_0   \,, \ \  \mu_1  \,, \ \ \mu_2  \,,  \ \  \Phi_1 \,, \ \  \Psi_1   \,, \ \  \Phi_2  \,, \ \ \Psi_2  \,, \ \  \Omega_0  \,, \ \  \Omega_1    \,, \ \  k \, \big\} \,.
  \label{functionlist}
\end{equation}
The Ansatz is consistent with the equations of motion and it is the one upon which we will focus in this paper. 

It is  useful (and an invaluable tool for checking the equations and numerics) to note that the ``single-mode superstratum'' with $F(\zeta) =\alpha_0 \, \zeta^n$, for some constant, $\alpha_0$,  in   (\ref{chi_superstrata}) corresponds to: 
\begin{equation}
\begin{aligned}
\nu   ~=~&\alpha_0 \,   \xi^n  \,, \qquad  \mu_1 ~=~ \coeff{1}{2}\,  \log \Big[ \, 1 - \coeff{1}{4}\, \alpha_0^2 \,  (1-\xi^2)\, \xi^{2n}  \, \Big]   \,,\qquad     \mu_0 ~=~ \mu_2 ~=~ 0 \,,  \\
  \Phi_1  ~=~  &\frac{1}{2}  \,, \qquad \Psi_1  ~=~ 0  \,,  \\  
  \Phi_2 ~=~ &\frac{1}{2} \,\bigg[\,1 ~-~  \frac{1}{\big(\,1 - \coeff{1}{4}\, \alpha_0^2 \, (1-\xi^2)\,  \xi^{2n} \, \big)}  \, \bigg]\,, \qquad  \Psi_2 ~=~
  - \frac{\alpha_0^2 }{8} \,\frac{ \xi^{2n+2}}{\big(\, 1 - \coeff{1}{4}\, \alpha_0^2 \,  (1-\xi^2)\, \xi^{2n} \, \big)}   \,,\\ 
  \Omega_0  ~=~ &  \sqrt{\, 1 - \coeff{1}{4}\, \alpha_0^2 \,  (1-\xi^2)\, \xi^{2n} }     \,, \qquad  \Omega_1 ~=~  1    \,, \qquad   k ~=~\xi^2  \,,
\end{aligned}
  \label{ssres1}
\end{equation}
where we have made a trivial gauge transformation of $ \tilde A^{34}$ in (\ref{gauge_superstrata}) so that $\Phi_2$ vanishes at $\xi =0$.

Note that with these choices, the coefficient of $d \psi^2$ in (\ref{genmet1}) is 
\begin{equation}
\frac{\xi^2}{(1- \xi^2)} \, \Big[ \, \coeff{1}{4} \alpha_0^2 \,  \xi^{2n} ~-~ 1\,\Big] \,,
  \label{CTCtest}
\end{equation}
This means that, to avoid CTC's one must have $| \alpha_0| \le 2$, and for  asymptotically AdS$_3$ space-time one must impose the strict inequality:
\begin{equation}
| \alpha_0| ~<~ 2 \,.
  \label{alpha0bound}
\end{equation}
The limit, $| \alpha_0| =2$, is usually thought of as the  ``extremal BTZ limit,''  but because of our formulation of the three-dimensional metric, this limit actually results in a three-dimensional metric that is asymptotic to AdS$_2$ $\times S^1$, and this is precisely the scaling limit described and analyzed in \cite{Bena:2018bbd}.

We also note that, modulo issues with gauge fixing that we will discuss later,  if one has $n > 0$ and $\omega >0$, then the excitation is traveling inside the light cone of the CFT, and so will consist of  both left-moving and  right-moving states.  This means that solutions with  $n, \omega >0$ will break all supersymmetries.

%%%%%%%%%%%%%%%%%%%%%%%%
\subsection{Other microstrata Ans\"atze}
\label{ss:Ansatze}
%%%%%%%%%%%%%%%%%%%%%%%%

The are several obvious variations on the Ansatz defined in Section \ref{ss:simpAnsatz} and it is useful to catalog some of them here for potential further study.  

First one can make a similar generalization of the $(1,1,n)$ superstratum by working in the $(3,4)$ sector, and requiring invariance in the $(1,2)$ directions.  This is a trivial flip $(1,2) \to (3,4)$ of the Ansatz above.  However, one can do both.  That is, one can take
\begin{equation}
 \chi_1 + i \chi_2 ~=~  \frac{a}{\sqrt{r^2 + a^2}}\ \,F_1(\xi, \psi) \,   e^{i \omega_1 t}\,, \qquad  \chi_3 + i \chi_4 ~=~  \frac{a}{\sqrt{r^2 + a^2}}\ \,F_2(\xi, \psi) \,   e^{i \omega_2 t}\,,  \label{genchi_microstratum}
\end{equation}
for two arbitrary functions $F_1$ and $F_2$.   The phase, $\sigma$, in (\ref{sigma_microstratum}) will now become $\sigma_1 =  \omega_1 t$. 
One will also need to introduce another scalar, $\mu_3$,  and another phase rotation, $\sigma_2$, analogous to   (\ref{SOmatform}) but in the $(3,4)$ direction. 
\begin{equation}
\sigma_2  ~=~ \omega_2 \, t \label{gensigma_microstratum}
\end{equation}
One should note that one can choose $\omega_2$ independently of $\omega_1$. 

The single-mode truncations follow in an analogous manner, and one can choose different single modes, $n_1, n_2$,  in the  $(1,2)$ and $(3,4)$ directions. This truncation can be characterized as  preserving a global $U(1) \times U(1)$ rotations by an angle $\alpha_1$ in the $(1,2)$-direction and an angle  $\alpha_2$ in the $(1,2)$-direction, combined with compensating phase shifts in $t$ and $\psi$. 

One particularly interesting family would be to look at ``co-existing'' superstrata.  That is, use a superstratum in the $(1,2)$-sector, and an ``anti-superstratum'' in the $(3,4)$ sector.  The former is the usual left-moving BPS solution, while the latter is the corresponding ``right-moving'' anti-BPS superstratum that can be obtained trivially by replacing the null coordinate, $v$, with the other null coordinate, $u$, in the superstratum solution.  The ``seed solution'' here would be to start with $F_1$ as a holomorphic function of $\zeta$, and $F_2$ as an anti-holomorphic function.  The interactions and scattering between these counter-moving superstrata will probably destroy the holomorphy properties of  $F_1$ and $F_2$ in the full solution, the attraction here is that the  momenta  and angular momenta of these solutions can, in principle, be adjusted to zero. 

Other variations on the theme could involve imposing discrete $\ZZ_2$ symmetries like $1 \to -1, 2 \to -2$, or $3 \to -3, 4 \to -4$.  If one imposes the latter, then one must set $ \chi_3 = \chi_4 =0$, but one retains $\mu_3$.  If one imposes both of these $\ZZ_2$ symmetries then all the $\chi_j$ must vanish, and yet one retains all the $\mu_j$, $j=0,1,2,3$.  This should be a relatively simple set of excitations, but perhaps cannot lead to scaling microstrata because  all the $\chi_j$ vanish.

While we have highlighted several extremely interesting options, we will not pursue them further here because we wish to study the generalizations of the superstratum, which requires  some non-zero $\chi_j$'s,  and we want to start with the smallest number of arbitrary functions.  

There is, however, one very interesting truncation worthy of note because of the connections it affords with other work:  One can use a form of generalized ``Scherk-Schwarz'' reduction of the three-dimensional supergravity to connect superstrata to a supersymmetric generalization of JT gravity.

%%%%%%%%%%%%%%%%%%%%%%%%
\subsection{The ``super-JT'' truncation}
\label{ss:superJT}
%%%%%%%%%%%%%%%%%%%%%%%%

The goal here is to define a reduction of the three-dimensional gauged supergravity described in Section \ref{Sect:3Dsugr},  to a two-dimensional supersymmetric extension of JT gravity and do it in such a manner that it retains a superstratum solution.  

We define the dimensional reduction on the $\psi$-direction by requiring that the fields are invariant under the $U(1)$ action:
\begin{equation}
\psi ~\to~\psi + \alpha \,, \qquad {\cal O}_{12} 
~=~ \begin{pmatrix}
\cos n_1  \alpha  & - \sin   n_1 \alpha  \\
 \sin  n_1 \alpha  &\cos n_1\alpha  
\end{pmatrix}  \,, \qquad {\cal O}_{34} 
~=~ \begin{pmatrix}
\cos n_2 \alpha  & - \sin   n_2  \alpha  \\
 \sin n_2  \alpha  &\cos n_2 \alpha  
\end{pmatrix} 
 \label{superJTaction}
\end{equation}
where the ${\cal O}_{IJ}$ are commuting global symmetries acting on the internal $(I,J)$ directions.    This will determine precisely how the scalars depend on the $\psi$-coordinate.  In particular, one will have 
\begin{equation}
 \chi_1 + i \chi_2 ~=~ F_1(t, \xi) \,   e^{i n_1  \psi}\,, \qquad  \chi_3 + i \chi_4 ~=~ F_2(t, \xi) \,   e^{i n_2  \psi} \,.  \label{chi_JT}
\end{equation}
The metric and gauge fields will thus be required to be independent of $\psi$.

Because it is defined by invariance under a symmetry action, this is manifestly a consistent truncation and will lead to a two-dimensional theory.  This theory will have both massless and massive scalars.  Indeed, the $\chi$'s will have masses set by $n_j$, and the $m_{IJ}$ will give two real massless fields ($\mu_1$ and $\mu_2$) and four complex fields whose masses are set by the eigenvalues $2 n_1$, $2 n_2$, $n_1 - n_2$ and $n_1 + n_2$.   The gauge fields and ``internal metric'' will dimensionally reduce to produce more scalars, and, in particular, the scale of the $\psi$-circle will yield the usual  scalar field of JT gravity.  The three-dimensional supersymmetries rotate under the action  of (\ref{superJTaction}) and so could possibly be broken depending on the mode choices. 

The important point is that this consistent truncation reduces the three-dimensional supergravity to a two-dimensional extension of JT gravity, and does so in precisely such a manner as to preserve and include two single-mode ($(1,0,n_1)$ and $(1,1,n_2)$) superstrata. This theory is certainly worthy of further study and, in particular, it would be very interesting to  construct the  two-dimensional action and investigate its supersymmetry.

%%%%%%%%%%%%%%%%%%%%%%%%%%%%%%%%%%%%%
\section{The action,  boundary conditions and gauge choices}
\label{sec:ActionBCGC}
%%%%%%%%%%%%%%%%%%%%%%%%%%%%%%%%%%%%%

The remainder of this paper will focus on the truncation defined in Section \ref{ss:simpAnsatz} and the eleven dynamical fields (\ref{functionlist}).  We will give the complete action that defines the dynamics of these  functions, and we pin down the gauge degrees of freedom and fix boundary conditions for the fields.   Before diving into the details it is useful to recall the origins of these fields in IIB supergravity and then review some of the details coming from holography. 

The scalar, $\nu$, encodes the NS flux field and is the primary degree of freedom that defines the superstratum and microstratum.  The other scalars,  $\mu_0$, $\mu_1$, $\mu_2$, represent shape modes of the $S^3$ of compactification.   In particular, the scalar, $\mu_0$, will play an important independent role in our microstratum solutions and it is interesting to note that within the six-dimensional theory this scalar generates elliptical deformations of the underlying supertube configuration.

The Maxwell fields,  $\Phi_j$,   $\Psi_j$, are Kaluza-Klein fields  that define how the $S^3$ is fibered over the three-dimensional space-time.  These fields also appear in the dilaton and fluxes of the higher-dimensional theory  \cite{Mayerson:2020tcl}.  From the superstratum construction, we also know that the field, $\nu$, is the leading driver of the flow in that one can choose its boundary data freely and the other fields are determined by the flow if one requires regularity.  As we will see, holography also tells us that this field is dominant in that it is is dual to the field with the lowest conformal dimension. 

The truncation in Section \ref{sec:Qball}  involves more fields than the superstratum and there are almost certainly other fields that drive the flow.  Indeed, based on our results, it also seems that the boundary conditions for $\mu_0$ allow additional independent families of solution.  Thus our microstrata solutions will involve more diverse families of solutions than merely a deformation of the superstratum.

%%%%%%%%%%%%%%%%%%%%%%%%
\subsection{Overview and lessons from holography}
\label{ss:Overview}
%%%%%%%%%%%%%%%%%%%%%%%%

While the superstratum, and all the backgrounds we construct in this paper, are non-trivial solutions to the supergravity equations, it is still very instructive to study them as perturbations around  the  AdS$_3$ vacuum.  This enables us to gain insight into the appropriate boundary conditions  on the fields, both from holography and from the perspective of higher dimensional geometries.  The mathematics of perturbation theory also elucidates the possible singular terms that can emerge at higher orders.

The massive scalar wave equation in global AdS$_3$ in the $(\tau, \xi, \psi)$ coordinates is:
\begin{equation}
\frac{ (1-\xi^2)}{\RAdSsq}\bigg[ \, \frac{(1-\xi^2)}{\xi} \, \partial_\xi \big( \xi \partial_\xi F \big) ~+~  (\hat \omega + \hat n)^2  \, F ~- \frac{\hat n^2 }{\xi^2 } \, F  \,  \bigg]  ~-~ m^2 \, F ~=~ 0
\label{scalar-massive1}
\end{equation}
where we have used (\ref{AdSvals}) and separated variables by seeking a solution of the form $F(\xi) e^{i \hat \omega \tau + \hat n \psi}$.  

The conformal dimension, $\Delta$, of holographic operators is related to the mass via:
\begin{equation}
 m^2~=~ \frac{\Delta(\Delta-2) }{\RAdSsq} \,.
\label{Deltavsm}
\end{equation}
This is invariant under $\Delta \to (2-\Delta)$, and the conformal dimension is the root with $\Delta \ge 1$.
The scalar wave equation becomes:
\begin{equation}
  \frac{1}{\xi} \, \partial_\xi \big( \xi \partial_\xi F \big) ~+~  \frac{ (\hat \omega + \hat n)^2 }{(1-\xi^2) }  \, F ~- \frac{\hat n^2 }{\xi^2 \,(1-\xi^2) } \, F  ~=~ \frac{\Delta(\Delta-2) }{(1-\xi^2)^2}  \, F \,.
\label{scalar-massive2}
\end{equation}

For generic $\Delta$, the solution to this differential equation is  given in terms of hypergeometric functions of $r^{2}$ (or $r^{-2}$), and the two solutions, $F_1(r)$ and $F_2(r)$, are asymptotic, at infinity, to  $r^{(\Delta-2)}$ and $r^{-\Delta}$,  where $r$ is related to $\xi$ via (\ref{xidef}).    The series solution around infinity for $F_1$ is a straightforward power series in $r^{-2}$, multiplied by $r^{-\Delta}$, but when $ \Delta \in \ZZ$ and $ \Delta \ge 2$,  the roots of the indicial equation differ by an integer and the series solution for $F_2$ contains a ``sub-leading logarithm'' at  $r^{-\Delta}$.  That is, the series solutions for $F_1$ and $F_2$  have the form:
\begin{equation}
 F_1(\xi) ~=~r^{-\Delta} \, \sum_{n=0}^ \infty \, b_n \, r^{-2n}     \,, \qquad F_2(\xi) ~=~r^{(\Delta-2)} \, \sum_{n=0}^ \infty \, a_n \, r^{-2n} ~+~ \,  F_1(\xi) \,  \log r   \,.
\label{hypersols1}
\end{equation}
For $\Delta = 1$, the indicial equation has a double root and the general solution has the form: 
\begin{equation}
r^{-1}  \,\bigg[ \,  \sum_{n=0}^ \infty \, a_n \, r^{-2n} ~+~  \log r   \, \sum_{n=0}^ \infty \, b_n \, r^{-2n}  \, \bigg] \,.
\label{hypersols2}
\end{equation}
This solution has a {\it ``leading logarithm''} in that it defines the leading ``non-normalizable'' solution\footnote{Neither solution is normalizable but the term ``non-normalizable'' is used here to connote the most divergent solution.}.

In holography, the standard lore tells us that normalizable modes, with series of the form $F_1$, are dual to  states of the system and non-normalizable modes, with series of the form $F_2$,  are dual to deformations of the Lagrangian.  Moreover, the sub-leading logs can be related to holographic renormalization and conformal anomalies \cite{Henningson:1999xi,Bianchi:2001kw,Skenderis:2002wp,Skenderis:2008dh,Skenderis:2008dg}.  Thus the leading log in (\ref{hypersols2}) plays a very different role from the sub-leading logs in  (\ref{hypersols1}).

The masses, or conformal dimensions of the scalar fields are easily read off from the potential, $V$. The eigenmodes around the supersymmetric critical point are:
\begin{equation}
\begin{aligned}
&  \hat \nu  :  \  m^2  =  -1 \,, \ \ \Delta = 1  \,;   \qquad  \qquad \qquad\quad \mu_0  :    \   m^2  =0  \,, \ \ \Delta  =  2  \,, \\
& \mu_- ~\equiv~  \mu_1 - \mu_2 :    \  m^2  = 0  \,, \ \ \Delta = 2 \,;     \qquad \mu_+  ~\equiv~  \mu_1 + \mu_2 :    \  m^2  = 8  \,, \ \ \Delta = 4  \,, 
\end{aligned}
\label{scalar-massive3}
\end{equation}
where 
\begin{equation}
\hat \nu ~\equiv~  \sqrt{1 - \xi^2} \,  \nu 
\label{hatnudefn}
\end{equation}
is the function that appears is the scalar field $ \chi_1 + i \chi_2$ without any factors, as seen in (\ref{singlemode1}).
One should also recall that the scalars $\mu_1$ and $\mu_2$ are not oscillating and so have $\hat \omega =  \hat n =0$, while the scalar $\nu$ has $\hat \omega =  \omega$ and $\hat n =n$ and $\mu_0$ has   $\hat \omega =  2\omega$ and $\hat n = 2 n$. 

At the origin, the solution is dominated by the centrifugal term, $\sim  \frac{\hat n^2 }{\xi^2 }$, and, as is familiar with Bessel's equation, the smooth solution falls of as $\xi^{\hat n}$ as $\xi \to 0$.   Requiring the solution to be smooth at $\xi =0$ and fall off as $\xi \to 1$ typically leads to an eigenvalue problem and a discrete spectrum for  $\omega$.

The normalizable modes of the fields  $\mu_0$, $\mu_-$ and $\mu_+$ are indeed normalizable, and fall off as $r^{-2}$ and $r^{-4}$ at infinity.  For these fields  we will only consider such excitations.  The field $\hat \nu$ has $\Delta =1$ and so we have
\begin{equation}
 \chi_1 + i \chi_2 ~\sim~ \hat \nu ~\sim~ \hat c_1 \, r^{-1} ~+~  \hat  c_2 \, r^{-1} \, \log(r) \qquad \Leftrightarrow \qquad  \nu ~\sim~ c_1 ~+~  c_2 \,  \log(r)  \,,
\label{nuasymp}
\end{equation}
as $r \to \infty$ for some constants  $\hat c_1, \hat c_2, c_1$ and $c_2$.  As we have noted,  $\hat \nu$ and all its radial derivatives, vanish at infinity for any choices of $\hat c_1$ and $\hat c_2$.

Since we are interested in microstates of a black hole, we are, {\it a priori}, interested in the normalizable modes, and hence the simple power series solutions.  We therefore  start out by seeking such solutions at leading order, before computing the back-reaction, perturbatively or numerically.  In particular, our solutions will start out with  $\hat c_2 =0$ in (\ref{nuasymp}).     However, we  find that, for some classes of microstrata, the back-reaction  generates logarithmic terms in the solutions, and at higher orders one is required to introduce leading logs in $\nu$.  See Section \ref{ss:higher_orders_perturbation} for more details.

This is most easily seen through our perturbative analysis, described in Section \ref{sec:Perturbative}, in which we have to solve  differential equations with sources. To find the required inhomogeneous solutions we need to use functions of the form:
\begin{equation}
F ~=~ F_1(r) \,  \log r  \,,
\label{subleadinglog}
\end{equation}
where $F_1$ is the ``normalizable solution'' of the homogeneous equation (\ref{scalar-massive2}). When  the operator in (\ref{scalar-massive2}) acts on a function of the form (\ref{subleadinglog}),  it produces a simple power series (without logarithmic terms) and precisely these terms appear in some of the sources.  Once one generates  such log terms in the solutions, they then appear as sources in the next order of perturbation theory,  and this can lead to  polylogarithms. One should note that, despite the appearance of the logs in (\ref{hypersols2}) and (\ref{subleadinglog}), all of these expressions, and their derivatives with respect to $r$,   vanish as $r \to \infty$.  This means that  such expressions, and the related polylogarithms, are consistent with vanishing boundary conditions at infinity. 

While one might have anticipated the appearance of sub-leading logs, and even the polylogarithms, as a renormalization effect, we also find that  the ``non-normalizable'' solution for $\nu$  emerges from the back-reaction.  This is surprising because it appears to represent  a non-conformal perturbation of the action of the CFT by a relevant operator.   We will discuss this in more detail  in Sections \ref{ss:higher_orders_perturbation}  and \ref{sec:Conclusions}.  Here we simply note that the perturbative analysis in supergravity presents some interesting challenges for the interpretation of microstrata in the dual holographic field theory.  Moreover, as we will also discuss in Section \ref{sub:series_expansions}, the generation of the log terms creates a technical complexity  in setting up the numerical analysis.

%%%%%%%%%%%%%%%%%%%%%%%%
\subsection{The action}
\label{ss:Action}
%%%%%%%%%%%%%%%%%%%%%%%%

The action was computed in \cite{Houppe:2020oqp}, and is given in (\ref{eq:3Daction}). It contains a gravitational term,  kinetic terms for the scalars and the gauge fields,  a Chern-Simons term,  a Chern-Simons-like coupling between the scalars and the gauge fields, and  a potential. We can rewrite it in terms of our Ansatz  in Section \ref{ss:simpAnsatz}. 

We decompose the Lagrangian into pieces:
\begin{equation}
    \cL ~=~ \cL_{\text{gravity}} + \cL_\chi + \cL_m + \cL_A + \cL_{CS} + \cL_Y - \sqrt{g}\, V \ ,
    \label{lagrangian_full}
\end{equation}
and we introduce a convenient shorthand that captures the mode dependence and minimal couplings: 
\begin{equation}
 \Gamma ~=~ \xi^2  \Omega_0^2 \, \big[\,\omega + 2\, \Phi_1   \big]^2 ~-~ \Omega_1^2 \, \Big[\,(1-\xi^2 ) \,  \big(n + 2\,\Psi_1) ~-~k\, \big(\omega + 2\, \Phi_1    \big) \,\Big]^2 \,.
    \label{gauge_term}
\end{equation}

We then find:
\begingroup
\allowdisplaybreaks
\begin{align}
\cL_{\text{gravity}} ~&\equiv~ -\frac14 \sqrt{g}\ R  \nonumber \\
~&=~  \frac{\Omega_1}{8\, g_0\, \xi}\, \bigg[ \,\frac{\Omega_1^2}{\Omega_0^2}\, \bigg(k' + \frac{2\,\xi}{1- \xi^2} \, k\bigg)^2  - 4\, \xi \,  \bigg(\partial_\xi\bigg( \xi\,  \frac{\Omega_0'}{\Omega_0}\bigg) + \frac{1}{\Omega_1} \,\partial_\xi\Big( \xi\,   \Omega_1' \Big)  \bigg)  -  \frac{16\, \xi^2}{(1- \xi^2)^2} \,    \bigg]
\\
\cL_\chi ~&\equiv~ \frac18 \sqrt{g}\, g^{\mu\nu}\, (\cD_\mu \chi_I)\, m^{IJ}\, (\cD_\nu \chi_J)  \nonumber \\
~&=~  \frac{e^{2(\mu_0 - \mu_1)}}{8\, g_0\, \xi \,(1- \xi^2) \,\Omega_1}\,  \,\bigg[\,  \Gamma \, \nu^2  ~-~   \xi^2\,(1- \xi^2)    \,\Omega_1^2 \, e^{-4\,\mu_0} \, \Big( \partial_\xi \Big (\sqrt{1- \xi^2}\, \nu \Big) \Big)^2    \, \bigg]
\\
\cL_m ~&\equiv~ \frac1{16} \sqrt{g} \, g^{\mu\nu} \, \Tr(m^{-1} (\cD_\mu m) m^{-1} (\cD_\nu m)) \nonumber   \\
~&=~  \frac{1}{2\, g_0\, \xi \,(1- \xi^2)^2 \,\Omega_1}\,  \,\bigg[\,  \Gamma \, \sinh^2 2\, \mu_0  ~-~   \xi^2\,(1- \xi^2)^2    \,\Omega_1^2 \,  \, \big( (\mu_0')^2 + (\mu_1')^2 +(\mu_2')^2 \big)   \, \bigg]
\\
\cL_A ~&\equiv~   -   \frac{1}{8}\, e \, g^{\mu \rho}  \, g^{\nu \sigma} \, m_{IK} \,m_{JL}\,  F_{\mu \nu }^{IJ}  \, F_{\rho \sigma }^{KL}  \nonumber  \\
~&=~  \frac1{2 g_0\, \xi\, \Omega_1} \Big[ \xi^2 \,\qty(e^{4 \mu_2}\ \Phi_1'^2 \,+\, e^{4\mu_1}\ \Phi_2'^2) \nonumber \\[-.3em]
& \qquad\qquad \qquad   -  \frac{\Omega_1^2}{\Omega_0^2}\,\qty(e^{4 \mu_2}\, \qty((1-\xi^2)\, \Psi_1' - k\, \Phi_1')^2 + e^{4\mu_1}\, \qty((1-\xi^2)\, \Psi_2' - k\, \Phi_2')^2) \Big]   
\\  
\cL_{CS} ~&\equiv~    \frac{1}{2}\,g_0 \, e  \, \varepsilon^{\mu \nu \rho} \,  \Big(A_\mu{}^{IJ}\, \partial_\nu  \widetilde A_\rho{}^{IJ}  ~+~\coeff{4}{3}\,  g_0 \, A_\mu{}^{IJ} \,  A_\nu{}^{JK}\, A_\rho{}^{KI} \,\Big)  \nonumber   \\
~&=~  \frac1{g_0} \qty( \Phi_1 \Psi_2' - \Psi_2 \Phi_1' + \Phi_2 \Psi_1' - \Psi_1 \Phi_2')  
\label{LCS}
\\
\cL_Y ~&\equiv~   \frac{1}{16}\,e  \, \varepsilon^{\mu \nu \rho} \,  {Y_\mu}{}^{IJ}  \, F_{\nu \rho}^{IJ}\nonumber \\
~&=~  \frac1{4g_0} \qty(1-\xi^2)\, \nu^2\,\qty((2 \Psi_1 + n)\, \Phi_2' \,-\, (2 \Phi_1 + \omega)\, \Psi_2')   \\
V ~&=~ \frac{g_0^2}{2}\, e^{-4(\mu_1 + \mu_2)} \Big[1 - 2\, e^{2(\mu_1 + \mu_2)} \cosh(2\mu_0) + e^{4\mu_1} \sinh^2(2\mu_0) \nonumber \\
&\qquad\qquad \qquad \qquad +   \frac1{16}  \nu^2\, \qty(1-\xi^2) \qty((1-\xi^2)\,\nu^2 + 4\, e^{2(\mu_0 + \mu_1)}  - 8) \Big]
\end{align}%
\endgroup
Here, $'$ indicates a differentiation with respect to $\xi$.  

One should note that because the action is gauge invariant,  the expression, $ \Gamma$,  contains the gauge invariant terms that result from either introducing $n$ and $\omega$ or shifting the gauge potentials,  $\Phi_1$ and  $\Psi_1$.  In particular, the gauge invariant combinations are
\begin{equation}
G_{\Phi_1} ~\equiv~ (\,\omega + 2\, \Phi_1 ) \,, \qquad G_{\Psi_1} ~\equiv~ (n + 2\,\Psi_1)  \,.
    \label{gauge_inv_tems}
\end{equation}
This means that the values of $\omega$ and $n$  only have meaning once we have specified the gauge for $\Phi_1$ and $\Psi_1$.  We will typically fix these gauges by specifying the asymptotics of the gauge potentials at $\xi =0$, or as $\xi \to 1$.   

It is important to note that for the superstratum (see (\ref{ssres1})) one has $G_{\Phi_1} = 1$ and $G_{\Psi_1} = n$.  Our earlier comments about supersymmetry breaking and $n, \omega > 0$ should be interpreted in this context. The gauge invariant way to express this is to consider the UV limit ($\xi \to 1$) and then the excitation will travel inside the light cone if   $G_{\Phi_1} (\xi=1) > 1$ and $G_{\Psi_1} (\xi=1)= n> 0$. We can only express this as $n, \omega >0$ {\it provided that we have chosen the gauge fields to limit to the superstratum values at $\xi =1$.}  When we make this gauge choice, we will denote the frequency by $\omega_\infty$.

One can always change the action by a total derivative, and it is convenient to do this with the gravitational action and with the Chern-Simons action, by defining
\begin{align}
  \widehat \cL_{\text{gravity}} ~&\equiv~   \cL_{\text{gravity}}   ~+~   \frac{1}{2\, g_0}\, \partial_\xi  \bigg[ \Omega_1 \bigg(  \frac{ \xi\, \Omega_0'}{\Omega_0} + \frac{\xi\,  \Omega_1'}{\Omega_1} +\frac{(1+ \xi^2)}{(1- \xi^2)}  \bigg) \bigg]   \nonumber \\
     ~&=~  \frac{\Omega_1}{8\, g_0\, \xi}\, \bigg[ \,\frac{\Omega_1^2}{\Omega_0^2}\, \bigg(k' + \frac{2\,\xi}{1- \xi^2} \, k\bigg)^2 ~+~  4\, \xi\, \frac{ \Omega_1'}{\Omega_1}  \, \bigg( \frac{ \xi\, \Omega_0'}{\Omega_0}  + \frac{(1+ \xi^2)}{(1- \xi^2)}\bigg)\, \bigg]
    \label{Lgrav-new}
\end{align}
\begin{equation}
  \widehat \cL_{CS} ~\equiv~       \cL_{CS} ~+~\frac1{g_0} \, ( \Phi_1 \Psi_2 ~-~ \Phi_2 \Psi_1)'  
     ~=~  \frac{2}{g_0} \,\qty( \Phi_1 \Psi_2'  ~-~ \Psi_1 \Phi_2')  
 \label{LCS-new}  
\end{equation}
Note that $\cL_{\text{gravity}}$  involves, at most,  first derivatives of metric quantities and that $  \widehat \cL_{CS}$ only involves the derivatives of $\Phi_2$ and $\Psi_2$.

The variation of the Lagrangian with respect to the eleven functions (\ref{functionlist}) yields the equations of motion. Note that the scale, $g_0^{-1}$ appears as a uniform overall factor in this Lagrangian, and so the equations of motion are scale free.  

We note that the standard supersymmetric minimum of the scalar potential corresponds to setting $\nu = \mu_0= \mu_1= \mu_2 =0$, and, as described in Section \ref{ss:sugra-action},  this truncation also contains a family of flat directions: $\nu = \mu_0= 0,  \mu_2= -\mu_1 $.

%%%%%%%%%%%%%%%%%%%%%%%%
\subsection{Integrals of the motion}
\label{ss:Integrals}
%%%%%%%%%%%%%%%%%%%%%%%%

The action given in Section \ref{ss:Action} can be used to obtain eleven second-order differential equations for the functions (\ref{functionlist}).  However, there are three elementary integrals of the motion.

The first comes from the usual  ``Hamiltonian constraint'' that emerges from the {\it four} non-trivial Einstein equations for the three metric functions. Define 
\begin{align}
H ~\equiv~ & \frac{\xi^2 \, (1- \xi^2)}{k} \, \bigg[\,  \frac{1}{\xi\,\Omega_1} \, \big(  \,  \widehat \cL_{\text{gravity}} + \cL_A  + \sqrt{g}\, V -   \cL_\chi -   \cL_m\,\big)  \nonumber\\
& \qquad \qquad\quad -\frac{1}{g_0 }\,  \,\bigg( (\mu_0')^2 + (\mu_1')^2 +(\mu_2')^2  ~+~\frac{1}{4}\,  e^{-2(\mu_0 + \mu_1)}\, \Big( \partial_\xi \Big (\sqrt{1- \xi^2}\, \nu \Big) \Big)^2  \bigg)   \, \bigg]  \,.
\label{Ham-constr}
\end{align}
One can then show, using the Euler-Lagrange equations derived from (\ref{lagrangian_full}),  that this is a constant of the motion.  

The other conserved quantities arise only for the truncation defined in Section \ref{ss:truncation}, where there are no fields that have minimal couplings to $\tilde A^{34}$.  This means that undifferentiated potentials, $\Phi_2$ and $\Psi_2$, only contribute  to the action via the Chern-Simons interaction, $\cL_{CS}$, in (\ref{LCS}), and so the equations of motion for $\Phi_2$ and $\Psi_2$ are total derivatives.  Indeed, replacing $ \cL_{CS}$ by $\widehat \cL_{CS}$  means that the complete action does not contain undifferentiated potentials, $\Phi_2$ and $\Psi_2$.  This leads to two integrals of the motion:
\begin{align}
\cI_1 ~\equiv~  & \frac{e^{4\,  \mu_1}\,(1- \xi^2)\,\Omega_1}{\xi\,\Omega_0^2} \, \big( (1- \xi^2) \, \Psi_2' - k\, \Phi_2' \big)~-~  \big(\omega + 2\, \Phi_1   \big)\, \big(1 - \coeff{1}{4}\, (1-\xi^2 ) \, \nu^2 \big) \,,  \label{Integral1}\\
\cI_2 ~\equiv~ &    \frac{e^{4\,  \mu_1}\, k\,\Omega_1}{\xi\,\Omega_0^2} \, \big( (1- \xi^2) \, \Psi_2' - k\, \Phi_2' \big) ~+~ \frac{e^{4\,  \mu_1}\,\xi \, \Phi_2' }{\Omega_1} ~-~  \big( n  + 2\, \Psi_1 \big)\, \big(1 - \coeff{1}{4}\, (1-\xi^2 ) \, \nu^2 \big) \,.\label{Integral2}
\end{align}

It is also useful to note that at infinity ($\xi =1$) the first of these conserved quantities captures the gauge invariant combination: 
\begin{equation}
\cI_1 ~=~  - (\,\omega + 2\, \Phi_1 ) \big|_{\xi =1}   \,,
    \label{I1_gauge_terms}
\end{equation}
and so if one fixes the gauge of $\Phi_1$ at infinity, this conserved quantity determines the frequency of the solution.

%%%%%%%%%%%%%%%%%%%%%%%%
\subsection{Boundary conditions, gauge fixing and coordinate choices}
\label{ss:BCGFCC}
%%%%%%%%%%%%%%%%%%%%%%%%

To solve the system  we need to fix all the gauge choices and impose boundary conditions.  Our choices of boundary conditions will be strongly influenced by the superstratum solution of Section \ref{ss:superstrata}, and while  there may be  broader options, the goal of the present work is to find examples of non-extremal microstrata, rather than classify more extensive families.  

The bottom line is that because our equations of motion are second order we need to specify two pieces of data for each field.   As is familiar from elementary physics problems, in seeking smooth solutions we are typically going to specify boundary data that requires each field to limit to finite values at both ends of the system:   $\xi =0$ and $\xi=1$.  For some fields, these choices will be influenced by coordinate choices, gauge fixing and the stipulation of the UV fixed point at $\xi=1$.

%%%%%%%%%%%%%
\subsubsection{Boundary conditions and gauge fixing}
\label{ss:BCGF}
%%%%%%%%%%%%%

The overarching condition on our solutions is that they must be smooth.  We will also require the solution to be asymptotic to the standard supersymmetric, AdS vacuum at infinity.  This means that the scalar fields will go to the standard supersymmetric critical point, and so will be required to vanish  at infinity ($\xi =1$), and  metric must limit to that of AdS$_3$ of radius $g_0^{-1}$.  

To be more specific, we take the metric to have the form  (\ref{genmet1}) and require:
\begin{equation}
\Omega_0\,, \Omega_1 ~\to~  \Omega_0^{(\infty)} \,,  \Omega_1^{(\infty)}\,,  \quad k  ~\to~ \frac{\Omega_0^{(\infty)}}{\Omega_1^{(\infty)}} \quad  {\rm as} \quad \xi ~\to~ 1\,,
\label{Omkasympinf}
\end{equation}
for some constants $\Omega_0^{(\infty)}> 0$ and  $\Omega_1^{(\infty)}> 0$.  Observe that the condition on $k$ at infinity is required in order to cancel the leading divergence in the $d\psi^2$ term of the metric.   One can also verify that, having chosen all the scalars to vanish at infinity, the equations of motion imply $|\Omega_0^{(\infty)}| =1$ and we take:
\begin{equation}
\Omega_0^{(\infty)} ~=~ 1 \,.
\label{Om0inf}
\end{equation}
This sets the scale of the metric at infinity to be that of AdS$_3$ of radius  (\ref{RAdSscale}).  Finally, observe that $\Omega_1^{(\infty)}$ can be rescaled by redefining the scale of the time coordinate, $\tau$, and without loss of generality we can take:
\begin{equation}
\Omega_1^{(\infty)} ~=~ 1 \,,
\label{Om1inf}
\end{equation}
and thus the metric at infinity limits to that of  (\ref{AdSvals}).  However, while imposing (\ref{Omkasympinf}) and (\ref{Om0inf}), we are, for technical reasons, {\it not 
going to impose} (\ref{Om1inf}) in our numerical analysis.  We will fix the freedom to re-scale $\tau$ at the origin.

At the other boundary, $\xi =0$, we allow more freedom.   The scalars can, in principle, go to any constant values, although the potential and dynamics constrains half of the scalars to vanish at $\xi =0$.   Interestingly enough, the scalar $ \mu_- $, defined in (\ref{scalar-massive3}), can, and does, limit to a non-zero constant value at $\xi =0$ in some of our solutions.  This is the flat direction in the scalar potential. 

Motivated by fact that  the superstratum caps off in a manner very similar to that of global AdS$_3$, we are going to impose similar AdS-like boundary conditions at the origin.  That is, we require
\begin{equation}
\Omega_0\,, \Omega_1 ~\to~  \Omega_0^{(0)},  \Omega_1^{(0)}\,, \quad k  ~\sim~ \cO(\xi^2)   \quad  {\rm as} \quad \xi ~\to~ 0\,, 
\label{Omkasymp0}
\end{equation}
where $\Omega_0^{(0)}> 0$ and  $\Omega_1^{(0)}> 0$ are  constants.   It is necessary that $k$ vanishes at least as fast as $\xi^2$ at the origin so as to avoid closed time-like curves and conical singularities at $\xi=0$ (because of the $\psi$-periodicity (\ref{psi-period})).  In our numerical analysis, we will fix the freedom to re-scale the time coordinate, $\tau$, by setting:
\begin{equation}
\Omega_1^{(0)} ~=~ 1 \,,
\label{Om10}
\end{equation}
(and we leave $\Omega_1^{(\infty)}$ as the free parameter).

Finally, the gauge potentials will be required to limit to constant values at both  $\xi =0$ and $\xi =1$.   These constants can be shifted uniformly through gauge transformations, and so to fix this gauge invariance we fix the constant values at $\xi =0$:
\begin{equation}
\Phi_1(0) ~=~ \frac{1}{2} \ , \qquad \Phi_2(0) ~=~ \Psi_1(0) ~=~ \Psi_2(0) ~=~ 0\,.
\label{gauge0}
\end{equation}
These choices are, of course, motivated by the superstratum solution  (\ref{gauge_superstrata}) and the form of   (\ref{gauge_ansatz}).    The constant values  at infinity are not necessarily the same as those in (\ref{gauge0}).  Indeed, the dynamics can create a physical electric and magnetic ``voltage'' differences between zero and infinity.  For example, (\ref{ssres1}) shows that for the superstratum there is a magnetic potential difference in $\Psi_2$ between $\xi =0$ and $\xi=1$.

As we noted in Section \ref{ss:Integrals},  constant shifts of $\Phi_2$ and  $\Psi_2$ do not change any of the equations of motion and so  the  constants in  $\Phi_2$ and  $\Psi_2$  can be chosen at will.  We have thus set $\Phi_2(0)$ and $\Psi_2(0)$ to zero in (\ref{gauge0}). 

On the other hand, the gauge fields $\Phi_1$ and  $\Psi_1$ minimally couple to non-vanishing scalar fields, and so their constant values play a role in the dynamics through  the gauge invariant term (\ref{gauge_term}).  By fixing the gauge as in  (\ref{gauge0}), we have thus given the choice of $\omega$ and $n$ a gauge-invariant meaning, with $\omega = 0$ corresponding to the superstratum.   Other choices of $\omega$ encode different, gauge inequivalent physics.
 
The gauge choices, (\ref{gauge0}), were made for convenience in the numerical shooting algorithm. As a result, the frequency, $\omega$, appearing in all our computations will be given in this gauge.  As we have already noted, it is more physically meaningful to fix the constant values of the gauge fields to those of the superstratum {\it at infinity}.  This is because we want the solution to limit, at infinity, to the gravity dual of the supersymmetric ground state of the UV CFT so that the frequency is measured relative to that UV fixed point.  In particular, this means that   the angular momentum terms in the $S^3$ fibration should limit, at infinity, to those of the supertube and so we should use a gauge where the three-dimensional gauge  fields limit to these that descend from the supertube, or superstratum, at infinity.  Thus the physical UV frequency, $\omega_\infty$, of the solution is determined by using this gauge.  We will return to this in Section \ref{ss:higher_orders_perturbation}, where we will make the necessary gauge shifts to compute the physical, UV frequency, $\omega_\infty$ from our numerical results. For future reference, we note that for the superstratum,  (\ref{ssres1}), the gauge fields take the following constant values at $\xi =1$:
\begin{equation}
\Phi_1(1) ~=~ \frac{1}{2} \ , \qquad \Phi_2(1) ~=~ \Psi_1(1) ~=~0\,, \qquad  \Psi_2(1) ~=~ - \frac{\alpha_0^2 }{8} \,.
\label{gauge1}
\end{equation}
%

%%%%%%%%%%%%%
\subsubsection{Residual coordinate choices: boosted frames}
\label{ss:Coords}
%%%%%%%%%%%%%

We must now consider whether we have fixed all the diffeomorphism invariance of the metric.  The ability to redefine $(\xi,\psi)$ has been fixed by making the spatial base flat, fixing the period of $\psi$ according to (\ref{psi-period}), and mapping infinity to $\xi =1$.  We have also used the ability to shift $\tau$ by a function of $\xi$ to remove $d \tau \, d \xi$ terms.  What remains is the possibility of mixing $\psi$ and $\tau$, and since these are both Killing directions this means one can only use constant linear combinations.   We have fixed the scales of $\tau$ and $\psi$ by the choices  (\ref{Om10}) and (\ref{psi-period}), and so this leaves the possible shifts:  $\tau \to \tau + c \psi$ and $\psi \to  \psi + \lambda \tau$ for some constants $c$ and $\lambda$.  The former is no longer an option because it conflicts with the asymptotics behavior in $k$ given in (\ref{Omkasymp0}).  The latter is a  more interesting possibility and we look at it in  more detail.

Consider the metric (\ref{genmet1}) and replace $\psi \to  \psi + \lambda \tau$.  One can now re-complete the square in the first term and re-write the metric in the form:
\begin{equation}
ds_{3}^{2}  ~=~  \RAdSsq \, \bigg[ \, \widehat \Omega_1^{2} \, \bigg(d \tau +   \frac{\hat k}{(1- \hat\xi^{2})} \, d\psi \bigg)^2~-~\,\frac{\widehat \Omega_0^{2}}{(1-\hat \xi^{2} )^{2}} \, \big( d \hat \xi^2 ~+~\hat  \xi^2 \, d \psi^2 \big) \, \bigg] \,,
\label{genmet3}
\end{equation}
where
\begin{equation}
 \frac{\widehat \Omega_1}{\Omega_1}  ~=~  \frac{1}{(1-\xi^{2} )}   \bigg[ \, \big(1+ \lambda k - \xi^{2} \big)^2 -  \lambda^2\,  \xi^{2}\,  \frac{\Omega_0^2}{ \Omega_1^2}\, \bigg]^{\frac{1}{2}}
\label{newOms1}
\end{equation}
and the  new coordinate $\hat \xi$ is defined by integrating
\begin{equation}
\frac{d \hat \xi}{\hat \xi} ~=~  \frac{\widehat \Omega_1}{\Omega_1}  \, \frac{d  \xi}{ \xi} 
\label{newvar}
\end{equation}
The remaining functions are then given by:
\begin{equation}
 \frac{\widehat \Omega_0}{\Omega_0}  ~=~ \frac{\Omega_1}{\widehat  \Omega_1}   \, \frac{\xi}{\hat \xi }   \, \frac{(1- \hat \xi^{2})}{(1- \xi^{2}) }    \,, \qquad   \hat k  ~=~  \frac{\Omega_1^2}{\widehat \Omega_1^2}\,\frac{(1- \hat \xi^{2})}{(1- \xi^{2})^2 }   \, \bigg[ \, k\, \big(1+ \lambda k - \xi^{2} \big) -  \lambda\,\xi^{2}\,  \frac{\Omega_0^2}{ \Omega_1^2}\, \bigg]\,.
\label{newOms2}
\end{equation}
It is evident from  (\ref{newvar})  that $\xi =0$  corresponds to $\hat \xi =0$, and one must then choose the constant of integration in (\ref{newvar}) so that   and  $\xi =1$  corresponds to $\hat \xi =1$, so as to retain the poles at $\hat \xi =1$ in the new metric.

Now suppose that the original metric obeys the asymptotics defined by  (\ref{Omkasympinf}) and (\ref{Om0inf}) and, for simplicity, choose the normalization of the $\tau$ coordinate that leads to (\ref{Om1inf}).  One then finds that as $\xi \to 1$, one has 
\begin{equation}
 \frac{\widehat \Omega_1}{\Omega_1}  ~\sim~ \sqrt{\lambda} \ \,\cO\bigg( \frac{1}{\sqrt{1-\xi^{2}}} \bigg) \,.
\label{newOmsasymp}
\end{equation}
Thus $\widehat \Omega_1$ becomes singular as $\xi \to 1$, except when $\lambda = 0$, which correspond to the identity transformation.  It follows that any such shift in the $\psi$ coordinate is excluded by our boundary conditions, (\ref{Omkasympinf}).

The significance of this for our analysis is that we are going to find solutions in which the natural frequency of oscillation undergoes a shift.  A coordinate change of $\psi \to  \psi + \lambda \tau$ can also change the frequency and so we need to make sure our frequency shifts are not  merely a coordinate artifact. Our boundary conditions  at infinity preclude such  coordinate  transformations and exclude the ``trivial'' frequency-shifted solutions swept out by them.

%%%%%%%%%%%%%%%%%%%%%%%%%%%%%%%%%%%%%
\section{Mass, charge and angular momenta}
\label{sec:Charges}
%%%%%%%%%%%%%%%%%%%%%%%%%%%%%%%%%%%%%

Once we have constructed our new solutions, we  need to read off their masses and charges and determine the extent to which they are non-BPS.  The challenge is that we need to extract black-hole data from an asymptotically-AdS solution, and so we need to understand the  issues in doing that.

%%%%%%%%%%%%%%%%%%%%%%%%
\subsection{Masses from the  five-dimensional black-hole perspective}
\label{ss:5dCharges}
%%%%%%%%%%%%%%%%%%%%%%%%

 It is useful to begin by reviewing how the mass is computed for supertubes and superstrata and see how the data of the asymptotically-flat solutions transitions into the asymptotically-AdS analogs.

%%%%%%%%%%%%%
\subsubsection{The mass for superstrata}
\label{ss:ssmass}
%%%%%%%%%%%%%

 Superstrata  are defined in terms of the six-dimensional metric:
\begin{equation}
d s^2_{6} ~=~ -\frac{2}{\sqrt{\cP}}\,(d v+\beta)\,\Big[d u+\omega + \coeff{1}{2}\, \cF\,(d v+\beta)\Big] ~+~ \sqrt{\cP}\,d s^2_4\,,
 \label{sixmet}
\end{equation}
and for most of the known solutions, the four dimensional spatial metric is simply flat $\mathbb{R}^4$ written in  spheroidal coordinates,  $(r, \theta, \varphi_1,  \varphi_2)$:
 \begin{equation}\label{ds4flat}
 d s^2_4 = \Sigma\, \Bigl(\frac{d r^2}{r^2+a^2}+ d\theta^2\Bigr)+(r^2+a^2)\sin^2\theta\,d\varphi_1^2+r^2 \cos^2\theta\,d\varphi_2^2\,.
\end{equation}
where
\begin{equation}
\Sigma ~\equiv~ r^2 + a^2 \cos^2\theta\,.
\end{equation}
To extract the mass one reduces this metric to five dimensions, which is canonically done \cite{Bena:2016agb} (given the coordinate choices (\ref{uvtyreln})) by defining:
\begin{equation}
 \label{eq:Z3k-2}
Z_3 ~=~ 1- \frac{\cF}{2} \,, \qquad  \mathbf{k} ~=~ \frac{\omega+\beta}{\sqrt{2}} \,, 
\end{equation}
and   completing the squares in the metric so as to write it as a $y$-circle fibered over the five-dimensional space-time asymptotic to $\IR^{4,1}$: 
\begin{equation}
ds_6^2 ~=  -\frac{1}{Z_3\sqrt{\cP} } \, (dt +  \mathbf{k})^2 \,+\, 
\frac{Z_3}{\sqrt{\cP}}\, \left[dy  +\left(1- Z_3^{-1}\right)  (dt +  \mathbf{k}) +\frac{\beta-\omega}{\sqrt{2}} \right]^2   +  \sqrt{\cP} \, ds_4^2(\cB)\,.
\label{sixmet-sqty-2}
\end{equation}
To dimensionally reduce on the $y$-circle and get the five-dimensional Einstein action, one must introduce the proper warp factors and this leads to the five-dimensional metric:
\begin{equation}
ds_5^2 ~=  -(Z_3 \, {\cP})^{-\frac{2}{3}} \, (dt +  \mathbf{k})^2 ~+~(Z_3 \, {\cP})^{\frac{1}{3}}\, ds_4^2(\cB)\,.
\label{five-met}
\end{equation}

At large $r$, one has\footnote{We have dropped the function $Z_4$ because it is sub-leading at infinity.}
\begin{equation}
\cP~\sim~ (Z_1 \, Z_5)  ~\sim~\bigg(1+ \frac{Q_1}{\Sigma} \bigg)  \bigg(1+ \frac{Q_5}{\Sigma}\bigg) ~\sim~ 1 ~+~ \frac{(Q_1 + Q_5)}{r^2}  \,,
   \label{cPexp}
\end{equation}
and 
\begin{equation}
\cF ~\sim~ -  \frac{2\, Q_P}{r^2}  \,.
   \label{cFexp}
\end{equation}
Thus the coefficient of $dt^2$  in the five-dimensional metric leads to:
\begin{equation}
(Z_3 \, {\cP})^{-\frac{2}{3}}   ~\sim~ 1 ~-~  \frac{2}{3}\, \frac{(Q_1 + Q_5+ Q_P)}{r^2}  \,.
   \label{cZ3Pexp}
\end{equation}
which gives the BPS result\footnote{A discussion of the dimension-dependent factors implicit in this result may be found in \cite{Gibbons:2013tqa} or  \cite{Peet:2000hn}.}:
\begin{equation}
M     ~=~ Q_1 + Q_5+ Q_P   \,.
   \label{MBPS}
\end{equation}
%

%%%%%%%%%%%%%
\subsubsection{Asymptotically AdS supertubes and superstrata}
\label{ss:AsympAdS}
%%%%%%%%%%%%%

To get superstrata that are asymptotically AdS, one simply ``drops the $1$'s'' in $Z_1$ and $Z_5$.  To see the effect of this, we first consider the round supertube  \cite{Bena:2015bea,Bena:2016ypk,Bena:2017xbt} and take:
\begin{equation}
 Z_1~=~  1~+~ \frac{Q_1}{\Sigma} \,, \quad  Z_2~=~1~+~ \frac{Q_5}{\Sigma} \,, \qquad \cF ~=~ 0   \,.  
   \label{BSTcharges}
\end{equation}
with the angular momentum and fibration vectors: 
 \begin{equation}
\beta ~=~  \frac{R_y \, a^2}{\sqrt{2}\,\Sigma}\,(\, \sin^2\theta\, d\varphi_1 - \cos^2\theta\,d\varphi_2\,)   \,.
 \label{betadefn}
\end{equation}
\begin{equation}
\omega ~=~ \omega_0 \,, \qquad \omega_0 ~\equiv~  \frac{a^2 \, R_y \, }{ \sqrt{2}\,\Sigma}\,  (\sin^2 \theta  d \varphi_1 + \cos^2 \theta \,  d \varphi_2 ) \,.
\label{angmom0}
\end{equation}
Smoothness at the supertube requires:
\begin{equation}
 Q_1 \, Q_5  ~=~  R_y^2 \, a^2  \,.  
   \label{stsmooth}
\end{equation}

After dropping the $1$'s in (\ref{BSTcharges}), the  metric (\ref{sixmet}) can then be recast as:
\begin{equation}
\begin{aligned}
d s^2_{6} ~=~  \sqrt{Q_1 Q_5}\, \bigg[  \, &   -\frac{ (r^2+ a^2)}{a^2 \, R_y^2 }\,dt^2 ~+~  \frac{dr^2}{(r^2+ a^2)} ~+~  \frac{ r^2}{a^2 \, R_y^2  }\,dy^2\\
&+ ~ d\theta^2 ~+~ \sin^2 \theta \, \bigg(d\varphi_1 - \frac{1}{R_y }\,dt\bigg)^2~+~ \cos^2 \theta \,\bigg(d\varphi_2 - \frac{1}{R_y }\,dy\bigg)^2   \bigg] \,.
\end{aligned}
\label{STmet}
\end{equation}

The important point here is that in the asymptotically-flat metric the mixed $d \varphi_j du$ and $d \varphi_j dv$ terms are folded into the $(du,dv)$ terms to define the $v$-fibration and the angular-momentum.  In the asymptotically-AdS metric, these cross-terms are folded into fibering and boosting the $S^3$ metric, leaving the intrinsic three-dimensional metric in  $(du,dv, dr)$.  Thus the $S^3$ of the consistent truncation is a boosted, fibered form of the $S^3$ at infinity in the five-dimensional space-time.  Since this re-writing of the metric involves a modification of the $dt^2$ and $dy^2$ terms, it will  shift the spectrum and energies of states.  This has a simple and well known interpretation in terms of the holographic theory:  the compactification on $S^3$ to AdS$_3$ naturally selects the NS vacuum, but to realize the geometry smoothly in an asymptotically-flat space one is required to do a spectral flow to the Ramond sector.  This spectral flow is implemented by the seemingly trivial shifts of $d\varphi_1$ and $d\varphi_2$ in (\ref{STmet}).

The situation for a superstratum, with a non-vanishing momentum charge, $Q_P$,  is rather more complicated. First, the smoothness relation (\ref{stsmooth}) is modified by by the superstratum modes in $\cF$, as in (\ref{ss-smoothness}).  In addition,  the  ``coiffuring conditions'' are slightly different for asymptotically flat and asymptotically AdS solutions \cite{Bena:2017xbt}.  Furthermore, the fibration and boosting of the $S^3$ now involves non-trivial Kaluza-Klein Maxwell fields, $\tilde A^{12}$ and $\tilde A^{34}$.   One could look at the asymptotics of these solutions and the fibrations and undo the asymptotic shifts and connect the compactification $S^3$ back to the $S^3$ at infinity in five dimensions, but there is a simpler way forward. One can use the superstratum as a baseline:   the superstratum is BPS, and so it represents the lowest mass state in a given charge sector with a given smooth topology.  Our new solutions are obtained as ``large'' deformations of the superstratum and so their ``mass above extremality'' can be determined by comparing the three-dimensional mass to the mass of a superstratum with the same charges.  So we need to look at the definition of mass in three dimensions.

%%%%%%%%%%%%%%%%%%%%%%%%
\subsection{The BTZ perspective}
\label{ss:3dCharges}
%%%%%%%%%%%%%%%%%%%%%%%%

The BTZ black hole has a metric of the form (see, for example, \cite{Peet:2000hn}):
\begin{equation}
d s^2_{3} ~=~ \frac{(\rho^2 - \rho_+^2)  (\rho^2 - \rho_-^2) }{\ell^2 \, \rho^2}  \, dt^2 ~-~ \frac{\ell^2 \, \rho^2}  {(\rho^2 - \rho_+^2)  (\rho^2 - \rho_-^2) }\, d\rho^2 ~-~\rho^2 \, \bigg(d \varphi ~+~\frac{ \rho_+ \, \rho_- }{\ell \, \rho^2}  \,  dt  \bigg)^2  \,.
\label{BTZmet}
\end{equation}
The horizons are at $\rho = r_\pm$ and the mass and angular momenta are\footnote{We have chosen to divide the angular momentum by an extra factor of $\ell$ so as to give it the same dimensions as the mass.  To get the more usual conventions one should replace $J$ here by $ \frac{J}{\ell }$.}
\begin{equation}
M  ~=~ \frac{(\rho_+^2 + \rho_-^2) }{8\, G_3 \ell^2}  \,, \qquad J  ~=~ \frac{\rho_+\, \rho_- }{4\, G_3 \ell^2}    \,,
\label{BTZmetM}
\end{equation}
and extremality corresponds to 
\begin{equation}
  \rho_+ ~=~ \rho_-    \quad \Leftrightarrow \quad M  ~=~  J\,.
\label{BTZmetrhos}
\end{equation}
A canonical choice is to take:
\begin{equation}
G_3 ~=~    \frac{1}{8 } \,.
\label{G3canon}
\end{equation}
and we convert to dimensionless variables:
\begin{equation}
\hat \rho ~\equiv~  \frac{\rho}{\ell} \,, \qquad   \hat \tau ~\equiv~  \frac{t}{\ell}\,.
\label{hatrho}
\end{equation}
At large $\hat \rho$ one then has:
\begin{equation}
d s^2_{3} ~\sim~\ell^2\, \bigg[\, \hat \rho^2  \bigg(1  - \frac{M}{\hat \rho^2}   + \cO(\hat \rho^{-2}) \bigg)   \, d\hat \tau^2 ~-~ \bigg(1  + \frac{M}{\hat \rho^2}  + \cO(\hat \rho^{-2}) \bigg)   \, \, \frac{d\hat \rho^2}{\hat \rho^2} ~-~\hat \rho^2 \,d\varphi^2 ~-~ J \, d \varphi \, d\hat \tau  \, \bigg] \,.
\label{BTZasympmet}
\end{equation}

From the BTZ perspective one sees that global AdS has a mass, $M =-1$ and $J=0$.  This reflects the energy of the NS vacuum and the fact that the BTZ black hole has no knowledge of the spatial angular momentum of the supertube and the flow to the Ramond sector.   

Extracting the details of the superstratum from the BTZ limit is a little more subtle and complicated.   One starts with the metric (\ref{genmet1}) taking $\RAdS =\ell =g_0^{-1}$ and uses the superstratum data:   (\ref{Omk_superstrata}) and (\ref{singlemode1}).   First one must define $\hat \rho$ so that the coefficient of $d \psi^2$ is exactly $\hat  \rho^2$.  Then one must complete  the square so as to absorb the $\hat  \rho^2  d \psi dt$ terms into a term of the form $ \hat  \rho^2  d \varphi^2  \equiv \hat  \rho^2  (d \psi + c dt)^2$, for some constant, $c$. One then collects the $d\tau ^2$ and $dr^2$ terms and rewrites them in terms of $\hat \rho$.  Finally one must rescale $\tau$ to $\hat \tau$ so that the leading term is  $\hat \rho^2  d \hat \tau^2$.  Doing this yields:
\begin{equation}
d s^2_{3} ~\sim~  \frac{1}{g_0^2}\, \bigg[\, \rho^2 \, \bigg(1  -  \frac{ \widetilde M}{\rho^2}   + \cO(\rho^{-2}) \bigg)   \, d\hat \tau^2 ~-~ \bigg(1  + \frac{ M}{\rho^2}  + \cO(\rho^{-2}) \bigg)   \, \, \frac{d\rho^2}{\rho^2} ~-~\rho^2 \, d \varphi^2  ~-~J  \, d \varphi \, d\hat \tau \, \bigg]\,,
\label{SSBTZmet}
\end{equation}
where
\begin{equation}
\hat \tau  ~\equiv~    \frac{\tau}{1 - \frac{1}{4} \, \alpha_0^2}  \,, \qquad \varphi  ~\equiv~    \psi -  \hat \tau     \,,
\label{hattaudefn}
\end{equation}
and 
\begin{equation}
\widetilde M ~\equiv~ - \bigg(1 - \frac{1}{4} \, \alpha_0^2  \bigg)^2 +  \frac{1}{2} \, n \, \alpha_0^2  \,, \qquad  M ~\equiv~   - 1 +  \frac{1}{2} \, n \, \alpha_0^2 +  \frac{1}{16} \, \alpha_0^4    \,, \qquad  J ~\equiv~ \frac{1}{2} \, n \, \alpha_0^2\,.
 \label{MMJdefn}
\end{equation}

The issue now is whether $M$ or $\widetilde M$ is the mass.   General Relativity tells us that the mass can be determined through the time-like Killing vectors, and hence from the expansion of $g_{tt}$.  This would also agree with the use of uplifts to six dimensions and then coupling to flat space.  From this perspective, the mass should be determined from $\widetilde M$.  We note that
\begin{equation}
 \widetilde M ~-~ |J|  ~=~   - \bigg(1 - \frac{1}{4} \, \alpha_0^2  \bigg)^2 ~=~   - \bigg(1 - \frac{J}{2\,n }  \bigg)^2     \,,
 \label{BPSbound-0}
\end{equation}
which, given  (\ref{alpha0bound}), is always negative and only vanishes at the  ``extremal BTZ limit:'' $| \alpha_0| =2$.

On the other hand, holographic renormalization in asymptotically-AdS$_3$  suggests  \cite{Balasubramanian:1999re} that the mass is to be read off from $g^{\rho \rho}$  and is therefore given by $M$.  One therefore has: 
\begin{equation}
 M ~-~ |J|  ~=~   - 1 ~+~  \frac{1}{16} \, \alpha_0^4 ~=~   - 1 ~+~ \bigg( \frac{J}{2\,n } \bigg)^2   \,.
 \label{BPSbound-1}
\end{equation}
and again, given  (\ref{alpha0bound}), this is always negative.  

The curve, (\ref {BPSbound-0}) for $\widetilde M - |J|$, and the curve, (\ref {BPSbound-1}) for $M - |J|$, considered as a function of $J$, are parabolae   that intersect at $J=0$ and  at $J=2 n$. Because these curves are defined for  superstrata, they should provide  BPS bounds.  We will remain agnostic about which represents the true mass, and we will compute both quantities for microstrata.  However,  we note here that $\widetilde M - |J|$ is based on a time-like Killing vector and so would seem better adapted to uplifting and coupling to flat space.

%%%%%%%%%%%%%%%%%%%%%%%%%%%%%%%%%%%%%
\section{Perturbative analysis}
\label{sec:Perturbative}
%%%%%%%%%%%%%%%%%%%%%%%%%%%%%%%%%%%%%

A very important aspect of the simplifications afforded by the Ansatz described in  Section \ref{ss:simpAnsatz} is that perturbation theory can be used to construct solutions to surprisingly high order.  In this section we will perform this perturbative analysis.  This will enable us to anticipate many of the features we find in the numerical analysis, and to confirm that our numerical methods are, in fact, converging on solutions to the system of equations.  Perhaps the two most significant  new features we uncover are the frequency shifts and the  emergence of non-normalizable modes.  Specifically,  we find microstata have normal modes whose frequencies depend on the amplitudes of the fields and we find that the field, $\nu$, develops a non-normalizable component at third order in perturbations.  

{\bf Note:} For simplicity we are henceforth going to restrict our study to the families of solutions with $n=2$ in  (\ref{singlemode1}) and (\ref{singlemode2}).    We choose this mode number  because lower mode numbers in superstrata have large and more extended bump functions. However, we choose $n=2$ rather than $n=1$ because we want the bump functions to vanish more strongly in the cap, leading to something approaching AdS geometry in the cap.   We will also focus primarily on modes with $\omega \approx 0$ and $\omega \approx 2$, but we will also present several results for  $\omega \approx 4$.

%%%%%%%%%%%%%%%%%%%%%%%%
\subsection{Setting up the perturbation theory}
\label{ss:perturbations}
%%%%%%%%%%%%%%%%%%%%%%%%

We will perturb around the global AdS$_3$  vacuum of the theory.  That is we take all the scalars and electromagnetic fields to vanish and the metric to be that of global AdS$_3$.  Specifically, one can think of the unperturbed background as that of Section \ref{ss:superstrata}  with $\chi_1=\chi_2=0$, of the solution  (\ref{ssres1})  with $\alpha_0 =0$.  Thus the  vacuum solution is given by
\begin{equation}
\begin{aligned}
    \nu^{AdS} ~&=~ 0 \ ,  \qquad\qquad \mu_0^{AdS} ~=~ \mu_1^{AdS} ~=~ \mu_2^{AdS} ~=~ 0 \ , \\
    \Phi_1^{AdS} ~&=~  \frac{1}{2} \ , \qquad\qquad \Phi_2^{AdS} ~=~ \Psi_1^{AdS} ~=~ \Psi_2^{AdS} ~=~ 0 \ ,\\
    \Omega_0^{AdS} ~&=~ \Omega_1^{AdS} ~=~ 1 \ , \qquad\qquad k^{AdS} ~=~ \xi^2 \ .
\end{aligned}
\label{AdSvac}
\end{equation}
We then introduce a small parameter, $\varepsilon$, and expand every field, ${\cal X}$, in  (\ref{functionlist}). as
\begin{equation}
{\cal X}~=~ {\cal X}^{AdS} \,+\, \varepsilon\, \delta {\cal X} \,+\, \varepsilon^2\, \delta^2{\cal X} \,+ \dots
\label{perturbativeExpansionFields}
\end{equation}
We will also expand the frequency in powers of $\varepsilon$~:
\begin{equation}
    \omega = \omega_0 \,+\, \varepsilon \, \delta \omega \,+\, \varepsilon^2 \, \delta^2 \omega  \,+ \dots \,.
   \label{omexp1}
\end{equation}

At linear order, the equations of motion naturally diagonalize on the fields:
\begin{align}
    \mu_+ &~=~ \mu_1 + \mu_2 \ , & \Phi_+ &~=~ \Phi_1 + \Phi_2 \ , & \Psi_+ &~=~ \Psi_1 + \Psi_2 \ , \\
    \mu_- &~=~ \mu_1 - \mu_2 \ , & \Phi_- &~=~ \Phi_1 - \Phi_2 \ , & \Psi_- &~=~ \Psi_1 - \Psi_2 \ .
\end{align}
The perturbation theory is also easier to organize in this basis. 

In order to find a unique solution, we need to fix the gauges and the asymptotics of the perturbed solution. As explained in Section \ref{ss:BCGFCC}, the geometry and the gauge fields are, at every order in perturbation, required to satisfy the following:

\begin{equation}
\begin{aligned}
    \Phi_1(0) ~&=~ \frac{1}{2} \ , \qquad \Phi_2(0) ~=~ \Psi_1(0) ~=~ \Psi_2(0) ~=~ 0\ , \\
    \Omega_0(1) ~&=~ \Omega_1(0) ~=~ 1 \ , \quad k(\xi\to 0) ~=~ \mathcal{O}(\xi^2) \ ,\quad k(1) = \Omega_1(1)^{-1} \ .
\end{aligned}
\label{gauge_fix}
\end{equation}

Note that we could have  chosen to fix the values of the gauge fields to any constant values at the origin.   As we have already remarked, shifting  $\Phi_2$ or $\Psi_2$ by a constant has no effect on the dynamics.     The explicit undifferentiated potentials,  $\Phi_1$ or $\Psi_1$,   only appear in (\ref{gauge_term}), where it is evident that their constant asymptotic values are gauge equivalent to shifts in $\omega$ and $n$.  As we will see, the value of $(n + 2\,\Psi_1)$ is easily fixed by regularity of series expansion about the origin.  The value of $(\omega + 2\, \Phi_1)$ is also fixed by regularity, but, as we will discuss in detail, this is a much more subtle phenomenon involving smoothness at both the origin and infinity. 

Since we are going to use the superstratum as a reference point, we use the asymptotic values of $\Phi_1$ or $\Psi_1$ for the superstratum to fix gauges.  This then gives an appropriately gauged fixed meaning to  the values of $n$ and $\omega$.   That is, the reference superstratum has (\ref{gauge_fix}) and $n=2$, $\omega=0$, and the microstrata will obey (\ref{gauge_fix}) with  $n=2$ and an $\omega \ne 0$.

Finally, we require that the solution in the UV goes to the supersymmetric critical point of the potential. This means that we impose:
\begin{equation}
    \mu_0(1) ~=~ \mu_1(1) ~=~ \mu_2(1) ~=~ 0 \,,  \qquad { \rm and}  \qquad \sqrt{1- \xi^2}\   \nu(\xi)\ ~\to~ 0 \,, \ {\rm as} \ \xi \to 1  \,.
    \label{susy_point_scalars}
\end{equation}
One should recall that $\nu$ is related to the  supergravity scalars via (\ref{singlemode1}), and hence  $|\chi(\xi)| \sim  \sqrt{1- \xi^2}\,   \nu(\xi) \to 0$ as $\xi \to 1$.

One should also note that requiring the $  \mu_j$ to vanish at infinity means that we are restricting them to their normalizable modes, and hence only allowing them to deform the state of the holographic system. 
 
%%%%%%%%%%%%%%%%%%%%%%%%
\subsection{Overview of the results}
\label{ss:overview}
%%%%%%%%%%%%%%%%%%%%%%%%

Our purpose here is three-fold.  First, we wish to evolve perturbative solutions as far as practicable to provide tests for our numerical solutions.  Second, we wish to see how the frequencies, $\omega$,  of the normal modes depend on the amplitudes of the fields.  This will be done by ensuring regularity of the perturbative solution, order by order, at $\xi=0$ and $\xi=1$  and using this to determine the expansion in  (\ref{omexp1}). We also want to track higher-order non-analytic terms,  the logs and poly-logs, as they appear in the solution.  These terms do not  affect the boundary conditions, but they  are  significant for holography and can affect the convergence of our numerics.  

Despite the complexity of the action defined in Section \ref{ss:Action}, it is remarkably straightforward to organize the perturbation theory.   The linearized analysis leads to a ``fundamental linear differential operator'' for each field in (\ref{functionlist}).   These differential operators are all hypergeometric operators, (\ref{scalar-massive2}), inherited from the linearized theory in the global AdS$_3$ background, and, as one would expect, at linear order these equations are all homogeneous and lead to  solutions written in terms of hypergeometric functions.  Imposing boundary conditions and smoothness also leads to oscillations that have the normal mode frequencies of global AdS.  As we will discuss in Section \ref{ss:linear}, this leads us to start the expansion in (\ref{omexp1}) with  even integers:
\begin{equation}
 \omega_0 ~=~ 0\,,  \ \pm 2\,, \  \pm 4\,, \dots 
   \label{omzero}
\end{equation}
and, as we commented earlier, we will focus on $\omega_0 =  0\,,  2\,, 4\,$.

At second and higher orders, the same linear differential operators appear with their corresponding functions from (\ref{functionlist}) and all the complicated non-linearities of the action  in Section \ref{ss:Action} appear only in the source terms.  That is, order by order, one only has to solve {\it linear} differential equations whose sources are made out of (non-linear) combinations of lower order solutions.  One then finds particular solutions to these equations and  adds homogeneous solutions  so that the final result does not diverge at $\xi=0$ or $\xi=1$, and obeys the conditions (\ref{gauge_fix}) and (\ref{susy_point_scalars}).  These conditions  lead to constraints on the frequency $\omega$, and to the non-trivial expansion (\ref{omexp1}).

In this process we find families of solutions that depend on three parameters:  the value of $\omega_0$ from  (\ref{omzero}), and two apparently independent continuum parameters, $\alpha$ and $\beta$. The parameter  $\alpha$ is analogous to  $\alpha_0$ in superstrata, and determines the overall scale of the scalar field, $\nu$.  The parameter $\beta$ determines the overall scale of $\mu_0$.  This second scalar vanishes identically in the superstratum but seems to be an independent degree of freedom in the microstratum.  It is possible that  $\alpha$ and $\beta$ become linked in some way at higher orders in perturbation theory, but to the order we have computed, these parameters remain independent. 

At second order in perturbations,  we find that the first-order shift in the frequency, $\delta\omega$, must be zero and it is only at third order  that we find a non-trivial frequency-shift. Indeed, we find an expression for $\delta^2 \omega$ as a quadratic in $\alpha$ and $\beta$.  In Section \ref{ss:normal} we will see that this expression  matches very well with our numerical results.

%%%%%%%%%%%%%%%%%%%%%%%%
\subsection{Linear perturbations}
\label{ss:linear}
%%%%%%%%%%%%%%%%%%%%%%%%

%%%%%%%%%%%%%
\subsubsection{The scalar fields}
\label{ss:scalarops}
%%%%%%%%%%%%%

The equation of motion for the scalar $\nu$ is given by
\begin{equation}
    \frac1\xi \, \partial_\xi \qty(\xi\, \qty(1-\xi^2)\, \delta\nu')  - \left(\frac{4}{\xi^2}- (\omega_0 + 2)(\omega_0 + 4) \, \right)\delta \nu ~=~ 0\,,
\end{equation}
or, equivalently:
\begin{equation}
    \frac1\xi \, \partial_\xi \qty(\xi\,  \delta \hat \nu') ~+~  \frac{ ( \omega_0 + 3)^2 }{(1-\xi^2) }  \, \delta \hat\nu ~- \frac{4 }{\xi^2 \,(1-\xi^2) } \, \delta \hat\nu ~+~ \frac{1}{(1-\xi^2)^2}  \,\delta \hat\nu  ~=~ 0\,,
  \label{nuhateqn}
\end{equation}
where  $\hat \nu =|\chi|$ is defined in (\ref{hatnudefn}).  This is precisely of the form  (\ref{scalar-massive2}) with $\Delta =1$.   The shift from $(\hat \omega + \hat n) =(\hat \omega + 2)$ to  $(\omega_0+ 3)$ occurs because we have a non-trivial gauge field, $\Phi_1$, in (\ref{gauge_fix}).
 
The linearized equations of the other scalars are:
\begin{spreadlines}{1em}% tweak
\begin{align}
    \frac1\xi \,\partial_\xi \qty(\xi\, \delta\mu_0') ~+~  \frac{ 4\, ( \omega_0 + 3)^2 }{(1-\xi^2) }  \, \delta\mu_0  ~- \frac{16}{\xi^2 \,(1-\xi^2) } \, \delta\mu_0    ~=~ 0 \\
    \frac1\xi \,\partial_\xi \qty(\xi\, \delta\mu_+') \,-\, \frac{8}{(1-\xi^2)^2} \delta\mu_+ ~=~ 0 \,, \qquad 
    \frac1\xi \,\partial_\xi \qty(\xi\, \delta\mu_-') ~=~ 0
\end{align}
\end{spreadlines}
Again these are of the form (\ref{scalar-massive2}).  The scalar, $\mu_0$, has $\Delta =2$ and, as noted in  (\ref{singlemode2}), it  oscillates with twice the frequency and mode numbers of $\nu$, which accounts for the extra factors of $4$ compared to  (\ref{nuhateqn}).   The scalars $\mu_+$ and $\mu_-$ only depend on $r$ and have $\Delta = 4$ and $\Delta = 2$ respectively.

The solutions for $\delta\mu_\pm$ are elementary:
\begin{equation}
  \delta\mu_+   ~=~   \frac{1}{(1-\xi^2) } \, \Big[ \, c_1 \, \big(1+\xi^2\big)~+~c_2 \, \big(2 +(1+\xi^2) \,\log \xi \big)\,\Big] \,, \qquad    \delta\mu_-   ~=~  c_3  ~+~c_4 \,\log \xi   \,,
  \label{mupmsol}
\end{equation}
for some constants $c_j$.  Regularity at $\xi= 0$ and  $\xi= 1$ implies $c_1 = c_2 = c_4 =0$, and the boundary conditions (\ref{susy_point_scalars}) means that $c_3 =0$.   So these scalars are trivial at this order.

The smooth solution for $\delta\nu$  can be written:
\begin{equation}
    \delta\nu ~\sim~    {}_2F_1\qty(3 + \frac{\omega_0}2,\, -\frac{\omega_0}2,\, 3 \,;\, \xi^2)\, \xi^2 \,.
 \label{nuseed}
\end{equation}
We also note that there are also solutions of the form 
\begin{equation}
    \delta\nu ~\sim~  {}_2F_1\qty(3 + \frac{\omega_0}2,\, -\frac{\omega_0}2,\, 3 \,;\, \xi^2) \,  \xi^2\  \log\bigg(\frac{\xi^2}{1- \xi^2}\bigg) ~+~\xi^{-2} \, P_{\omega_0}(\xi)  \,,
 \label{nuhomdiv}
\end{equation}
for some fully determined polynomial, $P_{\omega_0}(\xi)$.  This solution still leads to  $\hat \nu \to 0$ as  $\xi=1$, but it diverges as $\xi^{-2}$ as $\xi \to 0$. Its significance to our subsequent discussion is that it has the ``non-normalizable'' asymptotics at infinity: 
\begin{equation}
    \delta \hat  \nu ~=~   \sqrt{1- \xi^2}  \,  \delta   \nu  ~\sim~   \sqrt{1- \xi^2}    \log({1- \xi^2}) ~\to~ 0   \qquad {\rm as} \ \ \xi \to 1 \,,
 \label{nuhomdiv2}
\end{equation}
but it cannot be used because it is singular at the origin.

\begin{figure}[ht!]
\centering
\includegraphics[width=.75\textwidth]{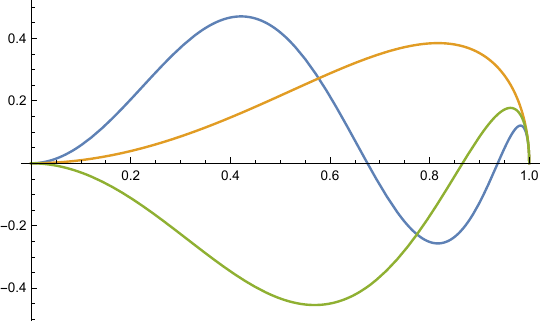}
\caption{\it Plots of $\delta\nu$ as given in (\ref{nuseedex}) with $\omega_0 =0, 2, 4$ and $\alpha =1$. As one would expect, the number of nodes  increases with frequency, and this can be used to distinguish the individual functions. }
\label{fig:nuharms}
\end{figure}

The solution (\ref{nuseed}) vanishes at $\xi=0$ and is generically log-divergent at $\xi=1$, as in (\ref{nuasymp}), and so $\hat \nu$ will vanish at $\xi=1$.  However, we want to start our solutions from the ``normalizable'' mode in the AdS$_3$, and so we need to avoid such log terms.   This means that we must choose $\omega_0$  to be an even integer,  as in (\ref{omzero}), and for $\omega_0 \in 2 \ZZ$ and $\omega_0 \ge 0$, these  solutions are all polynomial\footnote{They are actually the Jacobi polynomial of class $(2,0)$ and of order $\omega_0 / 2$, which is an integer since $\omega_0$ is even.}.  For $\omega_0 =0, 2, 4$ we will use:
\begin{equation}
    \delta\nu ~=~ \alpha\,   \xi^2 \,, \qquad      \delta\nu ~=~ \alpha\,   \xi^2\,( 4 \xi^2-3 ) \,, \qquad     \delta\nu ~=~   \alpha\,   \xi^2\,(15\,\xi^4 - 20\, \xi^2+6) \,,
  \label{nuseedex}
\end{equation}
for some constant parameter, $\alpha$.  Note that these solutions vanish at the origin and that $\hat \nu$ vanishes at $\xi=1$.  We have plotted these functions in Fig.\!~\ref{fig:nuharms} for later comparison with the perturbative corrections and numerical results. 

%%%%%%%%%%%%%%%%%%
\vspace{0.5mm}
\vspace{1mm}
%%%%%%%%%%%%%%%%%%

The story for $\delta \mu_0$ is very similar.  The values of $\omega_0$ are already fixed to be even integers and the smooth solutions are hypergeometric polynomials:
\begin{equation}
  \delta\mu_0    ~\sim~  {}_2F_1\qty(5+\omega_0,\, -1 -\omega_0,\, 5 \,;\, \xi^2)\, \xi^4  \,,
  \label{mu0seed}
\end{equation}
and  for $\omega_0 =0, 2$ we will use:
\begin{equation}
  \delta\mu_0 ~=~ \beta\,   \xi^4\,( 1- \xi^2 )  \,, \qquad      \delta\mu_0   ~=~  \frac{\beta}{5}\, \xi^4\,( 1- \xi^2 )  \, (12\xi^4  - 16 \xi^2+5) \,.
 \label{mu0seedex}  
\end{equation}
These solutions vanish at both $\xi=0$ and $\xi=1$.  
It is interesting to note that if $\nu \equiv 0$, or if one allows ``non-normalizable'' modes in $\nu$,  then $\mu_0$ is also polynomial for $\omega_0 \in \ZZ$.   

Our solutions involve the two free parameters, $\alpha$ and $\beta$, which appear independently at linear order.  These parameters also appear to remain independent at higher orders in perturbations. 

%%%%%%%%%%%%%
\subsubsection{The Maxwell potentials }
\label{ss:Maxops}
%%%%%%%%%%%%%

The equations for the gauge fields are similar to those of the scalar fields.  We find:
\begin{spreadlines}{1em}% tweak
\begin{align}
& \partial_\xi \qty( \xi \delta\Phi'_+)  ~=~ 0  \,, \qquad 
 \partial_\xi \qty(\frac{1-\xi^2}{2\,\xi} \delta\Psi'_+ \,-\, \delta\Psi_+ \,-\, \delta\Phi_+) ~=~ 0 \\
& \partial_\xi \qty(\frac{1-\xi^2}{2\, \xi} \delta\Psi'_- \,+\, \delta\Psi_- \,+\, \delta\Phi_-) ~=~ 0 \,, \qquad
    \partial_\xi \qty( \xi \delta\Phi'_- ) \,+\, 4 \, \delta\Psi'_- - 4\, \frac{\xi^2}{1-\xi^2}\, \delta\Phi'_-  ~=~ 0 \,.
\end{align}
\end{spreadlines}
The first two equations are trivially integrated to yield
\begin{equation}
    \delta\Phi_+  ~=~ c_5 \,+\, c_6\, \log\xi  \,, \qquad    \delta\Psi_+ ~=~ \frac{1}{1-\xi^2} \,\qty( c_7 \,+\, c_8\, \xi^2 \, +\, c_6\, \xi^2 \log\xi) ,
\end{equation}
for some constants $c_5$, $c_6$, $c_7$ and $c_8$.    One can eliminate $\delta\Phi_-$ from the fourth equation using the third equation to arrive at a third order equation for $\delta\Psi_-$.  This is trivially integrated to yield:
\begin{equation}
 \delta\Psi_- ~=~  \frac{1}{1-\xi^2} \,\qty(c_9 \,+\, c_{10}\, \xi^2 \,+\, c_{11}\, \xi^2 \log\xi ),
\end{equation}
for some constants $c_9$, $c_{10}$ and $c_{11}$. The third equation can then be used to determine the last gauge field:
\begin{equation}
 \delta\Phi_- ~=~ c_{12} \,-\, \frac{1}{1-\xi^2} \,\qty(2\, (c_9 \,+\, c_{10}) \,\xi^2 \,+\, c_{11}\, \qty(1 + \xi^2) \log\xi ),
\end{equation}
for some constant $c_{12}$.

The only smooth solutions are constant gauge potentials, and the conditions (\ref{gauge_fix}) then imply that the constants $c_5$ to $c_{12}$ all vanish. The gauge fields are then trivial at linear order.

%%%%%%%%%%%%%
\subsubsection{The metric functions }
\label{ss:Metops}
%%%%%%%%%%%%%

The equations for the metric functions can be reduced to a simple ``upper triangular'' linear system that can be solved successively:
\begin{spreadlines}{1em}% tweak
\begin{align}
    \partial_\xi \qty(\xi \,\qty(1-\xi^2)\, \delta\Omega_1' \,-\, 2 \, \delta\Omega_1) ~&=~ 0 \,,  \\
    \partial_\xi \qty( \xi \delta\Omega_0') \,-\, 8 \frac{\xi}{\qty(1-\xi^2)^2}\, \delta\Omega_0  ~&=~ -3 \, \frac{1+\xi^2}{1-\xi^2} \, \delta\Omega_1'  \,,  \\
    \partial_\xi \qty( \xi^{-1}\, \delta k')  ~&=~  \frac{1}{1-\xi^2} \,\qty( 4 \, \delta\Omega_0' \,-\, 6 \, \delta\Omega_1') \,.
\end{align}
\end{spreadlines}

The solution for $ \delta\Omega_1$ is 
\begin{equation}
 \delta\Omega_1  ~=~ c_{13} ~+~ \frac{c_{14} }{1-\xi^2}\,.
\end{equation}
The only smooth solution is the constant function, and (\ref{Om1inf}) means this must vanish. 

The equation for  $\delta\Omega_0$ becomes the same as the equation for  $\mu_+$, and the solution is given in  (\ref{mupmsol}).  Again the only smooth solution is the constant solution and this is set to zero by the boundary condition (\ref{Om0inf}).

The equation for  $\delta k $ now leads to
\begin{equation}
 \delta k ~=~ c_{15} ~+~ c_{16} \, \xi^2\,.
\end{equation}
The constant term is eliminated by the boundary condition at the origin, (\ref{Omkasymp0}), and while the term involving $c_{16}$ is consistent with (\ref{Omkasymp0}), this term is eliminated by (\ref{gauge_fix}).

%%%%%%%%%%%%%
\subsubsection{Summary of linear solution }
\label{ss:linsum}
%%%%%%%%%%%%%

At linear order, the only non-vanishing perturbation involves arbitrary linear combinations, with coefficients $\alpha$ and $\beta$,  of the two  ``seed solutions,''  (\ref{nuseed}) and (\ref{mu0seed}).  We will normalize these seeds in a somewhat arbitrary manner by taking $\alpha$ to be the value of $\delta\nu$ at $\xi = 1$, and $\beta$ to be the coefficient  of the $\xi^4$ term in the expansion of $\mu_0$ at $\xi = 0$.  Indeed, we have adopted this normalization in  (\ref{nuseedex}) and  (\ref{mu0seedex}).
The parameters, $\alpha$ and $\beta$, will also be taken to be the small parameters of the perturbation expansion. This makes $\epsilon$  redundant, and so we set $\epsilon = 1$.

We will see from our perturbative analysis that these seeds remain independent to very high orders and so lead to a two-parameter family of microstrata.

%%%%%%%%%%%%%%%%%%%%%%%%
\subsection{Results from higher orders  perturbation theory}
\label{ss:higher_orders_perturbation}
%%%%%%%%%%%%%%%%%%%%%%%%

We now use the zeroth order AdS vacuum solution (\ref{AdSvac}) and the linear seed solutions,  (\ref{nuseed}), (\ref{mu0seed}),  (\ref{nuseedex}) and  (\ref{mu0seedex}), as a starting point for the perturbative analysis for the equations of motion that follow from the action given in Section \ref{ss:Action}.   Our presentation  will focus primarily on the scalars. 

As we have remarked, at each order, we have the linear equations of Section \ref{ss:linear} sourced by the solutions at earlier orders.   As is typical in this sort of problem, we will  use the homogeneous linear solutions to adjust particular solutions  to remove divergencies at $\xi = 0$ or $\xi = 1$.  

%%%%%%%%%%%%%
\subsubsection{Overview}
%%%%%%%%%%%%%

There are two interesting physical phenomena that we wish to track.  The first is the shift in the ``resonant'' frequency, $\omega$, of the normal modes.   As one would expect, the frequency   depends on the smoothness at the origin and the boundary conditions at infinity.  We will typically impose the boundary conditions that drive the solution to the standard supersymmetric vacuum at infinity but one should note that other choices are possible and this leads to slightly different shifts in frequency.   These choices may well have significant implications for holography and for connecting our solutions to asymptotically-flat geometries.

The second feature that we will track are the ``leading-log'' terms that appear in $\nu$ at higher orders in perturbations, giving rise to asymptotic behavior of the form (\ref{nuhomdiv2})  as  $\xi \to 1$.   Indeed, while our seed solution has $\nu \to \alpha$ as $\xi \to 1$, we find that, at higher order in perturbation theory, the solution for $\nu$ necessarily generates a leading-log divergence that cannot be cancelled even with the tuning of integration constants and the shift in the frequency. These log terms are thus a feature of our solutions, but do not spoil the asymptotics because $\hat \nu$ still vanishes at infinity.   The fact that the superstratum, and some of the microstrata,   do not exhibit such leading logs reinforces the idea that such solutions represent states of the CFT.  The appearance of the leading logs in some microstrata suggests that the  microstratum  not only involves a perturbation of the state but may also involve a perturbation of the CFT Lagrangian.   

Interestingly enough, one can delay the onset of the leading logs by tuning the resonant frequency to a slightly different value, and therefore moving the scalars slightly away from  the supersymmetric critical point at infinity.  However, this merely delays the appearance of the log terms to higher order in the perturbation theory.  This observation may also have an interesting holographic interpretation, and so we will describe how it appears in supergravity.

We also find that sub-leading logarithmic terms  appear in the other fields at higher orders in the perturbation theory.  These logs are sub-leading in the sense described in Section \ref{ss:Overview}:  Terms of the form (\ref{subleadinglog}) play an essential role in solving the inhomogeneous equations but do not involve changes in the  leading asymptotics of the fields as $\xi \to 0$ or $\xi \to 1$.  As we also noted in Section \ref{ss:Overview}, such terms also do not lead to any singularities in the radial derivatives at infinity, but  they can lead to divergent  $\xi$-derivatives.  They are thus not an issue for physical boundary conditions but they can be  a source of some numerical instabilities.  We will not catalog these in any detail here and leave a more systematic analysis for future work \cite{GHW2}.

There are some very significant qualitative differences between microstrata with $\omega_0 =0$ and  $\omega_0 =2$.  If one sets $\beta =0$ and  solves the system with $\omega_0 =0$, one is led uniquely to the susperstratum solution.  One can therefore think of the $\omega_0 =0$ solutions, with general $\beta$, as being some  $\mu_0$-fluctuation about a superstratum background. (We will return to this in Section \ref{ss:Newterritory}.) Taking $\omega_0 \ne 0$ deforms the superstratum into a non-BPS microstratum.  We find that this physical distinction becomes manifest in the leading and sub-leading logs that appear in the perturbation theory.  For $\omega_0 = 0$, $\beta \ne 0$ we find that, once one makes the appropriate shifts in $\omega$ (or in the potential $\Phi_1$), the solution has no logs at all, leading or sub-leading.  We have confirmed this to eleventh order.  

However, for $\omega_0 = 2$ we find that even when frequency shifts are made, the log-terms, both leading or sub-leading, appear to be endemic.  By adjusting boundary conditions one can suppress the leading logs in $\nu$ at third order in perturbations, but they are irremovable at fourth order.  We find qualitatively similar results for $\omega_0 = 4$.

We also find that the physical, UV frequency, $\omega_\infty$, is shifted non-trivially for  $\omega_0 =2$, which means that these solutions break all supersymmetries.  On the other hand,  for $\omega_0 =0$ and any $\alpha, \beta$,  we find that $\omega_\infty \equiv 0$, which suggests that these solutions may well preserve supersymmetry.   We will discuss this in more detail in Section \ref{ss:Newterritory}.

While there is a complicated interplay between the perturbative corrections of all the fields, it is the scalar seeds, $\nu$ and $\mu_0$, that act as the bellwether excitations.  These fields not only control the asymptotic vacuum states but also encode the  essential features of the perturbation theory, and so we will focus on these scalars in our more detailed description of the perturbative analysis.

%%%%%%%%%%%%%
\subsubsection{The solution for $\omega_0 = 2$}
%%%%%%%%%%%%%

At second order, the equations for  $\delta^2\nu$ and $\delta^2\mu_0$ are:
\begin{equation}
\begin{aligned}
    \frac1\xi \, \partial_\xi  & \qty(\xi\, \qty(1-\xi^2)\, \delta^2\nu') - \left(\frac{4}{\xi^2}- 24 \, \right)\delta^2 \nu ~=~ \\  
    &  -10\, \alpha\, \delta\omega_1\, \xi^2 \qty(4\,\xi^2 - 3) +\, \frac45 \alpha \,\beta\, \xi^4 \qty(2400\, \xi^{10} - 7632\, \xi^8 + 9212\, \xi^6 - 5194\, \xi^4 + 1335\, \xi^2 - 120)
\end{aligned}
\end{equation}
and
\begin{equation}
\begin{aligned}
    \frac1\xi \,\partial_\xi  \qty(\xi\, \delta^2\mu_0') ~+~  \frac{ 100 }{(1-\xi^2) }  \, \delta^2\mu_0  ~-~ & \frac{16}{\xi^2 \,(1-\xi^2) } \, \delta^2\mu_0  \\
    & ~=~ -12\, \alpha^2 \,\xi^4 \qty(1-\xi^2) \,-\, 8\, \beta \,\delta\omega_1\, \xi^4 \qty( 12\, \xi^4 - 16\, \xi^2 + 5)  \\
\end{aligned}
\end{equation}
It is elementary to obtain the following solutions:
\begin{equation}
    \begin{aligned}
        \delta^2\nu ~=~ -\frac{\alpha\,\beta}5\, \xi^6\, (1 \,-\, \xi^2)\, & (1 \,-\, 2\xi^2)\,  (3 \,-\, 4\xi^2)\,  (5 \,-\, 6\xi^2)\, \\
        &\quad + \, \frac12 \, \alpha\, \delta\omega_1\, \qty(2 \,\xi^4  \,+\,   \xi^2\, ( 4 \,\xi^2-3) \, \log(1-\xi^2) )
    \end{aligned}
\label{delta2nu}
\end{equation}
and
\begin{equation}
    \begin{aligned}
        \delta^2\mu_0 ~=~ -\frac{\alpha^2}{10} &\xi^6 \, (1 \,-\, \xi^2) \, \qty(6  \,-\, 7 \,\xi^2) \\
        &\, -\, \frac{\beta\, \delta\omega_1}{105}\, \xi^6\, \qty(155 \,\xi^4 - 259 \, \xi^2 + 105) \,-\, \frac{\beta\, \delta\omega_1}{5}\, \xi^4 \,(1-\xi^2) \, (1-2\xi^2)\, \log(1- \xi^2)
    \end{aligned}
\end{equation}
where we have chosen homogeneous solutions so as to cancel  $\xi^{-2}$ divergences at the origin.  One can also add further amounts of the smooth homogeneous solutions described in Section \ref{ss:scalarops} but this is simply a re-definition of $\alpha$ or $\beta$, and so we ignore such additions.

First observe that 
\begin{equation}
        \delta^2\mu_0(1)  ~=~  -  \frac{\beta}{105}\, \delta\omega_1
\end{equation}
and so, to remain at the supersymmetric point in the UV we must take $\beta =0$ or $\delta\omega_1 =0$.  
Furthermore, the log term in (\ref{delta2nu}) is exactly of the form  (\ref{subleadinglog}) with $F_1$ being the middle term in (\ref{nuseedex}). Since $F_1(1) =1$, this term gives rise to a leading log correction to $\nu$.   

To preserve the generality of the solution ($\beta \ne 0$) and to remove this leading log, we set:
\begin{equation}
    \delta\omega_1 = 0 \,.
 \label{delta1omega}
\end{equation}

At third order, the sources become significantly more complicated.  We find:
\begin{equation}
\begin{aligned}
    \frac1\xi \, & \partial_\xi  \qty(\xi\, \qty(1-\xi^2)\, \delta^3\nu')  -  \left(\frac{4}{\xi^2}- 24 \, \right)\delta^3 \nu \\ &~=~ -10\, \alpha \, \delta \omega _2 \,\xi ^2 \,\left(4 \xi ^2-3\right)    ~+~ \frac{2}{5} \alpha ^3 \xi ^6 \left(1400 \xi ^8-4572 \xi ^6+5357 \xi ^4-2574 \xi ^2+405\right)  \\
    &- \frac{2\, \alpha\,  \beta ^2}{11025} \xi ^2 \big(76204800 \xi ^{22}-422739072 \xi ^{20}+1002999424 \xi ^{18}-1324731072 \xi ^{16}+1060451252 \xi ^{14} \\
    & \qquad\quad -523632522 \xi ^{12}+154429380 \xi ^{10}-24570315 \xi ^8+1587600 \xi ^6+360 \xi ^2-270\big) 
\end{aligned}
\end{equation}
\begin{equation}
\begin{aligned}
    \frac1\xi \,\partial_\xi & \qty(\xi\, \delta^3\mu_0') ~+~  \frac{ 100 }{(1-\xi^2) }  \, \delta^3\mu_0  ~- \frac{16}{\xi^2 \,(1-\xi^2) } \, \delta^3\mu_0  \\
   &  ~=~-8 \,\beta \, \delta \omega _2 \,\xi ^4 \,\left(12 \xi ^4-16 \xi ^2+5\right) \, - \, \frac{4}{5}  \alpha ^2 \beta \,\xi ^{8} \left(560 \xi ^8-1526 \xi ^6+1422 \xi ^4-525 \xi ^2+63\right) \\
    &-\frac{4 \,\beta ^3}{18375} \xi ^4 \big(16934400 \xi ^{24}-106884288 \xi ^{22}+292824000 \xi ^{20}-455704088 \xi ^{18}+442275240 \xi ^{16} \\
    &\qquad\qquad \qquad -276195666 \xi ^{14}+110209718 \xi ^{12}-26839537 \xi ^{10}+3575743 \xi ^8 \\ 
    & \qquad\qquad\qquad -196032 \xi ^6+688 \xi ^4-608 \xi ^2+140\big) 
\end{aligned}
\end{equation}
Solving this leads to:
\begin{equation}
    \delta^3 \mu_0 (1) ~=~ \frac{67 \alpha ^2 \beta }{21\,450} \,+\, \frac{11\,762 \beta ^3}{312\,687\,375} \,-\, \frac{\beta  \delta^2 \omega }{105}
\end{equation}
and setting this to zero, in order to keep the scalars at infinity at the UV critical point, yields
\begin{equation}
    \delta^2\omega ~=~ \frac{469\, \alpha ^2}{1\,430} \,+\, \frac{11\,762 \beta ^2}{2\,977\,975}
 \label{delta2omega}
\end{equation}
With this choice, the leading log in $\nu$ no longer vanishes and we find:
\begin{equation}
    \delta^3\nu ~=~ \frac{\alpha \, \left(371\,875 \,\alpha ^2\,+\, 52\,232 \,\beta ^2\right)}{178\,678\,500}\, \xi ^2\, \left(4\, \xi ^2\,-\, 3\right) \, \log(1\,-\,\xi ^2) \,+\, \text{polynomial terms}
\label{delta3nu}
\end{equation}

Alternatively, if one makes the choice:  
\begin{equation}
    \delta^2\omega ~=~ \frac{34\, \alpha ^2}{105} \,+\, \frac{8\,842 \beta ^2}{2\,627\,625}
\end{equation}
then this removes the leading log in $\nu$ at this order but displaces the scalars away from the supersymmetric critical point.  Indeed, with this choice, one finds:
\begin{equation}
    \delta^3 \mu_0 (1) ~=~  \frac{ (371\,875 \alpha ^2+52\,232 \beta ^2)\, \beta }{9\,380\,621\,250}
\end{equation}
We will, however, use  (\ref{delta2omega}) and keep the supersymmetric asymptotics at infinity.

At fourth order we find that the leading log in $\nu$ cannot be removed and that there are sub-leading logs in all the fields.   To preserve  supersymmetry at infinity we find a further correction to the frequency :
\begin{equation}
    \delta^3 \omega ~=~ \frac{7\,351\, \alpha ^2 \beta }{756\,756} \,.
\end{equation}
Thus perturbation theory to fourth order leads to the following third-order expression for the frequency of the normal mode: 
\begin{equation}
    \omega ~=~ 2 \,+\, \frac{469\, \alpha ^2}{1\,430} \,+\, \frac{11\,762\, \beta ^2}{2\,977\,975} \,+\, \frac{7\,351\, \alpha ^2 \beta }{756\,756} \,+\, \mathcal{O}\qty(\alpha, \beta)^4 \,.
\label{deltaomega2}
\end{equation}
As we observed in Section \ref{ss:BCGF}, this is not the physical, UV frequency of the solution.  To obtain this, we must change the gauge to that of (\ref{gauge1}) and this results in a further shift of the frequency because the  quantities (\ref{gauge_inv_tems}) are gauge invariant. 

The perturbative expansion  leads to the following expressions for the gauge fields at infinity:
\begin{align}
    \Phi_1(1) =& \frac12 + \frac{\beta^2}{1\,225} + \frac{211 \alpha^2 \beta}{264\,600} - \frac{\alpha^4}{7\,350} - \frac{9\,957\,284\,807 \alpha^2 \beta^2}{29\,827\,064\,767\,500} - \frac{170\,663\,576\,257 \beta^4}{32\,325\,081\,441\,778\,125} \nonumber  \\ 
    & \hskip 11.5cm  +\, \mathcal{O}\qty(\alpha, \beta)^5 \,,
 \label{Phi1infty2}    \\
    \Phi_2(1) =& \frac{\beta^2}{1\,225} + \frac{197 \alpha^2 \beta}{264\,600} + \frac{5\,874\,710\,946\,229\,125 \alpha^4 - 158\,400\,821\,198\,370 \alpha^2 \beta^2}{1\,034\,402\,606\,136\,900\,000} \,+\, \mathcal{O}\qty(\alpha, \beta)^5 \,,
 \label{Phi1_2_inf}    \\
    \Psi_1(1) =& 0 \,, 
    \\
    \Psi_2(1)= &- \frac{525 \alpha^2 + 8 \beta^2}{4\,200} - \frac{3 \alpha^2 \beta}{700}   \nonumber\\
    & +   \frac{1\,932\,099\,503\,959\,625\,  \alpha^4  + 80\,399\,006\,203\,530\, \alpha^2 \beta^2 + 
 743\,238\,512\,608\, \beta^4}{38\,311\,207\,634\,700\,000} \,+\, \mathcal{O}\qty(\alpha, \beta)^5 \,.  \label{Psi2infty2}   
\end{align}

The frequency, $\omega_\infty$, relative to the UV fixed point  is then given by:
\begin{align}
    \omega_\infty ~=~  &  \omega ~+~ 2 \, \Phi_1(1) ~-~ 1  \nonumber\\ 
~=~  & 2 ~+~ \frac{469 \, \alpha^2}{1\,430} ~+~ \frac{16\,624 \, \beta^2}{2\,977\,975} ~+~ \frac{283 \, \alpha^2 \beta}{25\,025}  ~+~ \mathcal{O}\qty(\alpha, \beta)^4  \,,
\label{ominfty2}
\end{align}
where  $\omega$ and $\Phi_1(1)$ are given by (\ref{deltaomega2}) and  (\ref{Phi1infty2}), and the $-1$ comes from $2\Phi_1(1)$ in the gauge  (\ref{gauge1}).   The quantity, $\omega_\infty$, is then the physical UV frequency relative to the supertube geometry.  The fact that it is positive means that the excitation propagates inside the light-cone of the dual CFT, and hence breaks the supersymmetry.

One can also combine  (\ref{Phi1infty2})--(\ref{Psi2infty2})  with  (\ref{gauge0}), to determine the potential differences between $\xi =0$ and $\xi = 1$.

%%%%%%%%%%%%%
\subsubsection{The solution for $\omega_0 = 0$}
\label{omega0perts}
%%%%%%%%%%%%%

This solution is significantly simpler than the one above, and it is not too much of a challenge to go to elventh order in perturbations.  As we have already remarked, if one sets $\beta =0$, the result is exactly the superstratum.  Thus one can think of this microstratum as $\mu_0$ oscillations, with scale set by $\beta$,  about a superstratum, whose scale is set by $\alpha$.  

Exactly as above, one finds  $\delta^2\mu_0(1)  \sim  \delta\omega_1$ and so we, once again impose (\ref{delta1omega}).   At third order we find that 
the value of $\mu_0$ at infinity is:
\begin{equation}
    \delta^3 \mu_0 (1) ~=~  - \frac{2 \beta ^3}{125} \,-\, \frac{\beta  \delta^2\omega }{5}
\end{equation}
which leads to the constraint
\begin{equation}
    \delta^2\omega ~=~ -\frac{2}{25} \beta^2 \,.
\end{equation}
The analysis  proceeds in a straightforward manner, order by order and by imposing $\delta^n\mu_0(1) =0$ up to tenth order we obtain:
\begin{equation}
\begin{aligned}
    \omega ~&=~ -\frac{2\, \beta ^2}{25}\,-\, \frac{\alpha ^2 \beta}{100} \,+\, \frac{6\, \beta ^4 }{625} \,+\, \frac{\alpha ^2 \beta ^3 }{500} \,+\, \frac{25\, \alpha ^4 \beta ^2\,-\,288 \, \beta ^6}{250\,000}
    \,-\, \frac{21\, \alpha^2 \beta^5}{62\,500} \\
    &\,-\, \frac{25\, \alpha^4 \beta^4 \,-\, 108 \, \beta^8}{781\,250} \,-\, \frac{25\, \alpha^6 \beta^3 \,-\, 1296 \, \alpha^2\beta^7}{25\,000\,000} \ +\ \frac{9 \, \beta^6 (125\, \alpha^4 - 288 \, \beta^4)}{156\,250\,000} \,+\,  \mathcal{O}\qty(\alpha, \beta)^{11}
\end{aligned}
\label{deltaomega0}
\end{equation}

While the intermediate solutions contain logs, we also find that  if one imposes $\delta^n\mu_0(1) =0$ then it also causes all the logarithmic terms to vanish.  Indeed, once one has imposed (\ref{deltaomega0}),  the complete solution at eleventh order is entirely polynomial in $\xi$.

The expression of the gauge fields, truncated to seventh order, are
\begin{align}
    \Phi_1(1) ~&=~ \frac12 + \frac{\beta^2}{25} + \frac{\alpha^2 \beta}{200} - \frac{3 \beta^4}{625} - \frac{\alpha^2 \beta^3}{1000} + \frac{-25 \alpha^4 \beta^2 + 288 \beta^6}{500\,000} \,+\,  \mathcal{O}\qty(\alpha, \beta)^{7}
  \label{Phi1infty0}  \\
    \Phi_2(1) ~&=~ \frac{\beta^2}{25} + \frac{\alpha^2 \beta}{200} - \frac{\beta^4}{625} - \frac{3\alpha^2 \beta^3}{5000} + \frac{25 \alpha^4 \beta^2 + 32 \beta^6}{500\,000} \,+\,  \mathcal{O}\qty(\alpha, \beta)^{7}
    \\
    \Psi_1(1) ~&=~ 0
    \\
    \Psi_2(1) ~&=~ -\frac{5 \alpha^2 + 8 \beta^2}{40} - \frac{\alpha^2 \beta}{20} - \frac{5 \alpha^2 \beta^2 - 8 \beta^4}{1000} - \frac{40 \alpha^2 \beta^4 + 640 \beta^6}{2\,000\,000} \,+\,  \mathcal{O}\qty(\alpha, \beta)^{7}  \label{Psi2infty0} 
\end{align}

One finds a very interesting result in making the gauge transformation to obtain the physical UV frequency, $\omega_\infty$,  relative to the supertube geometry at infinity.  One now has:
\begin{equation}
    \omega_\infty ~=~   \omega ~+~ 2\,\Phi_1(1)   -1 ~=~ 0  ~+~~ \mathcal{O}\qty(\alpha, \beta)^{11}         \,,
\label{ominfty0}
\end{equation}
where $\omega$ and $\Phi_1(1)$ are given by (\ref{deltaomega0}) and (\ref{Phi1infty0}).  This strongly suggests that for the $\omega_0=0$ solutions, the frequency relative to the supertube boundary conditions at infinity is identically zero:
\begin{equation}
    \omega_\infty ~=~ 0   \,,
\label{ominfty0res}
\end{equation}
for all $\alpha$ and $\beta$. Thus for $\omega_0 = 0$ there is {\it no shift in the physical UV frequency.}  This means that the excitations of the UV CFT are still purely left-moving. Therefore, these solutions could still be supersymmetric,.  However, it is also possible that other background fields still break the right-moving supersymmetry.  See Section \ref{ss:Newterritory} for further discussion of this issue.  

Combing (\ref{Phi1infty0})--(\ref{Psi2infty0})  with  (\ref{gauge0}), one can, once again, determine the potential differences between $\xi =0$ and $\xi = 1$.  

It is interesting to note that while the solution described above is symmetric under $\alpha\to-\alpha$, it is not symmetric under $\beta\to-\beta$. This symmetry is however restored when $\alpha = 0$.

While the fact that $\omega_\infty = 0$ perhaps represents the most remarkable feature of this family of solutions, we also catalog some other features  that will be compared against the numerical results.

The solution explores the $\mu_1 - \mu_2$ flat direction of the potential in the IR ($\xi \to 0$). Indeed, we find that the first terms in its expansion are :
\begin{equation}
    \mu_1 - \mu_2 ~\xrightarrow[\xi \to 0]{}~ -\frac{2\, \beta ^2}{25}\,-\, \frac{\alpha ^2 \beta}{100} \,+\, \frac{2\, \beta ^4 }{625} \,+\, \frac{\alpha ^2 \beta ^3 }{2\,500} \,-\, \frac{14 \, \beta ^6}{46\,875} \,-\, \frac{\alpha^2 \beta^5}{12\,500} \,+\, \mathcal{O}\qty(\alpha, \beta)^{8}
\end{equation}

We can compute the mass and angular momentum of the solutions using (\ref{BTZasympmet}), the first few terms are given by :
\begin{align}
    M + 1~&=~ \alpha^2 \,+\, \frac{12\, \beta^2}{5} \,+\, \frac{3\,\alpha^2 \beta}{5} \,+\, \frac{\alpha^4}{16} \,-\, \frac{32\,\beta^4}{125} \,-\, \frac{\alpha^2 \beta^3}{25} \,+\, \mathcal{O}\qty(\alpha, \beta)^{6}
    \\
    \widetilde M + 1~&=~ \frac32 \alpha^2 \,+\, \frac{12\, \beta^2}{5} \,+\, \frac{3\,\alpha^2 \beta}{5} \,-\, \frac{\alpha^4}{16} \,-\, \frac{\alpha^2 \beta^2}{5} \,-\, \frac{32\,\beta^4}{125} \,-\, \frac{\alpha^3 \beta^2}{20} \,-\, \frac{\alpha^2 \beta^3}{25} \,+\, \mathcal{O}\qty(\alpha, \beta)^{6}
    \\
    J ~&=~ \alpha^2 \,+\, \frac{8\, \beta^2}{5} \,+\, \frac{2\,\alpha^2 \beta}{5} \,-\, \frac{8\,\beta^4}{125} \,+\, \frac{\alpha^2 \beta^3}{25} \,+\, \mathcal{O}\qty(\alpha, \beta)^{6}
\end{align}

%%%%%%%%%%%%%%%%%%%%%%%%%%%%%%%%%%%%%
\section{Numerical analysis}
\label{sec:numerics}
%%%%%%%%%%%%%%%%%%%%%%%%%%%%%%%%%%%%%

%%%%%%%%%%%%%%%%%%%%%%
\subsection{Solving the boundary value problem}
\label{ss:boundary_value}
%%%%%%%%%%%%%%%%%%%%%%

We have to solve for the eleven functions given in the list, ${\cal F}$, given in (\ref{functionlist}). The equations of motion (\ref{lagrangian_full}) give us eleven second-order differential equations with three integrals of motion given in  Section \ref{ss:Integrals}.    To solve this system we need essentially $22$ pieces of data. Much of this data is encompassed by requiring that  the solution is smooth at $\xi =0$ and $\xi=1$, however, as we saw in Section \ref{ss:perturbations} this is not sufficient and so we will also impose the same boundary conditions that we imposed on the linear system:   The gauge fixing of the Maxwell potentials and metric functions, (\ref{gauge_fix}),  and the requirement that the scalars approach the supersymmetric critical point at infinity,  (\ref{susy_point_scalars}).   

Having done this, the linear system still had three degrees of freedom, $\omega$, $\alpha$ and $\beta$.  A canonical choice for the latter variables would be to take:
\begin{equation}
    \hat\alpha ~\equiv~\partial_\xi^2 \nu  (0) \qand \beta ~\equiv~ \partial_\xi^4 \mu_0 (0) \,.
\end{equation}
This choice has the advantage of not receiving higher order corrections in perturbation theory.  However, our numerical analysis is configured somewhat differently and we use $\alpha$ to parametrize the constant value of $\nu$ at infinity.  We also know that the value of $\omega$ is discrete  and we will solve for it in the vicinity of $\omega_0 \in 2 \ZZ$ and, as we have already indicated, we will consider $\omega_0 =0,2,4$.

This characterizes the families of solutions we seek, but, as always, one cannot simply plug these constraints into a numerical algorithm. The primary challenge is that $\xi =0$ and $\xi=1$ are regular singular points of all the differential equations, and  almost all these equations have singular branches.   This means that if one ``shoots'' from one end of the $\xi$-interval, $(0,1)$, then the numerical solution will, through numerical error, pick up one of the singular branches and diverge hopelessly at the other end of the interval. 

Our solution to this problem is to use a ``double-shooting'' method.  That is, we completely specify initial data  near $\xi=0$ and use standard algorithms (such as Runge-Kutta) to evolve it towards $\xi=1$.  Similarly, we completely specify initial data  near $\xi=1$ and numerically evolve the solution towards $\xi =0$.  We then examine both solutions at some intermediate point, which we take to be $\xi_{mid} \equiv \frac{3}{5}$, and  try to match the two solutions by adjusting the initial data at both ends while respecting the boundary conditions we wish to impose.

While seemingly simple, there is a further issue:  the fact that $\xi=0$ and $\xi=1$ are singular points of the differential equations means that we cannot simply specify the initial conditions at these points.  We have to determine the solution at an infinitessimal displacement away from the end points and then shoot from these displaced initial points.  Specifically, we  take the initial points for the numerics to be $\xi_0 = 1/100$ and $\xi_1 = 995/1000$.  We expand every one of the eleven functions in  series about $\xi =0$ and  in series about $\xi =1$ and impose the boundary data on these series and  choose values of $\alpha$ and $\beta$.  We then use the equations of motion to determine the series as much as as possible.  In this way we obtain  approximate solutions at $\xi_0$ and at $\xi_1$.  These approximate solutions still have  undetermined  coefficients and these  become the data that must be varied in order to find a matched solution  at the ``mid-point,'' $\xi_{mid} = \frac{3}{5}$.  Obviously the match will not be perfect, and we express the mismatch in terms of a {\it cost function}.  The complete numerical algorithm then involves the minimization of this cost function.

%%%%%%%%%%%%%%%%%%%%%%
\subsection{Series expansions at the boundaries}
\label{sub:series_expansions}
%%%%%%%%%%%%%%%%%%%%%%

Our purpose here is to use series expansions to generate  approximate solutions at $\xi_0 = 1/100$ and $\xi_1 = 995/1000$.  Motivated by our initial expectation that the normalizable modes would lead to simple power series solutions, we are going to take a short cut in this process: we will ignore all logs, both leading, and sub-leading. As we saw from the perturbation theory, this will be exact (at least to very high orders) for $\omega_0 =0$, but will introduce small systematic errors for $\omega_0 =2$.  Indeed, one can make an estimate of these errors from (\ref{delta3nu}), and they are typically $< 10^{-4}$, which will translate into $< 10^{-8}$ in the cost function.  The numerics actually performs better than this naive expectation:  the series solutions place us in the neighborhood of families of smooth solutions and the numerics is able to compensate for the small systematic errors in our series and converge on nearby, more accurate solutions.  As a result, our numerics  for $\omega_0 =2$ generally converge with cost function values between $ 10^{-8}$ and $ 10^{-12}$, depending on the size of $\alpha$.

We  also note that for $\omega_0 =0$, where are no log terms in the higher-order perturbation theory, our numerics converge with  cost function values less than $10^{-15}$.  We therefore find our numerical results to be well within range of an acceptable approximation for this first foray into  numerical solutions for these equations.  We will perform a much more careful series analysis in \cite{GHW2}.

At $\xi = 0$ we take:
\begin{equation}
    \begin{aligned}
    \nu ~&=~ \sum_{n\geq 2} \nu_n^{(0)}\, \xi^n \ , \qquad\qquad & \mu_0 ~&=~\beta \,\xi^4 + \sum_{n\geq 5} \mu_{0,n}^{(0)}\, \xi^n \ ,\\[.5ex]
    \mu_1 ~&=~ \sum_{n\geq 0} \mu_{1,n}^{(0)}\, \xi^n \ , \qquad\qquad & \mu_2 ~&=~ \sum_{n\geq 0} \mu_{2,n}^{(0)}\, \xi^n \ ,\\[.5ex]
    \Phi_1 ~&=~ \frac12 + \sum_{n\geq 1} \phi_{1,n}^{(0)}\, \xi^n \ , \qquad\qquad & \phi_2 ~&=~ \sum_{n\geq 1} \phi_{2,n}^{(0)}\, \xi^n \ ,\\[.5ex]
    \Psi_1 ~&=~ \sum_{n\geq 1} \psi_{1,n}^{(0)}\, \xi^n \ , \qquad\qquad & \Psi_2 ~&=~ \sum_{n\geq 1} \psi_{2,n}^{(0)}\, \xi^n \ ,\\[.5ex]
    \Omega_0 ~&=~ \sum_{n\geq 0} \omega_{0,n}^{(0)}\, \xi^n \ , \qquad\qquad & \Omega_1 ~&=~ 1 + \sum_{n\geq 1} \omega_{1,n}^{(0)}\, \xi^n \ ,\\[.5ex]
    k ~&=~ \sum_{n\geq 2} k_n^{(0)}\, \xi^n\ .
    \end{aligned}
\end{equation}
and at  $\xi = 1$ we take:
\begin{equation}
    \begin{aligned}
    \nu ~&=~ \alpha + \sum_{n\geq 1} \nu_n^{(\infty)}\, (1-\xi^2)^n \ , \qquad\qquad & \mu_0 ~&=~ \sum_{n\geq 1} \mu_{0,n}^{(\infty)}\, (1-\xi^2)^n \ ,\\[.5ex]
    \mu_1 ~&=~ \sum_{n\geq 1} \mu_{1,n}^{(\infty)}\, (1-\xi^2)^n \ , \qquad\qquad & \mu_2 ~&=~ \sum_{n\geq 1} \mu_{2,n}^{(\infty)}\, (1-\xi^2)^n \ ,\\[.5ex]
    \Phi_1 ~&=~ \sum_{n\geq 0} \phi_{1,n}^{(\infty)}\, (1-\xi^2)^n \ , \qquad\qquad & \Phi_2 ~&=~ \sum_{n\geq 0} \phi_{2,n}^{(\infty)}\, (1-\xi^2)^n \ ,\\[.5ex]
    \Psi_1 ~&=~ \sum_{n\geq 0} \psi_{1,n}^{(\infty)}\, (1-\xi^2)^n \ , \qquad\qquad & \Psi_2 ~&=~ \sum_{n\geq 0} \psi_{2,n}^{(\infty)}\, (1-\xi^2)^n \ ,\\[.5ex]
    \Omega_0 ~&=~ 1 + \sum_{n\geq 1} \omega_{0,n}^{(\infty)}\, (1-\xi^2)^n \ , \qquad\qquad & \Omega_1 ~&=~ \sum_{n\geq 0} \omega_{1,n}^{(\infty)}\, (1-\xi^2)^n \ ,\\[.5ex]
    k ~&=~ \frac1{\omega_{1,0}^{(\infty)}} + \sum_{n\geq 1} k_n^{(\infty)}\, (1-\xi^2)^n\ .
    \end{aligned}
\end{equation}

We then substitute these expansions in the equations of motions and solve for the coefficients order by order. 

The result is that all but 19 of the coefficients are fixed by the equations of motion. We can furthermore fix two of them using the conserved quantities defined in section \ref{ss:Integrals}.
Indeed, since these quantities are independent of $\xi$, they can be used to relate some of the coefficients at infinity with the coefficients at the origin:
\begin{equation}
    \phi_{1,0}^{(\infty)} ~=~ \phi_{1,0}^{(0)} \,-\, e^{4\mu_{1,0}^{(0)}} \, \frac{\psi_{2,2}^{(0)}}{\omega_{0,0}^{(0)}} \qand \psi_{1,0}^{(\infty)} ~=~ 0 \ .
\end{equation}

The constant terms in $\Phi_2$ and $\Psi_2$ are special in that they do not enter the dynamics, and so, {\it a priori}, $\phi_{2,0}^{(0)}$, $\psi_{2,0}^{(0)}$, $\phi_{2,0}^{(\infty)}$ and $\psi_{2,0}^{(\infty)}$ can be set to arbitrary values in the shooting process.  Indeed we start by setting them to zero. However, the {\it potential differences} $\phi_{2,0}^{(\infty)} - \phi_{2,0}^{(0)}$  and  $\psi_{2,0}^{(\infty)} - \psi_{2,0}^{(0)}$ do have physical meaning, and are determined by the dynamics.  What this means is that when we evolve the solutions from their zero initial values at  $\xi =0$ and $\xi =1$, the solutions from each end will have a constant offset relative to one another at the mid-point, $\xi_{mid} = \frac{3}{5}$. A smooth solution is then obtained by uniformly shifting either the solution from $\xi =0$, or the solution from $\xi =1$, by the constant offset.   We therefore determine the potential differences between $\xi =0$ and $\xi =1$ from these offsets at  $\xi_{mid}$.  The important point here is that the data  $\phi_{2,0}^{(0)}$, $\psi_{2,0}^{(0)}$, $\phi_{2,0}^{(\infty)}$ and $\psi_{2,0}^{(\infty)}$ is irrelevant to solving the shooting problems, but the potential differences are easily read off from the solutions.

We are thus left with 15 parameters that must be varied in order to find the solution:
\begin{equation}
\begin{aligned}
     \stP ~=~ &\Big\{ \nu_2^{(0)} \ ,\ \mu_{1,0}^{(0)} \ ,\ \mu_{2,0}^{(0)} \ ,\ \psi_{1,2}^{(0)} \ ,\ \psi_{2,2}^{(0)}  \ ,\ k_2^{(0)}\ ,\ \omega_{0,0}^{(0)} \ ,
     \\
     &  \quad \mu_{0,1}^{(\infty)}\ ,\ \mu_{1,1}^{(\infty)}\ ,\ \mu_{1,2}^{(\infty)}\ ,\ \phi_{2,1}^{(\infty)}\ ,\ \psi_{1,1}^{(\infty)}\ ,\ \omega_{1,0}^{(\infty)}\ ,\ k_{1}^{(\infty)}\ ,\ k_{2}^{(\infty)} \Big\} \ .
\end{aligned}
     \label{series_params}
\end{equation}
%

%%%%%%%%%%%%%%%%%%%%%%
\subsection{The minimization procedure}
\label{sub:minimizaton}
%%%%%%%%%%%%%%%%%%%%%%

The first step is to choose fixed  values of $\alpha$, $\beta$ and $\omega_0$.  We then  choose a set of values of the parameters, $\stP$,  (\ref{series_params}), and use them to fix the values of all the fields and their derivatives close to both ends of the segment, at $\xi_0 = 1/100$ and $\xi_1 = 995/1000$.  This provides  initial conditions for the shooting process from each end. Finally, we  select a value of $\omega$ close to $\omega_0$.  We apply the shooting algorithm from both ends, to get two solutions in the bulk. We denote these solutions respectively $\mathcal{S}_0$ and $\mathcal{S}_1$.

We then compare these solutions at a ``mid-point'' in the bulk, which we take to be $\xi_{mid} = \frac{3}{5}$. The comparison is  made by defining a cost function:
\begin{equation}
    C(\omega, \stP) ~=~ \sum_{v \in \mathcal{F}\setminus \{\Phi_2,\Psi_2\} } \qty(v_{\mathcal{S}_0}\qty(\xi_{mid}) - v_{\mathcal{S}_1}\qty(\xi_{mid}))^2 \ +\ \sum_{v \in \mathcal{F}} \qty(v'_{\mathcal{S}_0}\qty(\xi_{mid}) - v'_{\mathcal{S}_1}\qty(\xi_{mid}))^2
    \label{cost_fn}
\end{equation}
where $\mathcal{F} = \{ \nu, \mu_0, \mu_1, \mu_2, \Phi_1, \Phi_2, \Psi_1, \Psi_2, \Omega_0, \Omega_1, k \}$ is the set of all the fields. Note that, for the reasons explained above, we do not match on the values of $\Phi_2$ and $\Psi_2$. 

The goal is now to compute numerically the values of  $\omega$ and $\stP$ that minimize $C(\omega, \stP)$.  %This is sometimes referred to as  computing ``$\argmin C(\omega, \stP)$,'' which is a somewhat confusing shorthand for computing the  arguments that minimize the function.

We do this by using numerical algorithms built into {\it Mathematica,} and, in particular, we use {\it ParametricNDSolve} for the shooting and {\it FindMinimum} to compute the minimum of the cost function. Rather than simply treat the latter  as a  black box, we summarize what is going on inside the algorithm and how we adapted some of the options to make the solution technique more effective.

We use the Levenberg-Marquardt algorithm (a refinement of the Gauss-Newton algorithm), implemented in {\it FindMinimum}. This algorithm makes successively more accurate approximations of the minimum by using {\it quadratic} approximations to the cost function. This method is particularly well-adapted for minimization problems for which the cost function is written as a sum of squares, $C = \sum r_i^2$. The schematic process is then: 
\begin{enumerate}[label=\textbf{Step \arabic*}]
    \item Choose a first estimate of the solution $\omega$ and $\stP$. These values are used as a seed for the algorithm.  Such seeds can be based on other known solutions, or starting from the exactly known superstratum result.  The seed must be close enough to the solution, or at least not so far as to make the results from the shooting diverge before reaching $\xi_{mid}$.
    \item Use this data and {\it ParametricNDSolve}  to compute a solution and calculate  the value of the cost function, $C(\omega, \stP)$.  Next  compute numerically the Jacobian $J$ of the functions, $r_i$, at this point by computing the difference between the original value of the $r_i$,  and the 
 new values obtained using  small perturbations of the parameters.
        \item Ideally  a quadratic approximation would involve computing the Hessian of the cost function, but this is numerically very demanding and compounds numerical errors. Instead, the Gauss-Newton algorithm  uses the Jacobian to construct an approximation to the Hessian and makes a quadratic approximation based on this.  The displacements of the parameters, $\Delta \omega$ and  $ \Delta\stP$, that move the solution towards the minimum are thus estimated by computing the minimum of the quadratic approximation: 
    \begin{equation}
        \sum_{ij}\qty[r_i(\omega, \mathcal{P})\, J_{ji}\, \mqty(\Delta\omega \\ \Delta\stP)_j + \frac12 \mqty(\Delta\omega \\ \Delta\stP)_i (J J^\top)_{ij} \mqty(\Delta\omega \\ \Delta\stP)_j] \ .
    \end{equation}
    \item  The danger, as ever, with such an algorithm is that it might overshoot, or oscillate around, the minimum.  So  {\it FindMinimum}  actually treats  $(\Delta\omega\, ,\Delta\stP)$ as displacement in the parameters space and then finds a better estimate of the minimum of the cost function in the one-dimensional space along this direction.  This is the primary function of the ``step control'' within  {\it FindMinimum}.  The result is a new estimate of the parameters of the solution given by $ \omega + \lambda \Delta\omega$ and  $ \stP + \lambda \Delta\stP$, where $\lambda$ is a step in the minimizing direction deemed good enough by {\it FindMinimum}. 
  \item Then {\it FindMinimum}  repeats steps 2 to 4 with the new estimates, and does so until it achieves a good level of convergence\footnote{It can also generate errors where it has failed to converge adequately and then one must adjust the values of $ \omega$ and  $\stP$ and restart the search.}
\end{enumerate}

In this way our numerical methods converge not only on the  solution of the boundary value problem but also on the ``resonant frequency,'' that is the shifted frequency of the normal modes.

%%%%%%%%%%%%%%%%%%%%%%%%%%%%%%%%%%%%%
\section{Results}
\label{sec:results}
%%%%%%%%%%%%%%%%%%%%%%%%%%%%%%%%%%%%%

Here we present some representative examples of the results  obtained from the numerical method described in Section \ref{sec:numerics}. The examples are relatively typical and explore the moduli space parametrized by $\alpha$ and $\beta$.   We provide comparisons between the numerical solutions, the series solutions and the analytically--known superstrata.  For small to moderate $\alpha$ and $\beta$, the agreement between the numerics and the series solutions of Section \ref{sec:Perturbative} is exceptional.  This gives us great confidence in the accuracy  of our methods and provides further confirmation of the structure of our solutions.    

We use the series solutions and the numerical analysis to explore the physical properties of the moduli space of solutions.  First we track the shift in the frequencies of the normal modes as a function of $(\alpha, \beta)$.  Since our construction builds in smoothness at the outset, the standard superstratum ``smoothness conditions,'' like (\ref{ss-smoothness}), arise through the appearance of closed time-like curves (CTC's) at infinity ($\xi \to 1$).  We therefore track the appearance of CTC's and use this to constrain the moduli, $(\alpha, \beta)$.  We also  compute the masses and  momentum charge ($J=Q_P$)  of our solutions and compare them with the superstrata, and this will show that the new microstrata are indeed non-extremal. 

In terms of the practicalities of the numerics, we  track the behavior of the cost function,   (\ref{cost_fn}). We start to consider a solution reliable when the cost function is less than $10^{-8}$, and we fully trust it when it is less than $10^{-10}$. Since the cost function is a sum of squares, this means that we begin to trust the functions when the errors are less than  $10^{-4}$ and fully trust them when the errors are less than $10^{-5}$.

\begin{figure}[ht!]
\centering
\includegraphics[width=\textwidth]{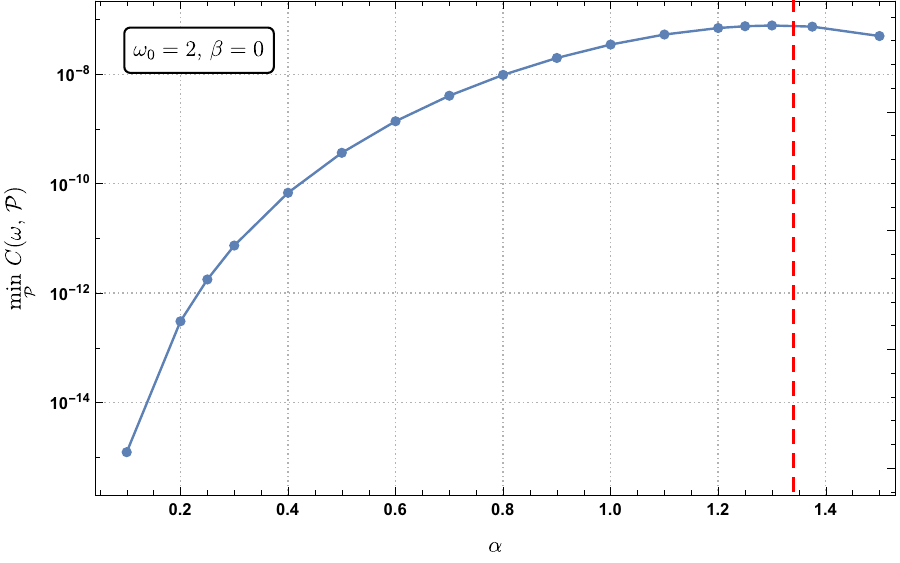}
\caption{\it Plot of the minimum value of the cost function for various values of $\alpha$, with $\beta = 0$ and $\omega_0 = 2$. The value of $\omega$ at each point grows with increasing $\alpha$, see Fig.~\ref{fig:shift_freq_perturb}. The vertical red dashed line corresponds to the CTC locus (see Section \ref{ss:CTCs}) and data points to the right of this line are unphysical.}
\label{fig:cost_function0}
\end{figure}

In practice, we have excellent accuracy for the solutions at $\omega_0 = 0$, where the cost function   (for $\alpha =1$) ranges from $10^{-24}$ at $\beta = \frac{1}{4}$, to $10^{-15}$ at $\beta = 2$. The solutions at $\omega_0 = 2$ are somewhat less accurate, almost certainly because of the appearance of log terms and the systematic errors they introduce into our series analysis. For $\omega_0 = 2$ , the cost function ranges from $10^{-15}$ at $\alpha=\frac{1}{5}$, to $10^{-7}$ at $\alpha \geq \frac{6}{5}$. It is below the limit of $10^{-8}$ when $\alpha \leq \frac{4}{5}$. In Fig.~\ref{fig:cost_function0}, we have shown the minimal values of the cost function for $\omega_0 = 2$, $\beta =0$ for seventeen solutions in the range $0 < \alpha < 1.6$.   As we will discuss in Section \ref{ss:CTCs}, the solutions with $\alpha \gtrsim \frac{4}{3}$ are unphysical because of the presence of CTC's.

%%%%%%%%%%%%%%%%%%
\vspace{0.5mm}
\vspace{1mm}
%%%%%%%%%%%%%%%%%%

In addition to the  smallness of the cost function, we have confidence in our solutions because of their excellent match with the series expansions, even at a relatively large range of values of the parameters $(\alpha, \beta)$.  We also find that the convergence of our numerics becomes dramatically better when the frequency approaches the  ``resonance'' for a normal mode, and the result is an extremely sharp, narrow valley  in the plot of the  cost function against frequency. See, for example, Fig.~\ref{fig:cost_function}, and note that this is a plot of the log of the cost-function. Not only does this add to our confidence in the numerics, but it also provides an effective search algorithm for the normal frequencies of oscillation.

%%%%%%%%%%%%%%%%%%%%%%
\subsection{Sample solutions}
\label{ss:Trophies}
%%%%%%%%%%%%%%%%%%%%%%

\begin{figure}[ht!]
	\centering
	\includegraphics[width=\textwidth]{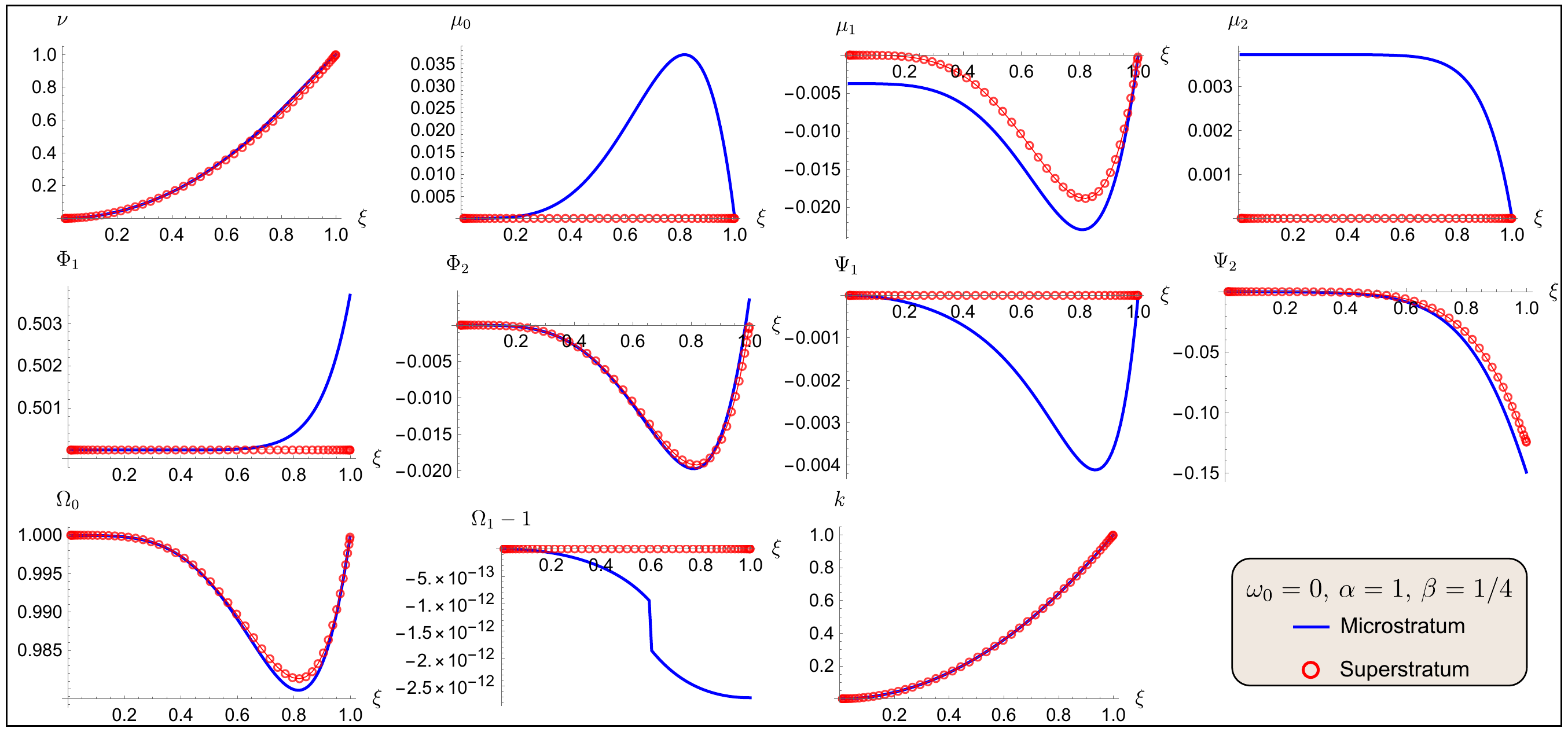}
	\caption{\it Numerical solution for all the fields at $\omega_0 = 0$, $\alpha = 1$ and $\beta = \frac{1}{4}$. The microstratum is the blue solid line, and the corresponding superstratum solution ($\alpha = 1$, $\beta = 0$) is the red line with circles. We note that there is a step discontinuity in $\Omega_1$ but its size much smaller than our numerical accuracy and is thus a numerical artefact.}
	\label{fig:fields_w_0}
\end{figure}

\begin{figure}[ht!]
	\centering
	\includegraphics[width=\textwidth]{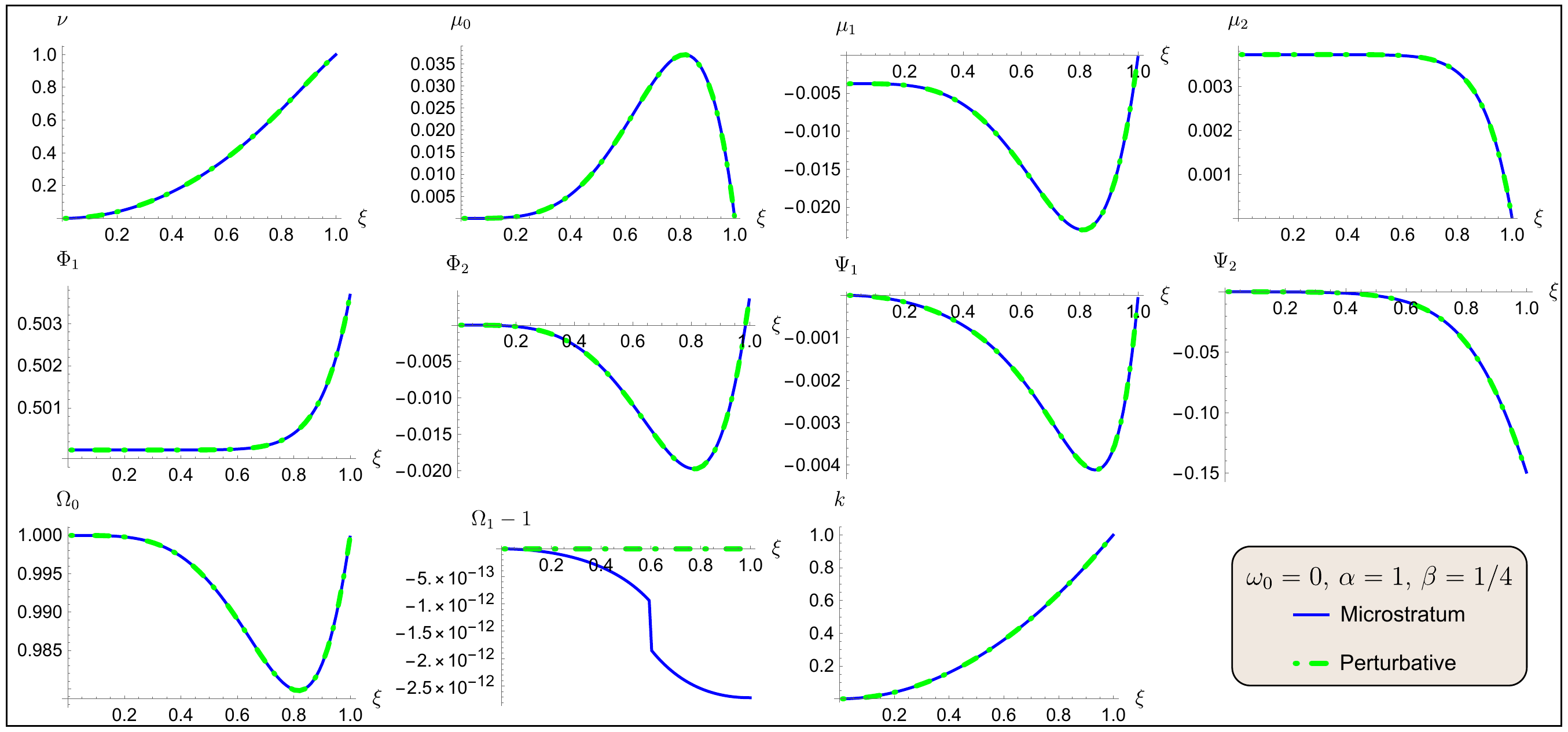}
	\caption{\it A comparison of the series solution and the numerical results for  $\omega_0=0$, $\alpha=1$ and $\beta=\frac{1}{4}$. The numerics are solid blue and the series solution is dashed-dotted green. We note that there is a step discontinuity in $\Omega_1$ but its size much smaller than our numerical accuracy and is thus a numerical artefact.}
	\label{fig:fields_perturb_w_0}
\end{figure}

For $\omega_0 =0$, and $\beta =0$, the series solution and the numerical solution both reduce to the analytically-known superstratum.  Thus the interesting families of microstrata correspond to taking $\beta \ne 0$.    Fig.~\ref{fig:fields_w_0} shows our numerical results for $\omega_0=0$, $\alpha = 1$ and $\beta =  \frac{1}{4}$, along with the superstratum for reference.    Fig.~\ref{fig:fields_perturb_w_0} shows a comparison of the numerical with the  series solution up to eleventh order.

First we note the prefect match between the numerics and the series solution.  The only somewhat anomalous plot is that of $\Omega_1$ (Fig.~\ref{fig:fields_w_0} and Fig.~\ref{fig:fields_perturb_w_0}).  There appears to be a step discontinuity but its size is $\sim 10^{-12}$, which makes a contribution of $10^{-24}$ to the cost function. Moreover the series solution shows that $\Omega_1 =1$, and so we are confident that this step is merely a numerical error, well below the level of the cost function.

There are also obvious differences between the microstratum and the superstratum  (see Fig.~\ref{fig:fields_w_0}).  These are most evident in the scalars and in the electromagnetic potentials.  However, as we will discuss in Section \ref{ss:Non-extremality}, the small differences in the metric coefficients lead to a different mass and momentum charge for the microstratum. 
Of particular note is the fact that $\mu_1$ and $\mu_2$  do not vanish at $\xi=0$:  this means that the scalars are not settling down to the supersymmetric minimum at $\xi =0$.  We will discuss this further in Section \ref{ss:Newterritory}.

\begin{figure}[ht!]
	\centering
	\includegraphics[width=\textwidth]{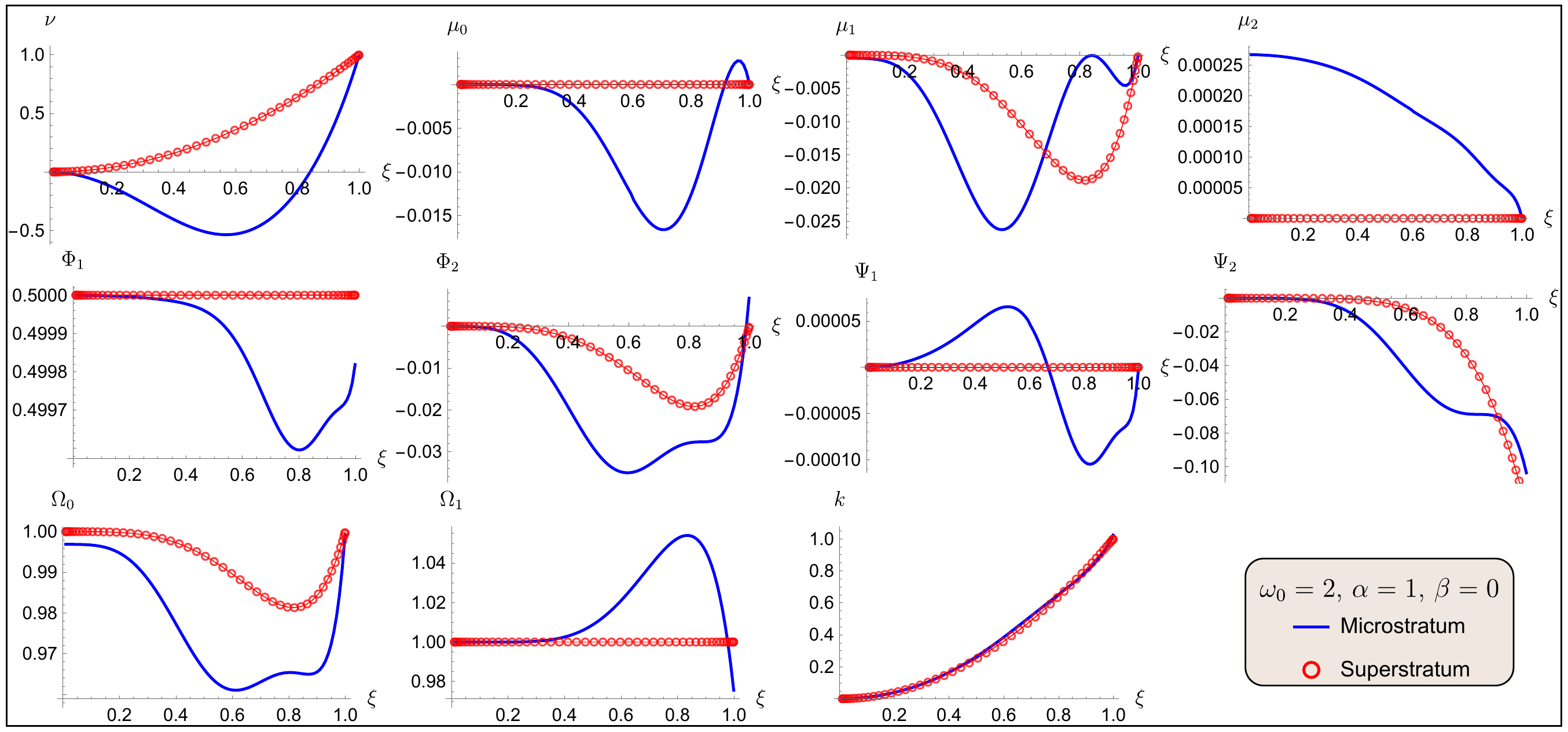}
	\caption{\it Numerical solution for all the fields at $\omega_0 = 2$, $\alpha = 1$ and $\beta = 0$. The microstratum is the blue solid line, and the corresponding superstratum solution ($\alpha = 1$, $\beta = 0$) is the red line with circles.}
	\label{fig:fields_w_2}
\end{figure}

\begin{figure}[ht!]
	\centering
	\includegraphics[width=\textwidth]{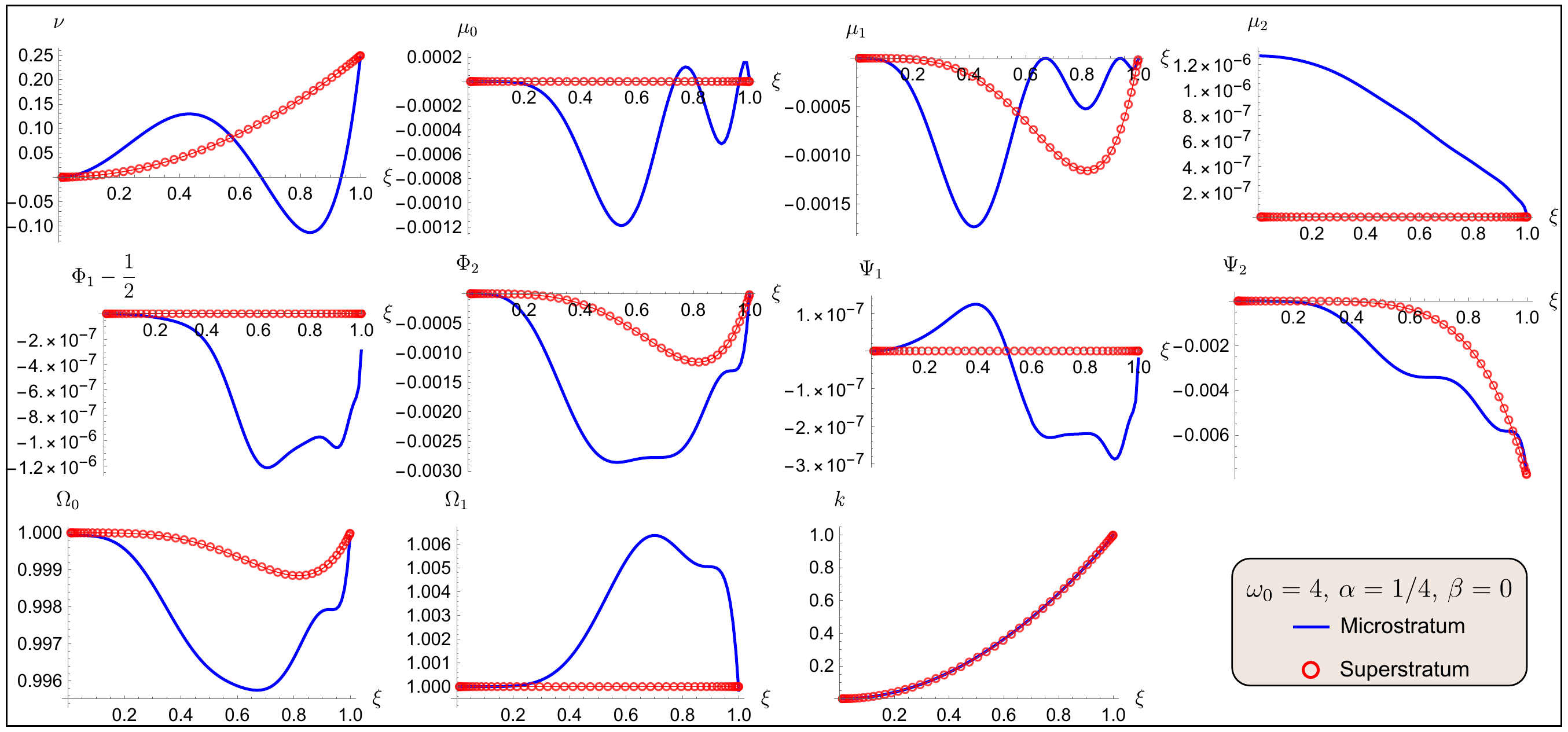}
	\caption{\it Numerical solution for all the fields at $\omega_0 = 4$, $\alpha = \frac{1}{4}$ and $\beta = 0$. The microstratum is the blue solid line, and the corresponding superstratum solution ($\alpha = 1$, $\beta = 0$) is the red line with circles.}
	\label{fig:fields_w_4}
\end{figure}

For $\omega_0 > 0$, there is no superstratum solution but there are solutions for generic $(\alpha,\beta)$ (so long as they are not too large). In Fig.~\ref{fig:fields_w_2} and Fig.~\ref{fig:fields_w_4} we show the numerical solutions $(\omega_0=2,\,\alpha=1,\,\beta=0)$ and $(\omega_0=4,\,\alpha=\frac{1}{4} ,\,\beta=0)$, respectively.  Again we have plotted the superstratum solution (with $\omega_0 = 0,\,\beta=0$ and same $\alpha$) for reference.   Note that for $\omega_0 =2$ and  $\omega_0 =4$,  the function $\nu$ has one and two nodes respectively, as one should expect from the linearized results in Fig.~\ref{fig:nuharms}. 

\begin{figure}[ht!]
	\centering
	\includegraphics[width=\textwidth]{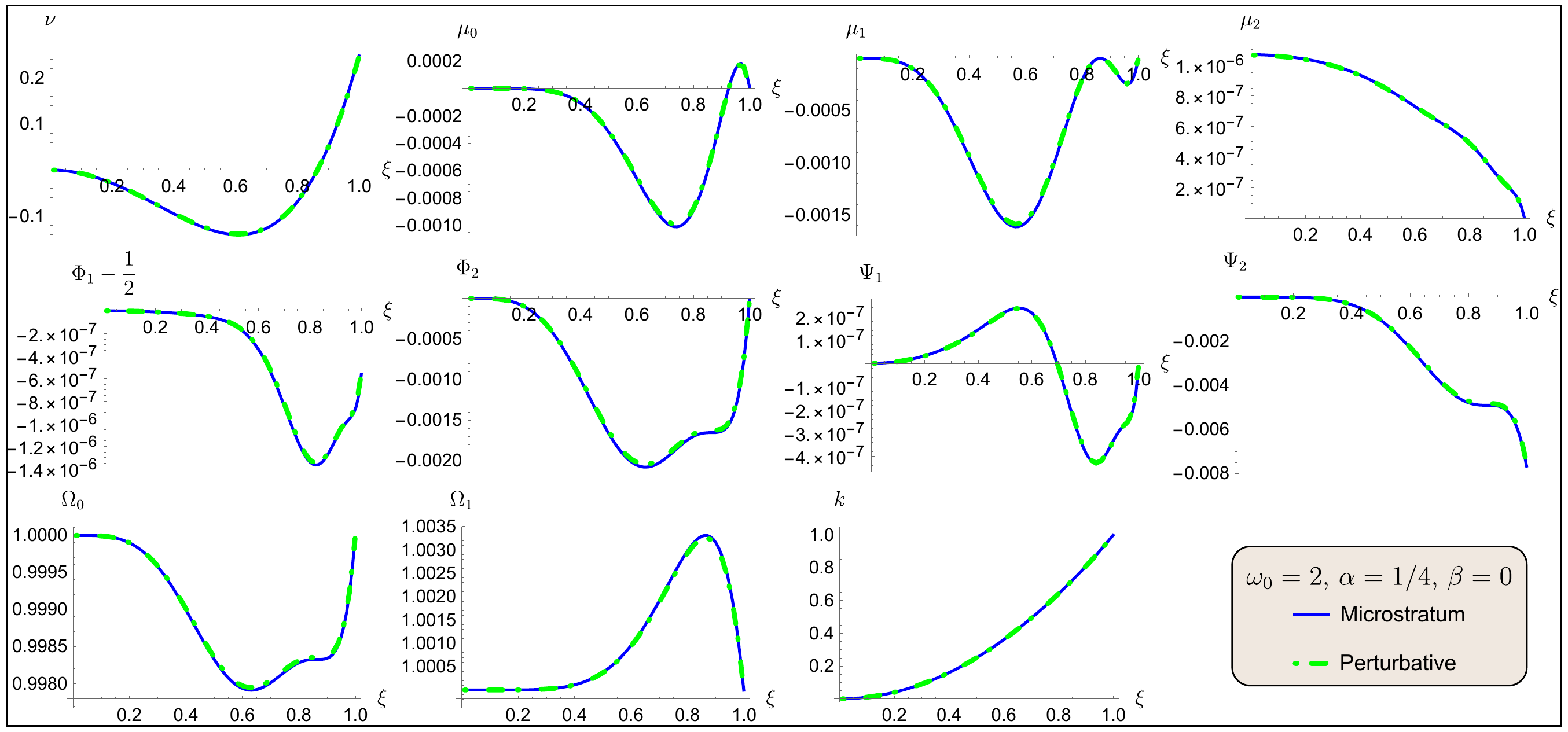}
	\caption{\it A comparison of the series solution and the numerical results for  $\omega_0=2$, $\alpha=\frac{1}{4}$,  $\beta = 0$. The numerics are solid blue and the series solution is dashed-dotted green. Note the close match between the two.}
	\label{fig:fields_perturb_w_2}
\end{figure}

\begin{figure}[ht!]
	\centering
	\includegraphics[width=\textwidth]{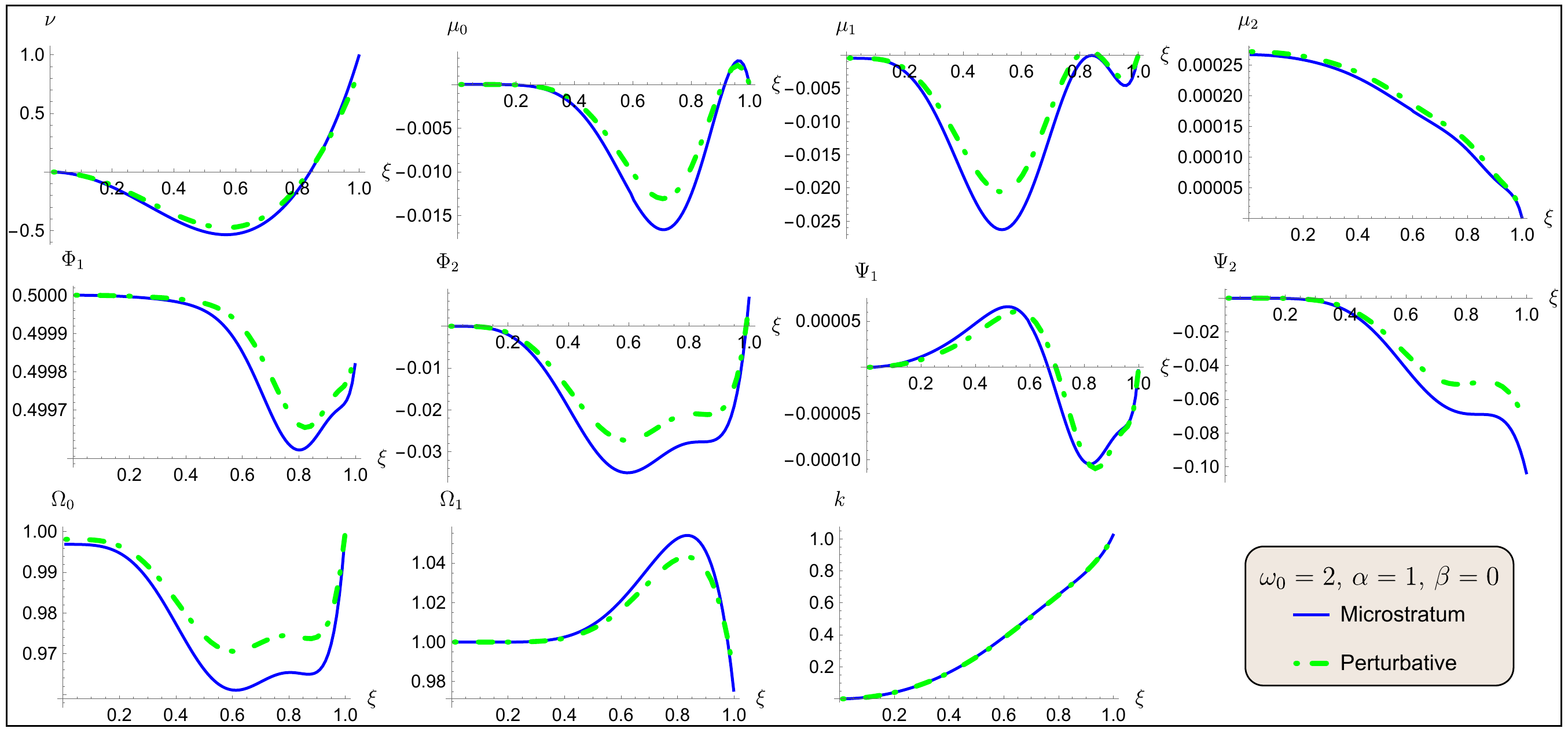}
	\caption{\it A comparison of the series solution and the numerical results for  $\omega_0=2$, $\alpha=1$,  $\beta = 0$. The numerics are solid blue and the series solution is dashed-dotted green. The two are schematically similar but the match is imperfect because $\alpha$ is not small.}
	\label{fig:fields_perturb_w_2_alpha_1}
\end{figure}

We have also generated comparison plots of the numerical and fourth-order series solutions for $\omega_0=2$. The results for $\alpha=\frac{1}{4},\,\beta=0$  are shown in Fig.~\ref{fig:fields_perturb_w_2}, where it is evident that agreement is essentially perfect.   We  show similar plots with $ \alpha=1,\,\beta=0$ in Fig.
\ref{fig:fields_perturb_w_2_alpha_1}.  The agreement is schematically similar, but is not perfect because $\alpha=1$ is far from being a ``small parameter,'' and in this regime the accuracy of the numerics is slightly below our threshold of reliability: the cost function is of order $10^{-7}$.

%%%%%%%%%%%%%%%%%%
%%%%%%%%%%%%%%%%%%

%%%%%%%%%%%%%%%%%%
%%%%%%%%%%%%%%%%%%

%%%%%%%%%%%%%%%%%%
%%%%%%%%%%%%%%%%%%

%%%%%%%%%%%%%%%%%%
%%%%%%%%%%%%%%%%%%

%%%%%%%%%%%%%%%%%%
%%%%%%%%%%%%%%%%%%

%%%%%%%%%%%%%%%%%%
%%%%%%%%%%%%%%%%%%

%%%%%%%%%%%%%%%%%%%%%%
\subsection{Finding the normal modes}
\label{ss:normal}
%%%%%%%%%%%%%%%%%%%%%%

In Section \ref{sec:Perturbative}, we showed that the solutions depend on the choice of a ``zeroth-order'' frequency $\omega_0$, which, for our choice of boundary conditions, will be an even integer, and we showed that the excitations produce a shift in this frequency of the normal mode, (\ref{deltaomega2}) and  (\ref{deltaomega0}). This is to be expected because the excitations generate a change in the shape and depth of the geometry.

%%%%%%%%%%%%%%%%%%
\vspace{0.5mm}
\begin{figure}[ht!]
\centering
\includegraphics[width=\textwidth]{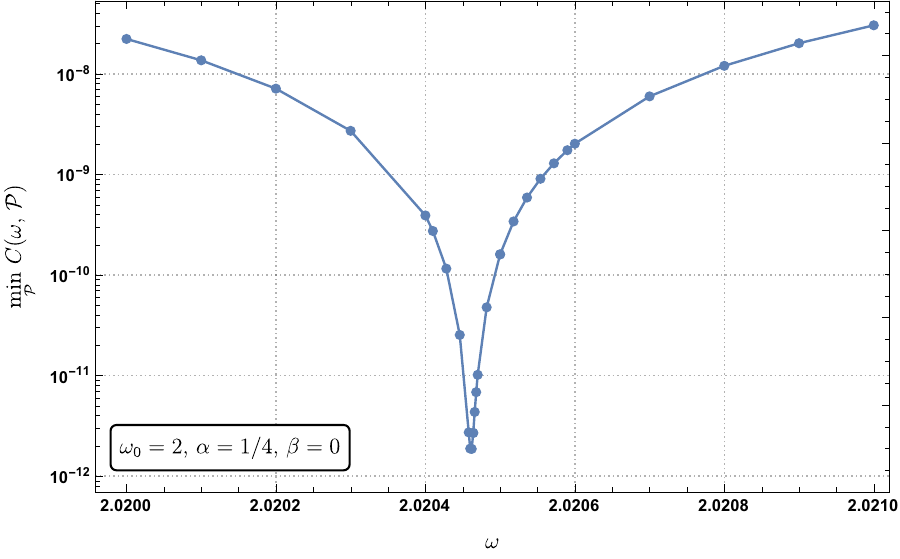}
\caption{\it Plot of the cost function $\min_\mathcal{P} C(\omega, \mathcal{P})$ as a function of $\omega$, with parameters $\alpha = \frac{1}{4}$, $\beta = 0$ and $\omega_0 = 2$. It shows the frequency shift   away from $\omega_0 = 2$.}
\label{fig:cost_function}
\end{figure}
\vspace{1mm}
%%%%%%%%%%%%%%%%%%

The numerical solutions also exhibit the same shift. To track this shift, we compute the minimum of the cost function $\ref{cost_fn}$ at fixed values of $\omega$ around the ``zeroth-order'' frequency  and look for a ``resonance,'' at which the convergence and accuracy of the numerical solution improves dramatically.   A typical result of our search algorithm is shown in Fig.~\ref{fig:cost_function}. The steep dip in the cost function provides a sharp signal of the normal frequency of oscillation.
 
\begin{figure}[ht!]
\centering
\includegraphics[width=.49\textwidth]{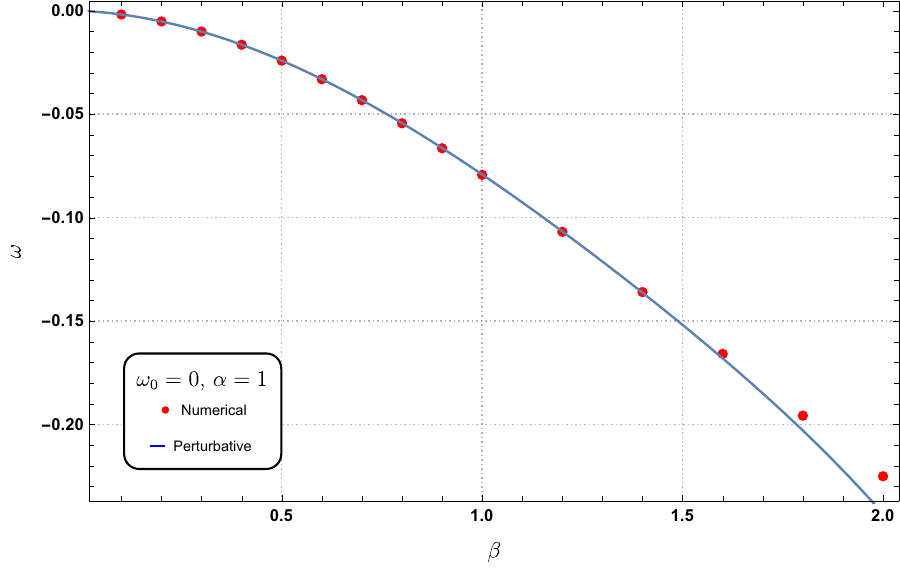}
\includegraphics[width=.49\textwidth]{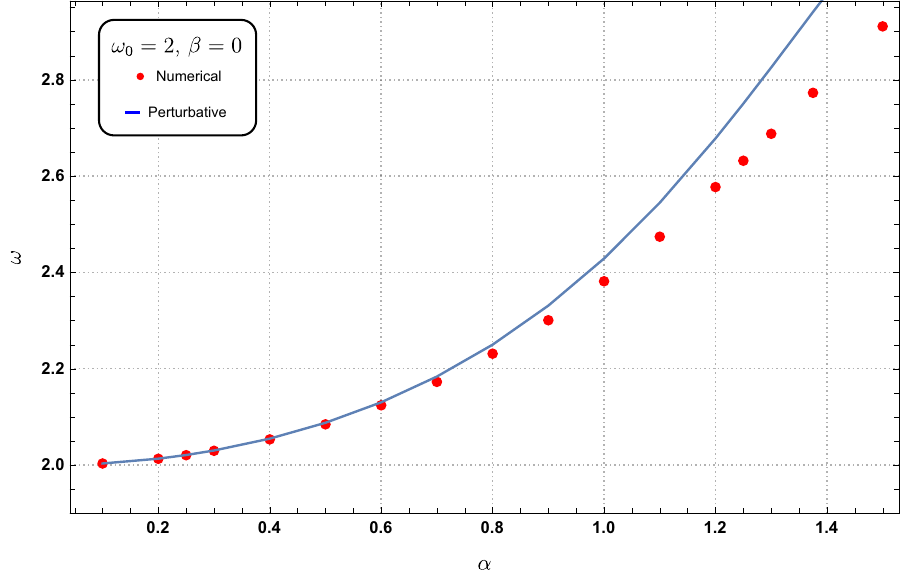}
\caption{\it Plots of the frequencies of the normal modes. The red dots represent the numerical results, while the blue curves are perturbative results, at eleventh order when $\omega_0=0$ (\ref{deltaomega0}), and at fourth order when $\omega_0=2$ (\ref{deltaomega2}). In the first graph we have taken $\omega_0=0$, $\alpha = 1$, and $\beta$ going from $0$ to $2$. The second graph corresponds to $\omega_0= 2$, $\beta = 0$, and $\alpha$ going from $0$ to $\frac{3}{2}$.}
\label{fig:shift_freq_perturb}
\end{figure}

To test that the numerical algorithm is indeed correctly identifying the normal modes, we compared the results of the numerical searches for ``resonances''  with the predictions from the series expansions (\ref{deltaomega2}) and  (\ref{deltaomega0}).  These comparisons are shown in Fig.~\ref{fig:shift_freq_perturb}.  The numerical results   closely match the perturbative computations for a surprisingly large range of $\alpha$ and $\beta$, which means that the numerical search algorithm does indeed provide an effective method of determining the normal modes  of  microstata.

%%%%%%%%%%%%%%%%%%
\vspace{0.5mm}
\vspace{1mm}
%%%%%%%%%%%%%%%%%%

%%%%%%%%%%%%%%%%%%%%%%
\subsection{Closed time-like curves and scaling limits }
\label{ss:CTCs}
%%%%%%%%%%%%%%%%%%%%%%

In superstrata there is a familiar constraint, like  (\ref{ss-smoothness}), on the amplitude of the supertube and superstratum modes that is usually characterized as a regularity or smoothness condition.      However, this condition arises from the combined requirement of well-behaved asymptotics, with no CTC's at infinity, and smoothness at the center of the solution.  In the solutions we construct here, we have built in the smoothness at $\xi =0$ and so this standard smoothness condition will emerge from requiring the absence of CTC's at infinity.    This condition also places a bound on the amplitude of the superstratum modes, and in the limit in which this bound is saturated (and the original supertube modes have vanishing amplitude), the geometry approaches a scaling limit in which it develops  an infinitely long AdS$_2$ $\times S^1$ throat.

For the $(1,0,n)$ superstratum the bound arises from taking the $a \to 0$ limit in (\ref{ss-smoothness}), and, as we will discuss below, this leads to the condition  $\abs{\alpha} \le 2$.  However the scaling limit, $\alpha \to 2$, can be taken in two ways.  The standard limit  is most simply expressed in terms of the conventional superstratum metric written in terms of the radial coordinate, $r$.  One then keeps $r$ finite and takes $a \to 0$.  This produces  an asymptotically AdS$_3$ geometry   and the metric   becomes that of extremal  BTZ black hole. The other way to take this limit was discussed in \cite{Bena:2018bbd}:  One also sends  $a \to 0$ but one keeps $r/a$ finite.  In this  limit one scales with the cap, and the asymptotic AdS$_3$ now goes to an infinite distance: The geometry limits to a smooth cap that is asymptotic to AdS$_2$ $\times S^1$ at infinity.   It is this second limit that appears  in our formulation of microstrata and superstrata:  In using the coordinates $\xi$ and $\tau$, the  parameter, $a$, has been scaled out and thus $\alpha \to 2$ limit of the superstratum will yield the asymptotically AdS$_2$ $\times S^1$ geometry of  \cite{Bena:2018bbd}.  Similarly, the corresponding  limit of microstata will result in asymptotically AdS$_2$ $\times S^1$ geometries, and thus generate a capped, semi-infinite  AdS$_2$ $\times S^1$  throat.

Put differently, we will find a range of parameters for which the coefficient of $d\psi^2$ is negative and diverges as $(1-\xi^2)^{-1} \sim r^2/a^2$ as $\xi \to 1$.  These geometries are asymptotic to AdS$_3$.  At the edge of this range of parameters the  coefficient of $d\psi^2$ is negative and limits to a constant:   These geometries are asymptotic to AdS$_2$ $\times S^1$.  Outside this range of parameters, the coefficient of $d\psi^2$ becomes {\it positive} (and diverges as $(1-\xi^2)^{-1} \sim r^2/a^2$) as $\xi \to 1$.  These contain CTC's and such metrics are unphysical.

\begin{figure}[ht!]
\centering
\includegraphics[width=.49\textwidth]{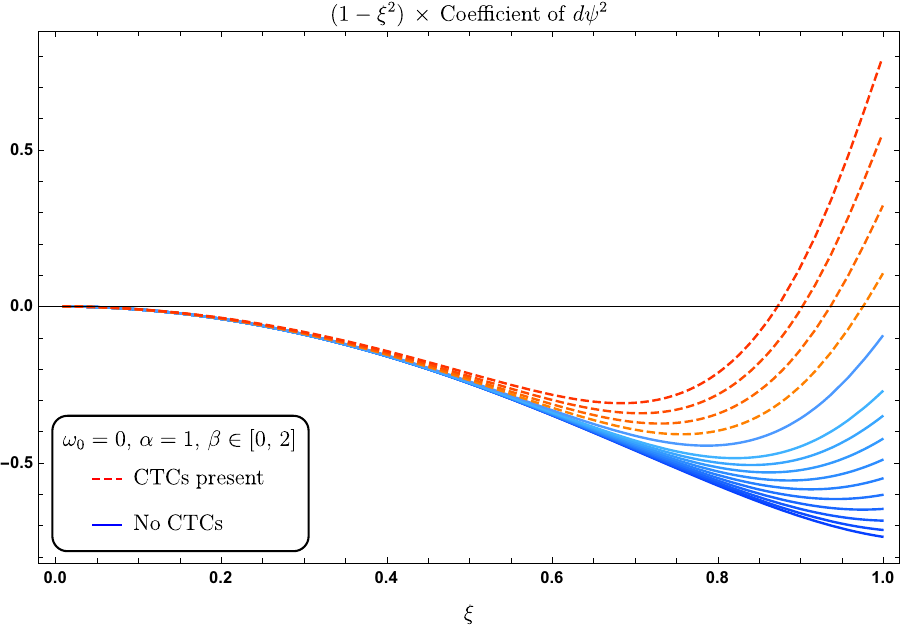}
\includegraphics[width=.49\textwidth]{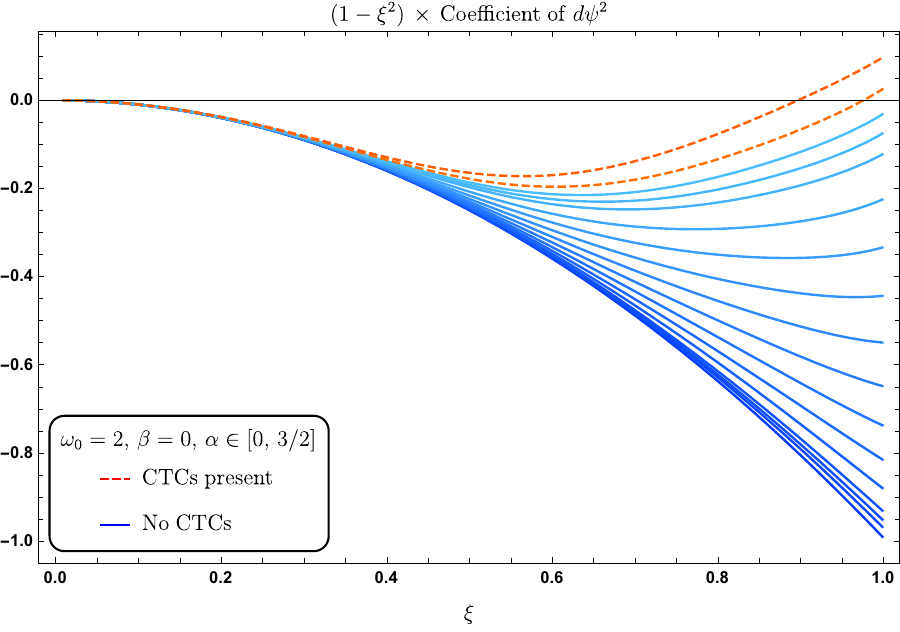}
\caption{\it Plot of $(1-\xi^2)$ times the coefficient of $d\psi^2$ as a function of $\xi$. The curves in solid blue correspond to solutions with no CTC's. The curves in dashed red have CTC's. In the first figure we have taken $\omega_0 = 0$, $\alpha = 1$, with $\beta$ going from $0$ to $2$. In the second figure we have taken $\omega_0 = 2$, $\beta = 0$, with $\alpha$ going from $0$ to $\frac{3}{2}$. (Note that the curves  do not represent regularly spaced values of the parameters.)}
\label{fig:dpsi_sq}
\end{figure}

In Fig.~\ref{fig:dpsi_sq}, we have plotted  $(1-\xi^2)$ times the coefficient of $d\psi^2$ as a function of $\xi$ for two different ranges of $\alpha$ and $\beta$.  We see that the solutions are CTC-free when the parameters $\alpha$ and $\beta$ are sufficiently small and develop CTC's if $\alpha$ or $\beta$ become too large. We thus have a  range of physical solutions. For $\omega_0 = 2$ and $\beta =0$ we find that the solutions are CTC-free for $\alpha \lesssim 1.3$.  For $\omega_0 = 0$ we will discuss the CTC limit in detail below.

As one transitions from the CTC-free families to the families with CTC's, one sees from Fig.~\ref{fig:dpsi_sq} that all the curves have a positive slope near $\xi = 1$.  This means that, in the limiting geometry, in which the coefficient of $d\psi^2$ limits to a constant, this constant is negative (because $-(1-\xi^2)$ has positive slope) and so the $S^1$ in these geometries remains space-like. For $\omega_0 = 0$, we have used the eleventh-order perturbative solution to estimate that this slope is $\sim4$ to very high accuracy.  

It follows that these limiting microstrata are also good physical solutions, and it is in this sense that we mean that microstrata exhibit the same scaling behavior as superstrata.   We similiarly expect infinite red-shifts between the cap and the top of the throat, and so in the scaling limit, excitations that localize in the cap will have vanishingly small energies.

%%%%%%%%%%%%%%%%%%
%%%%%%%%%%%%%%%%%%

We can also use the series analysis  of Section \ref{ss:perturbations} to estimate the limits placed by CTC's on physical ranges of the parameters.  In particular, in Section \ref{ss:higher_orders_perturbation}  we computed  analytic results to  high orders in the perturbation theory for  $\omega_0 = 0$. 
From the series expansion, we can obtain an expression  for the coefficient of $d\psi^2$ that is reliable up to eleventh order, and we suspect that it is probably valid at all orders\footnote{We found a startlingly simple pattern in the series expansion, which we then fit to the rational function (\ref{infinite_throat_limit}).  We then  checked that (\ref{infinite_throat_limit}) gives the  correct result to eleventh order in perturbations.}:
\begin{equation}
    \lim_{\xi\to1}\, (1-\xi^2) g_{\psi\psi} ~=~ -1+\frac{25}{(25-\beta ^2)}\frac{\alpha^2}{4}+\frac{10 \beta ^2}{25+\beta ^2}+\frac{125\, \alpha ^2 \,\beta }{2 (25-\beta ^2)(25+\beta ^2)} 
    \label{infinite_throat_limit}
\end{equation}

\begin{figure}[ht!]
\centering
\includegraphics[width=\textwidth]{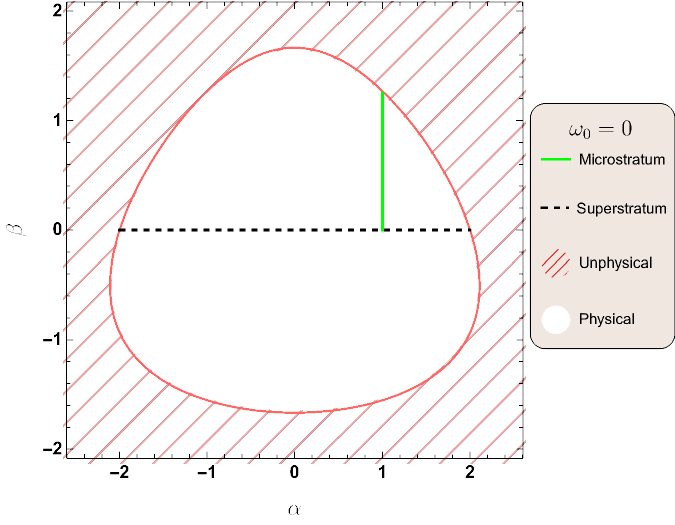}
\caption{\it The physical region of the $(\alpha, \beta)$ parameter space  at $\omega_0 = 0$ based on the (\ref{infinite_throat_limit}). The part of the space we have investigated numerically is given by the horizontal dashed black line and the vertical solid green line. The former corresponds to superstrata, whereas the latter represents microstrata with $\alpha=1$. The solid, egg-shaped, red line delimits the CTC-free region of space, with the hatched region corresponding to unphysical solutions containing CTC's.}
\label{fig:ctc_region}
\end{figure}

The solutions are CTC-free when this coefficient is negative. We have depicted the CTC-free region of $(\alpha, \beta)$-space  in Fig.~\ref{fig:ctc_region}. The region with CTC's is hatched. The limit where this coefficient goes to zero corresponds, as for the superstrata, to the infinite-throat limit of the geometry. It is represented by a red line on the figure.

It is interesting to note that  (\ref{infinite_throat_limit}) is not an even function of $\beta$ and it seems that for small $\beta < 0$, one can have $\alpha >2$, where one should recall that $\alpha = 2$ is the limit set by superstrata.  It would be most interesting to understand what this represents in terms of the physics of microstrata.

% %%%%%%%%%%%%%%%%%%
% \begin{figure}[ht!]
% %
% \centering
% \includegraphics[width=.49\textwidth]{figures/CTC_alpha1_0.pdf}
% \includegraphics[width=.49\textwidth]{figures/CTC_beta0_2.pdf}
% \caption{\it Plots of the size of the compact $y$ direction at infinity, rescaled by a $1- \xi^2$ factor, as a function of the parameters of the model. The right plot correspond to $\omega_0 = 0$, and we have fixed $\alpha = 1$ while varying $\beta$. The left plot correspond to $\omega_0 = 2$, and we have fixed $\beta = 0$ while varying $\alpha$. A positive value means that the solution has CTC's.}
% \label{fig:ctc}
% \end{figure}
%%%%%%%%%%%%%%%%%%

%%%%%%%%%%%%%%%%%%
%%%%%%%%%%%%%%%%%%

More generally, the putative exact formula (\ref{infinite_throat_limit}) raises an interesting question about the general smoothness condition.  Superstratum  smoothness requires  (\ref{ss-smoothness2}) and, for our particular superstratum, only $b_2$ is non-zero.  This condition therefore reduces to:
\begin{equation}
    a^2 R_y^2 g_0^4 ~=~ 1 - \frac{\alpha^2}{4} \,.
    \label{expr_a_superstrata}
\end{equation}
where $\alpha$ is related to $b_2$ via: 
\begin{equation}
    \alpha  ~=~ \frac{2 \, b_2}{\sqrt{b_2^2 + 2 \, a^2}}\,.
    \label{alphareln}
\end{equation}
Note that the superstratum bound $\alpha < 2$ comes from the limit $a \to 0$.
  
Equation (\ref{CTCtest}) shows how the right-hand side of (\ref{expr_a_superstrata}) emerges from
\begin{equation}
    \lim_{\xi \to 1} (1- \xi^2) g_{\psi\psi}
\end{equation}
We also know that the limit $a\to 0$ corresponds to the infinite throat limit, where the size of the $\psi$ circle becomes constant at infinity.

Given the analytic expression for the CTC bound, (\ref{infinite_throat_limit}), one can make an educated guess for the smoothness condition for general values of $a$, $\alpha$ and $\beta$ with $\omega_0 =0$:
\begin{equation}
    a^2 \, R_y^2 \, g_0^4 ~=~1 - \frac{25}{(25-\beta ^2)}\frac{\alpha^2}{4}- \frac{10 \beta ^2}{25+\beta ^2}- \frac{125\, \alpha ^2 \,\beta }{2 (25-\beta ^2)(25+\beta ^2)} \,.
\end{equation}
We note that this reproduces the correct result for $\beta =0$, and  in the limit $a \to 0$.

%%%%%%%%%%%%%%%%%%%%%%
\subsection{Non-extremality}
\label{ss:Non-extremality}
%%%%%%%%%%%%%%%%%%%%%%

As we discussed in Section \ref{sec:Charges}, there are two important notions of non-extremality.  The first, and most basic, is non-extremality relative to the superstratum: namely, how much more mass do our microstrata have relative to superstrata with the same charges.  The other, more stringent, notion of non-extremality is to measure it relative to the  BTZ solution and ask where the microstrata lie with respect to the mass and angular momentum ($J = Q_P$) of a BTZ black hole.  

\begin{figure}[ht!]
\centering
\includegraphics[width=.49\textwidth]{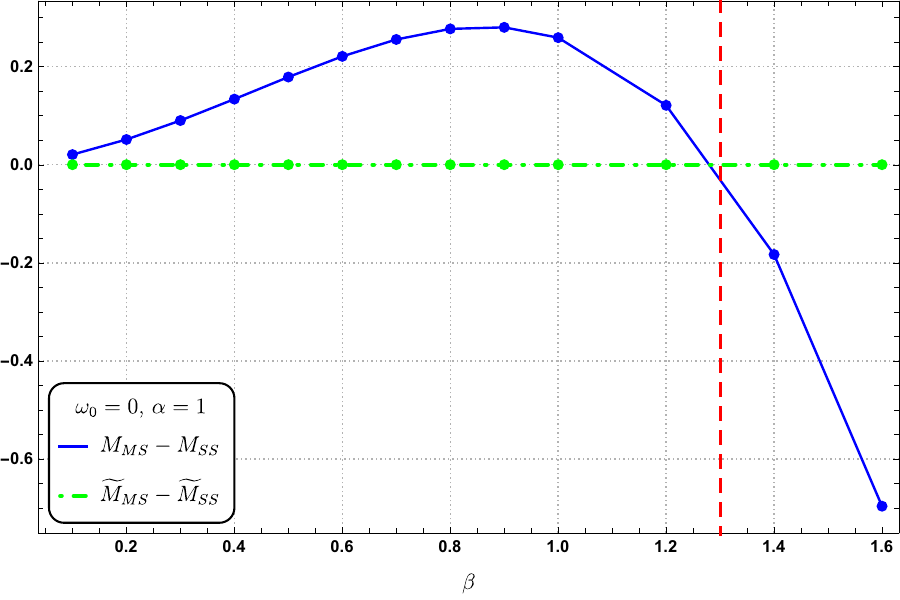}
\includegraphics[width=.49\textwidth]{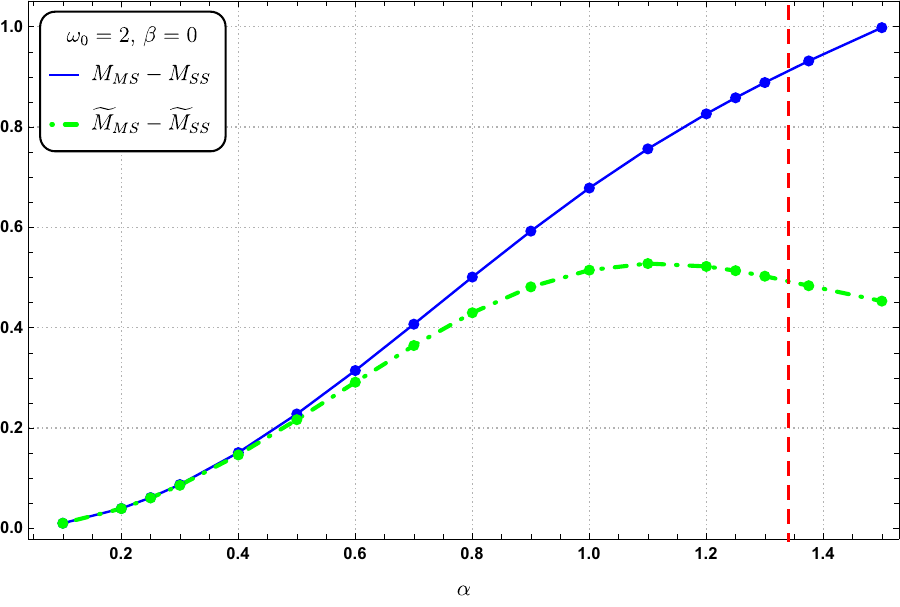}
\caption{\it Plot of the difference, $M_\text{MS} - M_\text{SS}$ and $\widetilde M_\text{MS} - \widetilde M_\text{SS}$, at the same values of $J$. In the first plot we have taken $\omega_0 = 0$, $\alpha = 1$, with varying $\beta$.  In the second plot we have taken  $\omega_0 = 2$, $\beta = 0$, with varying $\alpha$. The vertical red dashed lines correspond to the CTC locus and data points to the right of this line are unphysical.}
\label{fig:mass_comparison}
\end{figure}

We therefore used our numerical solutions to make an asymptotic expansion of the metric at infinity as in (\ref{SSBTZmet})  and extracted the parameters    
$M$, $\widetilde M$ and $J$.   We then plotted the difference between the masses, $M_{MS}$ and $\widetilde M_{MS}$, of the microstratum and the corresponding masses, $M_{SS}$ and  $\widetilde M_{SS}$, of the superstratum with the same charges. The results are shown in  Fig.~\ref{fig:mass_comparison}. 

For $\omega_0 =2$ we find that for all the physical, CTC-free solutions these mass differences are always positive.  (The only solutions where this conclusion fails are unphysical in that they have CTC's.)  The non-extremality of the physical, CTC-free solutions is completely consistent with the supersymmetry breaking and the fact that the microstratum excitations propagate inside the light cone of the dual CFT. 

For $\omega_0 =0$, we find a mixed message. The holographic mass, $M$, suggests non-extremality, while the mass, $\widetilde M$, coming from the time-like Killing vector leads to a vanishing mass difference (to within the accuracy of the approximations).  We will discuss this in Section \ref{ss:Newterritory}.

\begin{figure}[ht!]
\centering
\includegraphics[width=.49\textwidth]{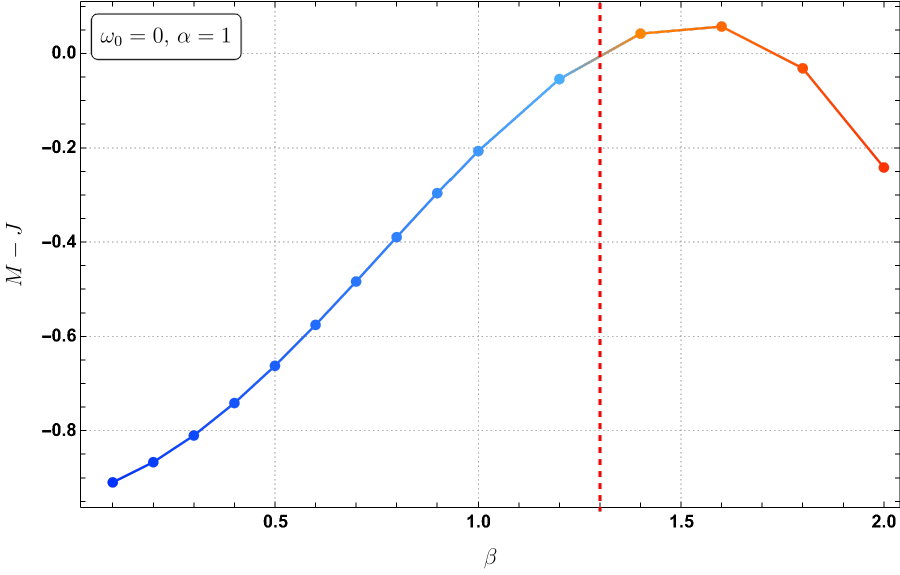}
\includegraphics[width=.49\textwidth]{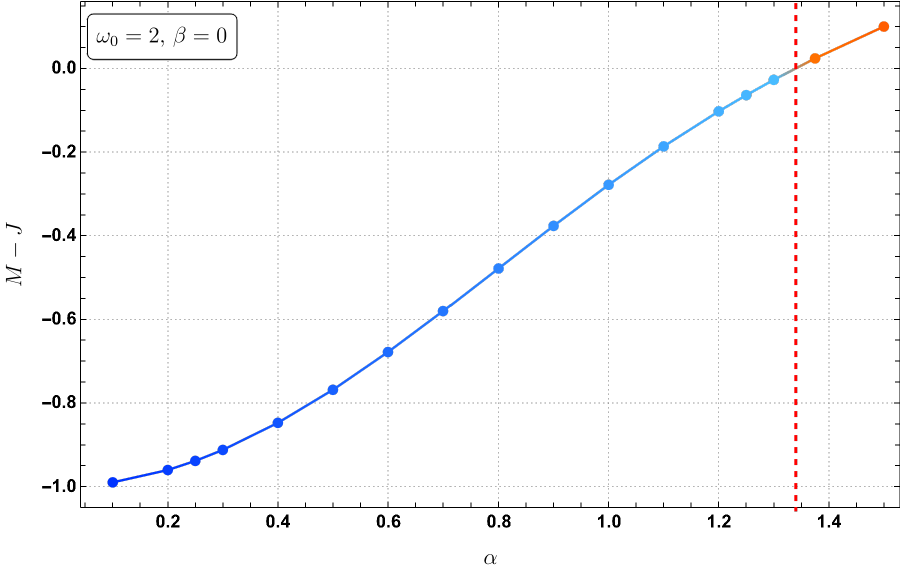}
\includegraphics[width=.49\textwidth]{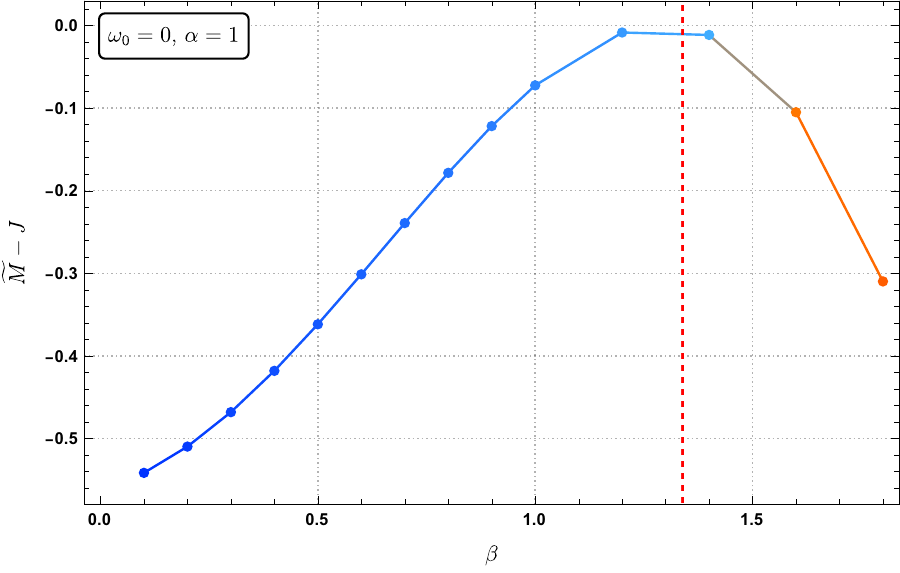}
\includegraphics[width=.49\textwidth]{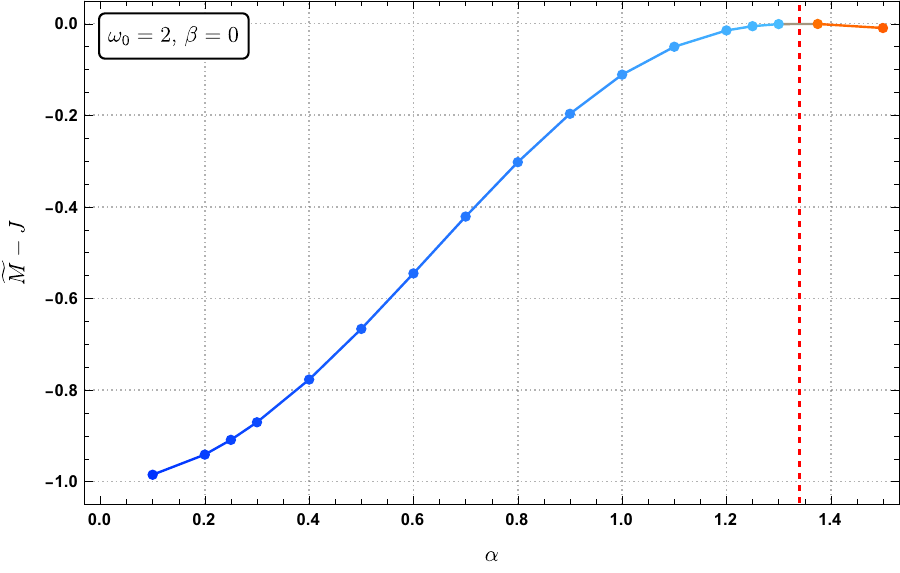}
\caption{\it Plots of $M-J$ and $\widetilde M-J$ for various families of microstrata. In the first plot we have taken $\omega_0 = 0$, $\alpha = 1$, with varying $\beta$.  In the second plot we have taken  $\omega_0 = 2$, $\beta = 0$, with varying $\alpha$. The vertical red dashed lines correspond to the CTC locus and data points to the right of this line are unphysical.}
\label{fig:m_minus_j}
\end{figure}

To track non-extremality relative to the BTZ black-hole metric, we plotted $M -J$ and $\widetilde M -J$ against the parameters for two families of numerical solutions. The results are shown in Fig.~\ref{fig:m_minus_j}.   One should remember that $M-J < 0$ for superstrata and has the value $-1$ for the supertube.  

We find that for both families of microstratum solutions one has $-1 < M-J < 0$ and $-1 < \widetilde M-J < 0$.  Indeed, these quantities were only positive in  solutions  with CTC's.  

%%%%%%%%%%%%%%%%%%
%%%%%%%%%%%%%%%%%%

%%%%%%%%%%%%%%%%%%
%%%%%%%%%%%%%%%%%%

\begin{figure}[ht!]
\centering
\includegraphics[width=\textwidth]{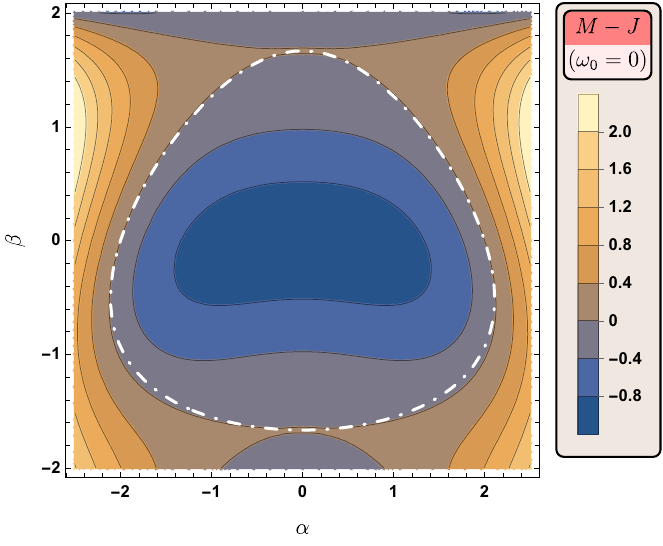}
\caption{\it Contour plot of $M-J$ in the $(\alpha,\beta)$ plane at $\omega_0 = 0$, using eleventh order in perturbations. The white dashed-dotted line delimits the CTC-free region. To within the accuracy of the approximation, the CTC-free region corresponds with $M-J \le 0$.}
\label{fig:m-j_region}
\end{figure}

\begin{figure}[ht!]
	\centering
	\includegraphics[width=\textwidth]{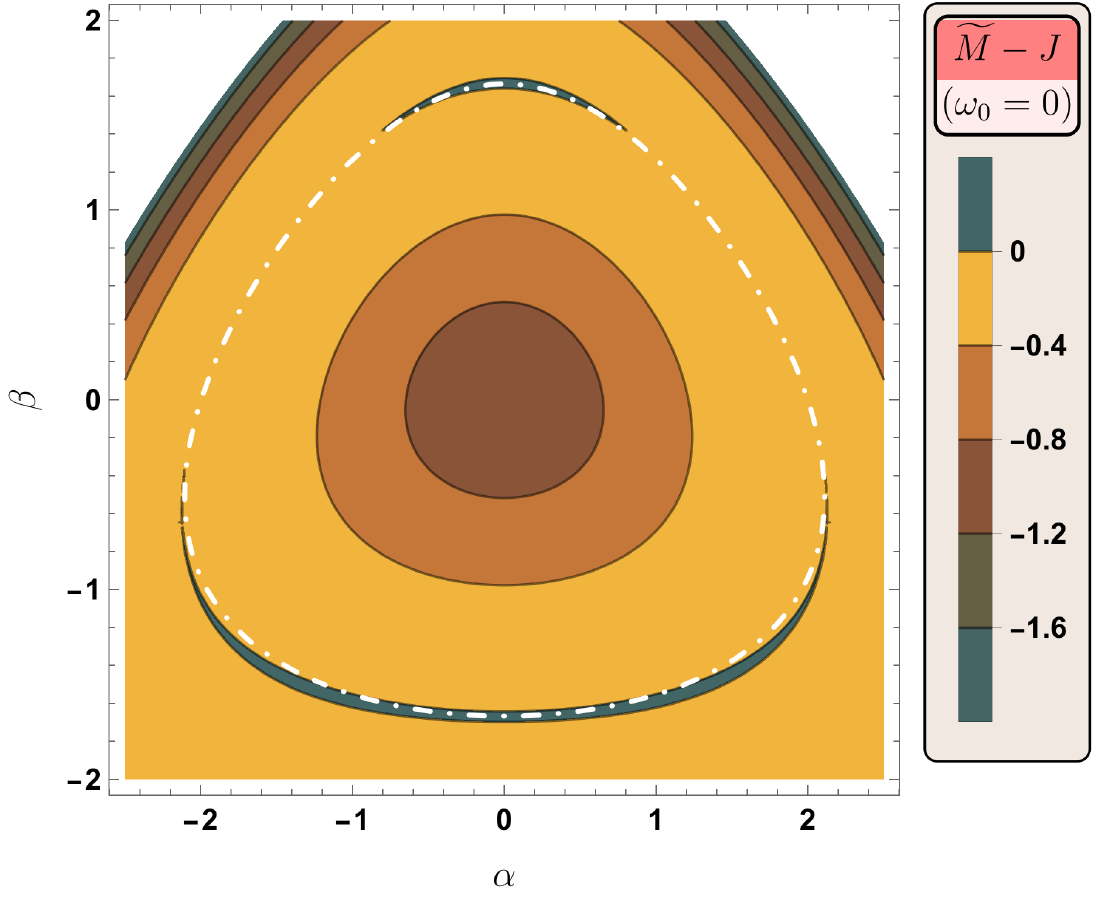}
	\caption{\it Contour plot of $\widetilde{M}-J$ in the $(\alpha,\beta)$ plane at $\omega_0 = 0$, using eleventh order in perturbations.The green dotted line delimits the CTC-free region. To within the accuracy of the approximation, it seems that $\widetilde M-J \le 0$ everywhere, and that the boundary of CTC-free region corresponds with $\widetilde M-J = 0$}
	\label{fig:mTilde-j_region}
\end{figure}

For $\omega_0 = 0$ we have also used the eleventh order perturbative analysis of Section \ref{sec:Perturbative} to compute contours of $M-J$ and $\widetilde M-J$ on the $(\alpha, \beta)$ plane and on these plots we have overlaid the CTC locus determined from (\ref{infinite_throat_limit}).  The results are shown in Fig.~\ref{fig:m-j_region} and Fig.~\ref{fig:mTilde-j_region}.  We see that the  interior of the CTC-free domain obeys $M-J < 0$ and $\widetilde M-J < 0$, and the boundary coincides with $M-J = 0$ and $\widetilde M-J = 0$ to within the accuracy of the perturbation theory. 

It therefore seems that, with our simple microstratum Ansatz, at least for $\omega_0 \geq 0$,  we are unable to break into the non-extremal region of the BTZ phase diagram.

%%%%%%%%%%%%%%%%%%
%%%%%%%%%%%%%%%%%%

%%%%%%%%%%%%%%%%%%
%%%%%%%%%%%%%%%%%%

%%%%%%%%%%%%%%%%%%%%%%
\subsection{Superymmetry breaking and generalized superstrata}
\label{ss:Newterritory}
%%%%%%%%%%%%%%%%%%%%%%

For $\omega_0 >0$ it is evident that supersymmetry is broken and the solutions are non-extremal.  First, the positive frequency, $\omega_\infty$, means that the microstratum momentum lies inside the light cone of the CFT and the supergravity solution explicitly depends upon $t$.  Moreover, both mass parameters, $M$ and $\widetilde M$, of the microstrata exceed those of the corresponding superstrata.

The situation for $\omega_0 = 0$ is a little more ambiguous.  First, these solutions have $\omega_\infty \equiv 0$ for all $\alpha$ and $\beta$. From the perspective of the UV CFT, this means that the excitations are purely left-moving.  The right-moving sector could therefore still be in the Ramond ground state and thus preserve right-moving supersymmetries.  The fact that the energy, $\omega_\infty$, is not modified or lifted by perturbations is highly suggestive of some supersymmetric protection.

Indeed, the deformation by $\beta$ gives the supertube a fundamentally elliptical shape and we know that supertubes of any shape are \nBPS{4}. Naively one would expect that adding left-moving momentum excitations, starting from  $\omega_0 =0$, would produce a  \nBPS{8} configuration.  Thus one might reasonably expect that the $\omega_0 =0$ microstrata could actually be generalized superstrata.

The mass, $\widetilde M$, of these microstrata is consistent with the states being BPS, and hence supersymmetric.  We expect  this mass  would be the one that is relevant to the uplift to six dimensions and the coupling to flat space, and so these solutions may indeed be supersymmetric.  The discrepancy with the holographic mass remains a puzzle.  It is possible that we have not used a suitably general holographic formula  that takes into account the gauge fields and scalars in our rather complicated families of solution\footnote{We are grateful to Rodolfo Russo for suggesting this possibility.}.
 
Independent of the supersymmetry of the  $\omega_0 = 0$ family for general $\alpha$ and $\beta$, we also would like to draw attention to a special locus of solutions:
\begin{equation}
   \beta ~=~ -\frac{1}{8} \, \alpha^2 \,,
    \label{special_locus}
\end{equation}
which exhibits properties that are even more closely matched to the supersymmetric signatures of the superstratum and which we will explore in more detail in (\ref{ss:genss}).

%%%%%%%%%%%%%
\subsubsection{Supersymmetry breaking}
\label{ss:susybreaking}
%%%%%%%%%%%%%

We begin by noting some of the  other interesting features of the asymptotics of microstratum solutions that also signal supersymmetry breaking.  By construction, the ones we have considered here asymptote, at infinity, to the standard supersymmetric AdS$_3$ vacuum.   The original superstrata also asymptote to the supersymmetric AdS$_3$ vacuum in the center of the cap.  Thus the holographic ``flow''  from the UV to the IR goes between supersymmetric vacua and the holographic state changes the intervening geometry.  

This raises the obvious question as to whether microstrata also flow to the supersymmetric  vacuum in the cap.  The answer is generically no, and the best way to see this is to consider the behavior of the scalars fields, $\nu$, $\mu_0$, $\mu_1$ and $\mu_2$, as  $\xi \to 0$.   A quick look at Figures \ref{fig:fields_w_2} and \ref{fig:fields_w_4} suggest that microstrata with $\omega_0>0$ and $\beta =0$ are asymptotic to the superstrata behavior as $\xi \to 0$, however one can see that $\mu_2$ does not  vanish at  $\xi = 0$, and this will probably break supersymmetry.  For $\omega_0 =4$, depicted in Fig.~\ref{fig:fields_w_4}, the value of $\mu_2$ at the origin is of the same order as the numerical errors.  However, the perturbative solutions shown in Figures \ref{fig:fields_perturb_w_2} and  \ref{fig:fields_perturb_w_2_alpha_1} confirm that $\mu_2$ is non-vanishing at $\xi = 0$, and so this also suggests that supersymmetry is broken in the cap.

\begin{figure}[ht!]
\centering
\includegraphics[width=\textwidth]{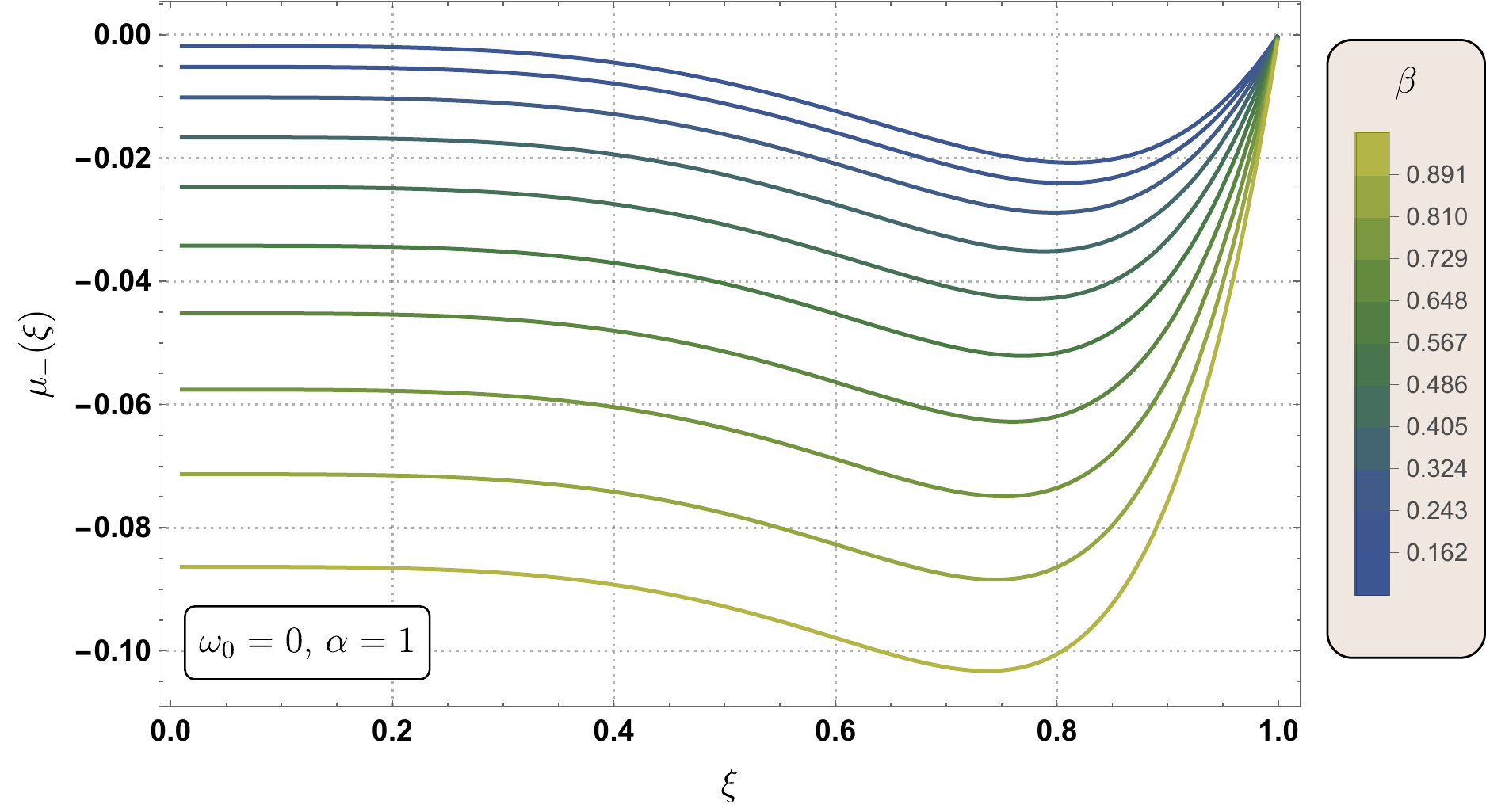}
\caption{\it Exploring the non-supersymmetric vacuum in the cap. Plot of $\mu_-$ as a function of $\xi$, at $\omega_0 = 0$, $\alpha = 1$, and $\beta$ varying from $0.1$ (dark blue curve at the top of the graph) to $1$ (yellow-green curve at the bottom of the graph).}
\label{fig:mu1-mu2}
\end{figure}

\begin{figure}[ht!]
\centering
\includegraphics[width=\textwidth]{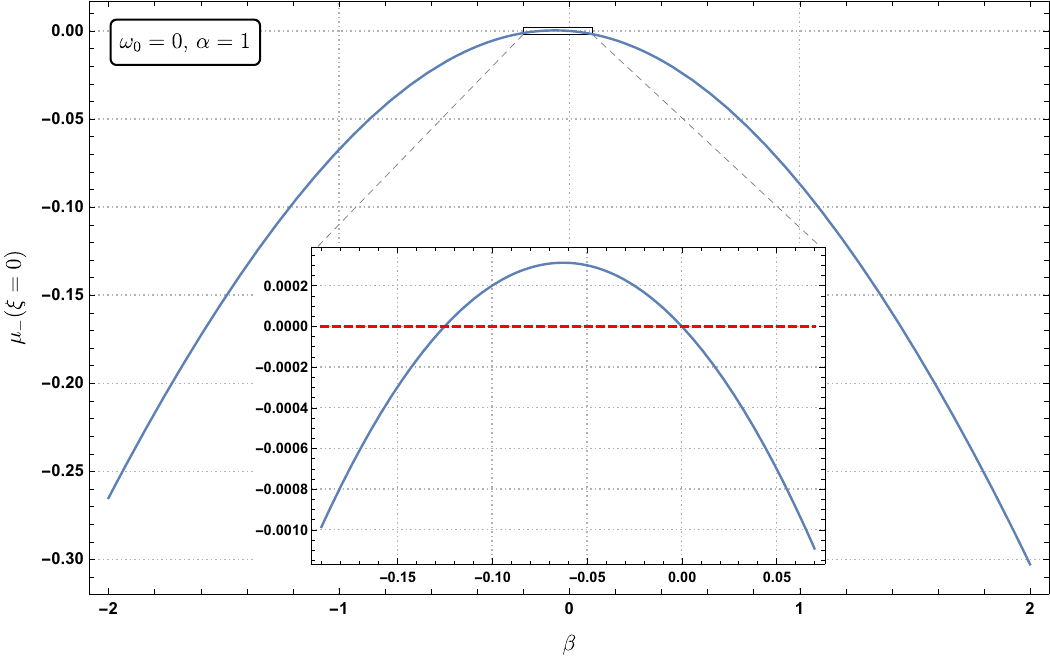}
\caption{\it Plot, based on eleventh order in perturbation theory, of the value of $\mu_-$ at $\xi =0$ for  $\omega_0 = 0$, $\alpha = 1$ and varying $\beta$ .  The main plot shows $\mu_0(0)$ for $-2 < \beta <2$, while the inset zooms on the range $-0.19< \beta <0.07$. Observe that the curve is not invariant under $\beta \to - \beta$ and that $\mu_-$ vanishes at $\beta =0$ and $\beta =-\frac{1}{8}$.}
\label{fig:mu-minus-plots}
\end{figure}

As evident from Figures \ref{fig:fields_w_0} and \ref{fig:fields_perturb_w_0}, the microstrata with $\omega_0 =0$ and $\beta \ne 0$  have non-vanishing $\mu_1$ and $\mu_2$  at $\xi =0$.  Indeed, we find that for $\omega_0 =0$, $\alpha =1$  and $\beta \ne 0$, the scalar  $\mu_- = \mu_1-  \mu_2$ limits to a finite, non-zero value.  We have plotted the behavior of $\mu_-(\xi)$ as a function of $\xi$ for various values of $\beta$ in Fig.~\ref{fig:mu1-mu2} and we have also plotted the value of $\mu_-$ at $\xi =0$ for a larger range of $\beta$ in Fig.~\ref{fig:mu-minus-plots}. 

While it is hard to see from the plots, we also find that the scalar   $\mu_+ = \mu_1+ \mu_2$, for $\omega_0 =0$ and $\beta \ne 0$, does not vanish at $\xi =0$ but is very much smaller than $\mu_-$.  For $\alpha= 1$, and $\beta =0.1$, $\mu_+$ is $1000$ times smaller than $\mu_-$ and as  $\beta$ approaches the CTC region, $\mu_+$  is still about 10 times smaller than $\mu_-$.  In this sense, the solutions with  $\omega_0 =0$ appear to be exploring the neighborhood of the flat direction (described in Section \ref{ss:sugra-action} and defined by $\nu =0$, $\mu_0 =\mu_+ =0$) of the supergravity potential. 

Naively, such behavior would suggest that supersymmetry is broken, but such a conclusion is only valid for purely scalar and metric excitations.  The non-trivial Maxwell fields modify the BPS conditions and this may allow supersymmetry to survive.  

Independent of the possible supersymmetry at general values of $\beta$,   one can see from  Fig.~\ref{fig:mu-minus-plots} that $\mu_-$ not only vanishes at $\beta =0$ (the superstratum solution) but also vanishes at\footnote{We computed this to eleventh order in perturbation theory and obtained  $\beta \approx -0.12500000000054966$  and made the obvious choice.}  $\beta = -\frac{1}{8}$, $\alpha=1$.  This lies on the special locus,    (\ref{special_locus}), and we now investigate this locus in detail.

%%%%%%%%%%%%%%%%%%
%%%%%%%%%%%%%%%%%%

%%%%%%%%%%%%%%%%%%
%%%%%%%%%%%%%%%%%%

%%%%%%%%%%%%%
\subsubsection{Generalized superstrata?}
\label{ss:genss}
%%%%%%%%%%%%%

In this section we focus entirely on the solution with $\omega_0 =0$, and we base our discussion on the results from eleventh order in perturbation theory.   As we have discussed, it is possible that the microstrata with $\omega_0 =0$ are all supersymmetric, and thus represent generalized microstrata.  However, the locus  (\ref{special_locus}) has many additional  features that match those of the original superstrata.  

First, consider the  potential difference:
\begin{equation}
 V_{\Phi_1} ~\equiv~ \Phi_1(\xi =1) ~-~ \Phi_1(\xi =0)   \,,
 \label{Vdefn}
\end{equation}
between $\xi=0$ and $\xi =1$.  The superstratum has a constant electromagnetic potential, $\Phi_1$, and so this difference is zero.  In Fig.~\ref{fig:omega-contours} we have plotted this potential difference,  as  a function of $(\alpha, \beta)$, for $\omega_0=0$.  The superstratum is, of course, $\beta =0$, but the curve (\ref{special_locus}) also stands out as a second branch of the vanishing locus of $V_{\Phi_1}$. 

We  find it intriguing that between the superstratum locus and the new special locus the solution has  $V_{\Phi_1} < 0$, while  outside these loci the solution has $V_{\Phi_1} >0$. These regions therefore seem to define two very different families of microstrata and it would be extremely interesting to study them from the CFT perspective.

\begin{figure}[ht!]
\centering
\includegraphics[width=\textwidth]{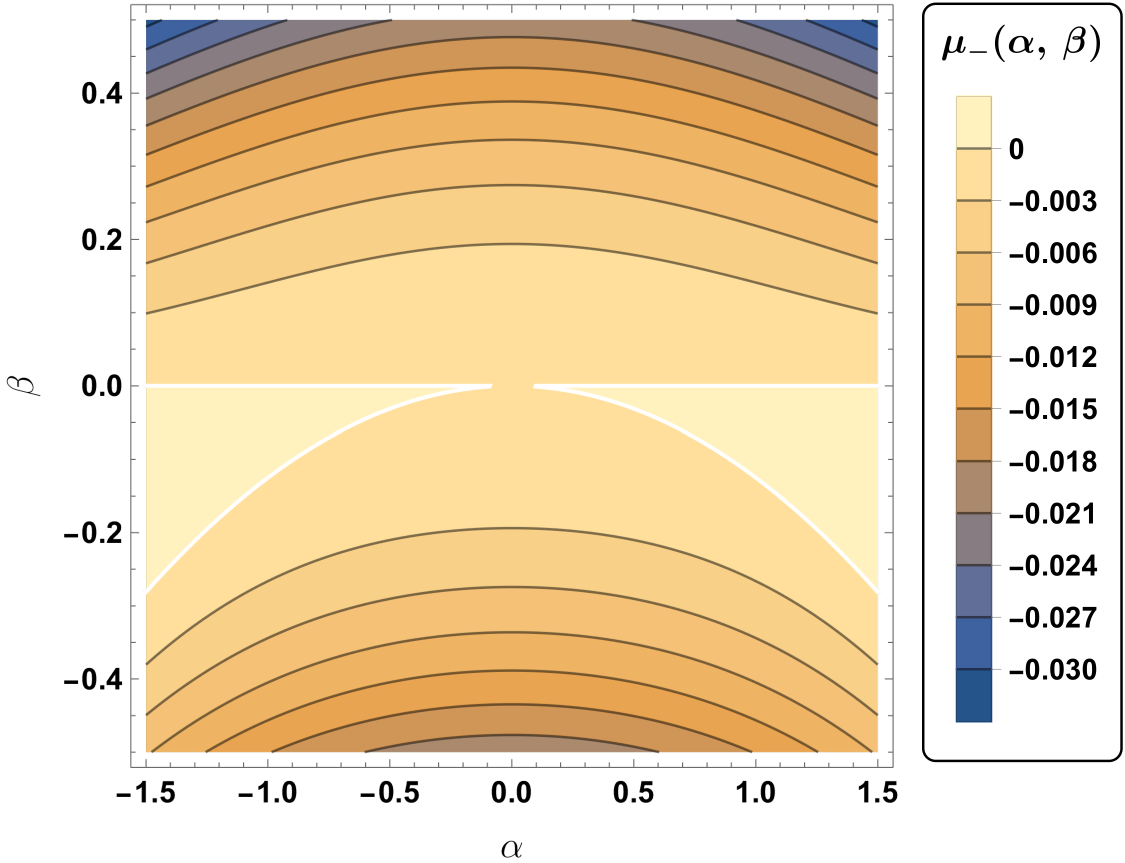}
\caption{\it  Contours of $\mu_-(\xi =0)$ as a function of $(\alpha, \beta)$.  The vanishing loci are highlighted in white, and are defined by $\beta =0$ and (to high perturbative accuracy) by $\beta =-\frac{1}{8} \, \alpha^2$. }
\label{fig:muminus-contours}
\end{figure}

We also find an almost identical plot coming  from the values of $\mu_-(\xi=0)$  as a function of $(\alpha, \beta)$.  This is shown in Fig.~\ref{fig:muminus-contours}.   This suggests that at least for the special locus, the solution in the cap is limiting to the standard supersymmetric vacuum.

To substantiate this further, and to help with future investigation, we catalog the similarities between the special microstratum (for $\alpha =1$, $\beta = -\frac{1}{8}$) and the superstratum (for $\alpha =1$, $\beta = 0$):

%%%%%%%%%%%%%%%%%%
\begin{figure}[ht!]
	\centering
	\includegraphics[width=\textwidth]{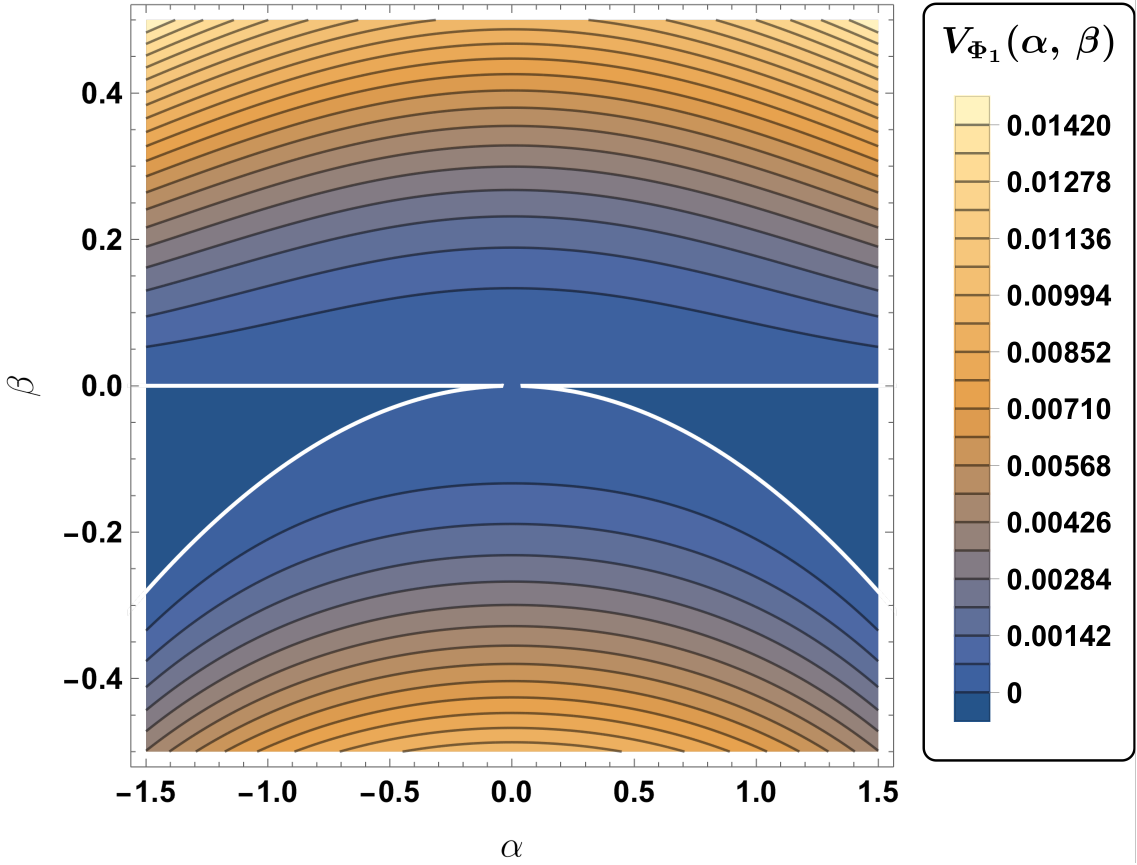}
	\caption{\it  Contours of $V_{\Phi_1}$ as a function of $(\alpha, \beta)$. The  loci $V_{\Phi_1}\,=0$ are highlighted in white, and are defined by $\beta =0$ and, to high perturbative accuracy, by $\beta =-\frac{1}{8} \, \alpha^2$. (These loci meet at $(0,0)$ but this detail is not properly resolved by the plot.)  Between these loci, one has $\omega >0$ and outside these loci one has $\omega < 0$.}
	\label{fig:omega-contours}
\end{figure}
%%%%%%%%%%%%%%%%%%

%%%%%%%%%%%%%%%%%%
%%%%%%%%%%%%%%%%%%

%
\begin{itemize}
	\setlength\itemsep{0pt}
	\item  The scalar function, $\mu_2(\xi)$ vanishes to an accuracy of $10^{-13}$.  It is identically zero for the superstratum, (\ref{ssres1}). Varying $\alpha$ keeps the $\mu_2(\xi)$ amplitude vanishingly small, strengthening our expectations that it should vanish for the special microstratum locus.
	\item  The metric function, $\Omega_1(\xi) = 1$  everywhere, exactly as in the superstratum, (\ref{ssres1}).
	\item  The potential functions, $\Phi_1$ and $\Psi_1$ are given by the superstratum (\ref{ssres1}) values to an accuracy of $10^{-13}$. As with $\mu_2$, varying $\alpha$ does not change this picture qualitatively and we expect that $\Phi_1$ and $\Psi_1$ should vanish identically.
	\item  The scalars $\nu$ and $\mu_j$ all  vanish to better than $3.1 \times 10^{-13}$ at $\xi =0$.  For general $\alpha$ we have verified that these scalars  vanish  to $\cO(\alpha^{12})$  at $\xi =0$.
\end{itemize}
while the differences are
\begin{itemize}
\setlength\itemsep{0pt}
\item  The primary distinction is driven by the scalar, $\mu_0(\xi)$, which is identically zero in the superstratum. The scalar, $\nu(\xi)$, also exhibits a slight difference between the microstratum and superstratum profile.
\item   The scalar, $\mu_1(\xi)$,  potential functions, $\Phi_2(\xi)$, $\Psi_2(\xi)$,   and the metric functions, $\Omega_0(\xi)$, $k(\xi)$ all show small differences between the superstratum and microstratum profiles, but despite the smallness of the difference, these are well within the accuracy of the perturbation theory, and so they are indeed real differences.
\end{itemize}

Another very important property of the microstratum with $\omega_0 =0$ and  satisfying (\ref{special_locus}) is that 
\begin{equation}
  \big |  \mu_1(\xi) ~-~ \mu_0(\xi)  \big|  ~<~ 4 \times 10^{-10}  \,,
\end{equation}
for  $\alpha < 1$.  That is, to high perturbative accuracy, one has:
\begin{equation}
\mu_1 ~=~ \mu_0 \,.
\label{mudegen}
\end{equation}
Combining this with $\mu_2 =0$ and (\ref{mmatformq}) implies that the matrix, $m_{IJ}$, has three eigenvalues equal to $1$ and one eigenvalue of $e^{4 \mu_0}=e^{4 \mu_1}$.  This makes the solution fundamentally distinct from the superstratum. In particular, the degeneracy of the eigenvalues of $m_{IJ}$ mean that  the shape of the $S^3$ preserves  an $SO(3)$ invariance, as opposed to the $U(1) \times U(1)$-invariance of the superstratum.  Amongst other things, this means that the special locus is fundamentally distinct from the original superstratum locus.

Naively, the triple degeneracy of eigenvalues and the $SO(3)$ symmetry suggests that the special locus involves the flattening of the supertube to a line.  However, we note that this $SO(3)$ does not commute with the combined gauge and $U(1)$ symmetry in the $(1,2)$ directions.  This means that the supertube is still spiraling around the $\psi$-direction, and the flattening may be an artifact of the projection along the $\psi$ direction.  We will investigate this further in \cite{GHW1}.

%%%%%%%%%%%%%%%%%%
\begin{figure}[ht!]
\centering
\includegraphics[width=\textwidth]{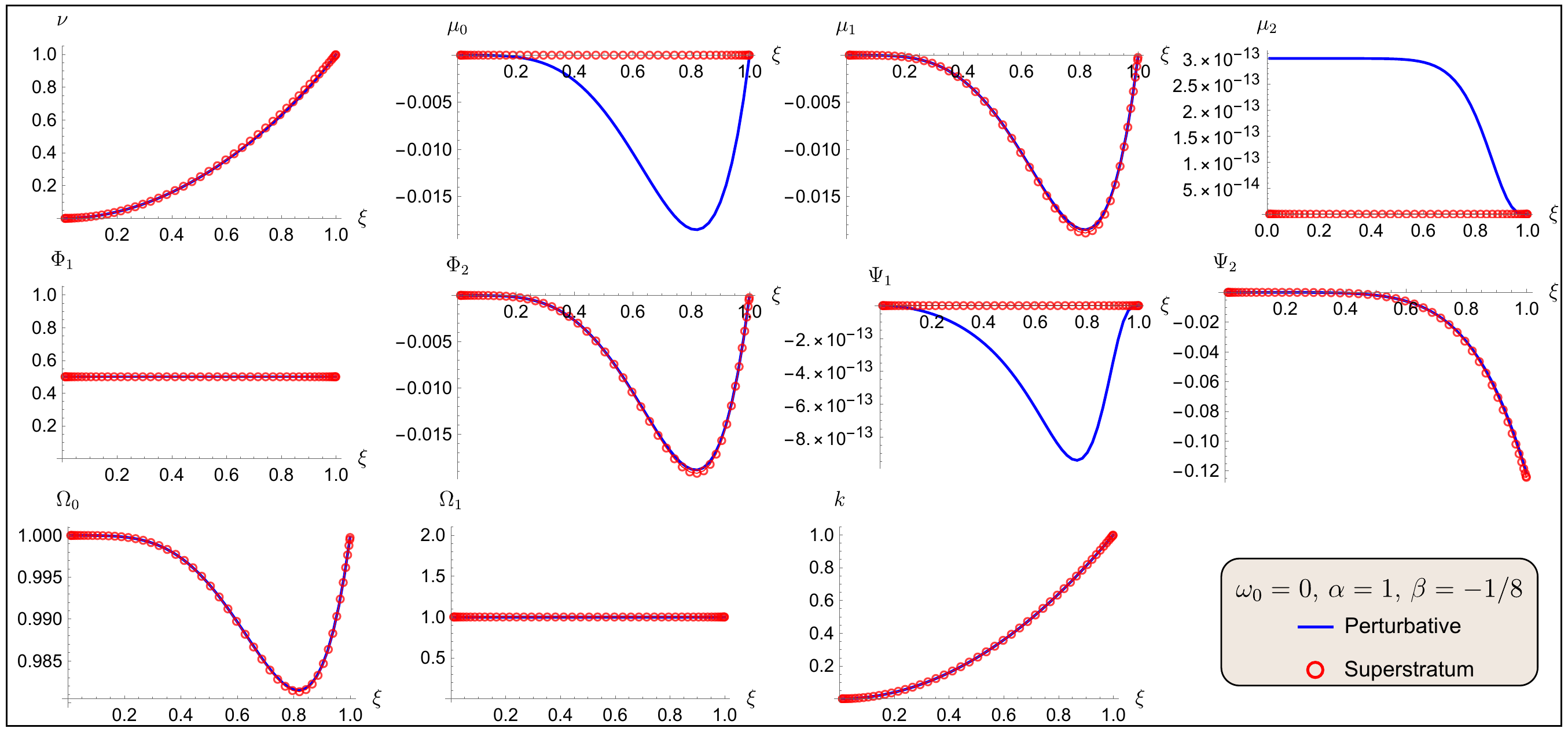}
\caption{\it  The solution for all the fields of the microstratum at $\omega_0 = 0$, $\alpha = 1$ and $\beta = -\frac{1}{8}$ are shown as solid blue lines.  The corresponding superstratum solution ($\alpha = 1$, $\beta = 0$) is the red line with circles. The apparent differences for $\mu_2$ and $\Psi_2$ are due to insignificant errors in the perturbation expansion.  The most significant difference is in $\mu_0$, and there are also some similar differences in some of the other functions: see Fig.~\ref{fig:gen-ss-diff}. }
\label{fig:gen-ss-comp}
\end{figure}
%%%%%%%%%%%%%%%%%%

%%%%%%%%%%%%%%%%%%
\begin{figure}[ht!]
\centering
\includegraphics[width=\textwidth]{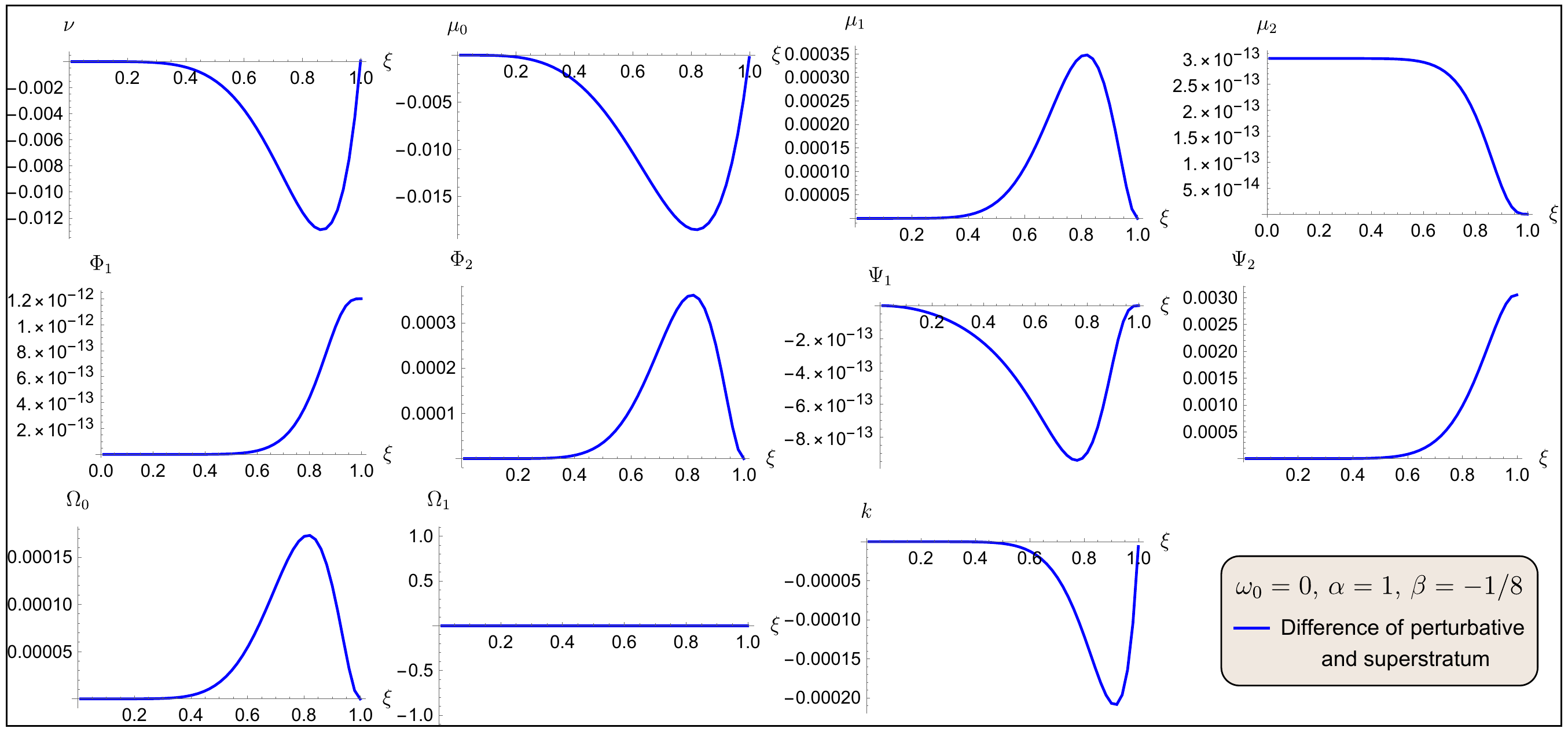}
\caption{\it  The differences between all the functions for the  microstratum at $\omega_0 = 0$, $\alpha = 1$ and $\beta = -\frac{1}{8}$  and the corresponding superstratum solution ($\alpha = 1$, $\beta = 0$). There are small, but significant differences in $\nu$, $\mu_0$, $\mu_1$, $\Phi_2$, $\Psi_2$, $\Omega_0$ and $k$.}
\label{fig:gen-ss-diff}
\end{figure}
%%%%%%%%%%%%%%%%%%

%%%%%%%%%%%%%%%%%%%%%%%%%%%%%%%%%%%%%
\section{Final comments}
\label{sec:Conclusions}
%%%%%%%%%%%%%%%%%%%%%%%%%%%%%%%%%%%%%
%\vspace{0.8cm}

We have constructed, both in perturbation theory and numerically,  families of non-extremal microstrata.  There is excellent accord between the numerical and perturbative results and this gives us a high level of confidence in the accuracy of our solutions and in our predictions for the microstrata frequencies.   It is important to stress that the microstrata with $\omega_0>0$  are not just non-supersymmetric, or non-BPS, but are genuinely non-extremal solutions, whose mass exceeds that of a BPS superstratum with the same charges.  Our primary goal in this paper has been proof of concept: showing that  such microstrata exist and can be constructed using the ``Q-ball/coiffuring'' trick.  In achieving this we found a new families of microstrata involving fluctuations in $\mu_0$, parametrized by $\beta$, and, for  $\omega_0 =2,4$,   we also have non-trivial, non-extremal deformations of the underlying superstratum parametrized by  $\alpha$. 

The existence of non-extremal microstrata demonstrates, once and for all, that superstrata are not merely isolated supersymmetric freaks whose slightest perturbation will result in singularities.  However, our successful microstrata constructions are not immune to becoming singular or suffering from the presence of CTCs at diverse edges of their moduli spaces.  Indeed we saw that, just like superstrata, if the amplitudes become too large then the solution develops closed time-like curves. From the perspective of microstate geometries and fuzzballs, the presence of singularities or CTCs within the geometry is not a pathology that implies one should go back to working with black holes. Such problematic edges of moduli space are a consequence of either too much symmetry, or suppressing low-mass degrees of freedom that are becoming important and that will resolve the singularity.  Thus, while this paper represents  a very significant advance for the microstate geometry program, the goal posts will now move, and continue to move, until we have mapped out every branch of every moduli space of every family microstate of geometries and show how they provide a network of coherent structures throughout the  space of fuzzballs. 

One of the other important features of the microstrata constructed here, and their generalizations based on our approach, is that they have been constructed in a gravity theory for which the holographic dictionary has been mapped out and thoroughly tested.  This makes these microstrata especially interesting as one already has the tools and background necessary to relate these geometries to excitations of the D1-D5 CFT. 
 
In retrospect, the dependence of microstratum frequencies on their amplitudes should not be very surprising.  The normal modes of oscillation will naturally depend on the shape of the geometry and particularly upon the depth, or red-shift, between the cap and the top of the AdS throat.  Since we know that these geometric features  depend on the amplitudes, $\alpha$ and $\beta$, of the microstratum, it must follow that these amplitudes lead to non-linear shifts in the frequencies of the normal modes.   It is, however, gratifying to see this explicitly and it also suggests how microstate geometry fluctuations will lead to the development of a chaotic spectrum.  As we noted in the introduction,   the non-linear dynamics   will re-shuffle the spectrum,  and may well result in the eigenvalue repulsion that is characteristic of non-integrable field theories. 

Apart from the questions about the holography of microstrata, there are a multitude of possible future directions arising from our results within supergravity. 

First, we constructed our microstrata by making reductions, consistent truncations, implementing a Q-ball Ansatz,  focussing on a single mode number, $n=2$, and restricting to three fundamental frequencies, $\omega_0 = 0,2,4$.  We  have thus mapped out a tiny  corner of the possible families of microstrata within the scope of our analysis.  There are thus some simple generalizations that are potentially very interesting.  

When the  frequencies of the microstrata are positive we  expect that they represent solitonic configurations propagating inside the light cone of the CFT. Indeed, for $n=2$  the linearized solutions have what one would naturally associate with a collective velocity (with regards to the propagation of such configurations), whose Minkowski norm is proportional to the quadratic form\footnote{The shift is created by a mixture of $n$ and the electrostatic potential, but the spectrum is invariant under $\omega_0 \to - \omega_0 - 6$ and so this is the correct quadratic form.} $\omega_0(\omega_0 + 6)$. This is negative for  $-6 < \omega_0 <0$ and positive outside this range.  The  frequency $\omega_0 =-6$ corresponds to the right-moving anti-superstratum, and  $\omega_0 =-4$  is mapped to $\omega_0 =-2$ by sending $y \to -y$.  This leaves  $\omega_0 =-2$ as an interesting new possibility (and there will be more such possibilities for $n >2$).    Linearized analysis shows that these solutions, while smooth, involve mixed series  involving $\log (1 - \xi^2)$ and $\log \xi$ and  powers of $\xi$.    The series expansions needed to initiate the numerical analysis are significantly more complicated and so we restricted our attention here to the  sector with $\omega_0 >0$.    We will examine the negative values of $\omega_0$ in future work \cite{GHW2}.  One  might hope that because microstrata with $-6 < \omega_0 <0$  have smaller, or even vanishing $Q_P =J$,  this ``space-like'' range of momenta could lead to microstrata with $M -|J| >0$. 
 
We also note that the discussion of these qualitatively different microstata are based on the linear seed solutions and their unperturbed frequencies, $\omega_0$. The fully back-reacted solutions have metric deformations and shifted frequencies that will modify our discussion of the Minkowski norm.  Small perturbative corrections will not push microstrata across a light cone, and so  the linearized solution will determine whether the velocity is space-like or time-like for small parameters.  If the parameters become too large, it is possible that the velocity could move between being space-like and time-like, but we expect that this would involve the emergence of CTC's,  as described in Section \ref{ss:CTCs}, and the solution will become unphysical.

On a more physical level, by focussing on $\omega_0 \ge 0$, we have restricted our attention to ``time-like'' microstrata.    There should also be families of space-like microstrata that consist primarily of space-like excitations.  These could emerge  as  a final  state of some superstratum ``collision'' that has come to equilibrium with the excitations of the graviton multiplet.   

The solutions $\omega_0 = 0$ and $\beta \ne 0$  are particularly interesting.  First, recall that $\beta =0$ corresponds to the superstratum and that turning on $\beta$ makes an elliptical deformation of the underlying supertube.   In Section \ref {ss:Newterritory},  we gave several pieces of evidence that these solutions are still supersymmetric, \nBPS{8} backgrounds.  First, the supersymmetry of the superstrata should be robust against such elliptical deformations.  As further evidence for this, we found that the frequency of the solution relative to the supertube at the UV fixed point, $\omega_\infty$, is identically zero.  Equivalently, we found that the gauge invariant quantity, $G_{\Phi_1}$, defined in  (\ref{gauge_inv_tems}) is exactly the same as for the supertube.  This  means that the CFT excitations are all purely left-moving. In addition, the mass $\widetilde M$, suggests that these solutions saturate the BPS bound.

Independent of the status of the solutions with  $\omega_0 = 0$ and general $\beta$, we also identified a very interesting special locus (\ref{special_locus})  on which many of the fields vanish, as they do in the original superstratum.  We also saw that the scalar matrix degenerates in a different way from the superstratum, which means that the underlying supertube has a  very different configuration.   We are currently investigating the supersymmetry of all of these solutions as well as their underlying geometries.

As we described in Section \ref{ss:Ansatze}, there are slightly more general Ans\"atze that result in equations and actions that also only depend on the radial coordinate, $r$. These include the possibility of interacting and colliding superstrata in different sectors of the theory, and perhaps will allow us to see how such a collision could develop into a more ``relaxed'' ground state.  
In another vein, there are  going to be time-independent families of single-variable solutions that represent oscillating end-states of microstate geometry interactions.  

Investigating all these solutions will require a more careful series analysis than the approximate series we used in Section \ref{sub:series_expansions}, especially because we need to get control of the log terms in the more general problems.   Indeed, if one improved some of the systematic errors arising from the series expansion, it would be really interesting to  construct and investigate, with high accuracy, the deep, scaling microstrata that arise at the limiting values of $\alpha$ and $\beta$ discussed in Section \ref{ss:CTCs}.  Thus, there is much to be done in a more systematic investigation of the ideas explained here. 

Moving in a more ambitious direction, one can try to find solutions that depend on more that one variable. There are trivial extensions of our ``Q-ball'' truncations that  suppress only one variable, leaving an action that depends on two variables.  Perhaps the best choice for a first attempt at this would be to look for time-independent ground states of the microstratum system.  Another possibility is to use perturbative methods to explore two-variable solutions in the vicinity of the one-variable solutions considered here.  Put differently, starting from the kinds of microstrata we have constructed here, there are evidently going to be vast families of microstrata solutions ``nearby.'' That is, by turning on additional excitation modes, or allowing dependence on more variables, one will be able to access  many microstrata that have non-trivial dependence on all the variables in the three-dimensional supergravity.   

The extension of super-JT gravity discussed in Section \ref{ss:superJT} has similar mathematical underpinnings, but has an entirely different goal.  One of the surprising features of JT gravity is that it is a very simple ``bottom-up'' model whose Euclideanization can capture some of the universal aspects of the spectrum of black-hole microstates.  It would be very interesting to connect this approach to ``top-down'' stringy models  and  use this connection to generalize JT gravity in a way that might capture ``microstratum-inspired'' features of black-hole microstructure.   
 
Returning to the search for new microstrata,  it may be possible  to distill the core elements of the ``Q-ball'' trick we found in three dimensions and  implement it directly in six dimensions.  Doing this might enable one to get beyond the limitations imposed by the consistent truncation Ansatz. One of the issues with this Ansatz is that it restricts generic $(k,m,n)$ superstrata to $k=1$, and these have rather weak fall-off at infinity.  Making the ``Q-ball'' trick work for higher values of $k$ would be most interesting.  

By construction, the superstrata we have considered here asymptote, at infinity, to the supersymmetric AdS$_3$ vacuum.  We have seen that there are also limiting microstrata that asymptote to AdS$_2$ $\times S^1$ and generalize the limiting superstrata studied in \cite{Bena:2018bbd}.  As we noted in  Section \ref{ss:CTCs}, these microstrata seem to develop infinitely long throats and exhibit the same scaling behavior as superstrata.  As we remarked above, it would  be interesting to construct accurate scaling microstrata, along with their limits whose geometries asymptote to AdS$_2$ $\times S^1$,  and verify that they do indeed share the same features as scaling superstrata.

Then there is the very important problem of microstrata in asymptotically-flat space.  This has to be done in the six-dimensional formulation.  For superstrata, the extension from asymptotically-AdS to asymptotically-flat space is a straightforward computational process.  Interestingly enough, the coupling to flat space breaks the coiffuring so that the geometry depends on the same variables  as the fluctuating fields, except that these non-coiffured terms can be made small in deep-scaling superstrata \cite{Bena:2020yii}.   One might hope to make a similar approach to that of   \cite{Bena:2020yii}, and treat the coupling to flat space perturbatively, using it to compute the Hawking radiation. Indeed, probably the simplest first attempt at this difficult, but extremely important, problem would be to use the $\mu_0$ excitation, parametrized by $\beta$, in the microstratum with $\omega_0 = 0$, and generalize the perturbation theory of Section \ref{omega0perts}.   This has the advantage of being very close to the starting point of  \cite{Bena:2020yii}.

More broadly, it is valuable to consider the microstrata constructed here in wider context of constructing smooth, solitonic solutions in supergravity.  The key new ingredient in using the ``Q-ball trick'' is to break the supersymmetry by introducing a time-dependence in some of the fields while cancelling it in the energy-momentum tensor and the electromagnetic currents. This is how such configurations dodge the limitations of the theorem in  \cite{Gibbons:2013tqa}:  this theorem about solitons specifically excludes time-dependent matter fields.  Bose stars are entirely supported by such scalar dynamics, but are extremely finely-tuned with essentially no structure and no moduli. What makes microstrata far richer is that they are supported by both topological fluxes and  by  scalar dynamics that provides additional energy and momentum to source the ``classical lump,'' or soliton.  The resulting hybrid has all the rich structure of superstrata combined with the non-supersymmetric scalar dynamics of Bose stars.  This suggests that the study of Bose stars, and their properties, would be hugely enriched  by incorporating higher-dimensional physics, and especially the fluxes,  that are central to the construction of microstate geometries. 
 
At the more formal level, our ``Q-ball'' Ansatz involves a new scalar field, $\mu_0$, whose amplitude is parametrized by $\beta$, and this field has not, so far, played a major role in the construction of superstrata.  As we noted in Section \ref{sec:ActionBCGC} this generates elliptical deformations of the supertube and the holographic dual of such deformations is part of the standard lexicon \cite{Kanitscheider:2006zf,Kanitscheider:2007wq,Taylor:2007hs,Giusto:2015dfa,Bombini:2017sge,Giusto:2019qig, Tormo:2019yus, Rawash:2021pik}.  Such elliptical supertubes were also  examined from the perspective of the world-sheet CFT in \cite{Martinec:2020gkv}. Given our extensive knowledge of the properties of such supertubes, it seems likely that there are \nBPS{8} momentum excitations on them, and these should lead to generalized  superstrata.  The results of Section \ref{ss:Newterritory} suggest that these generalized  superstrata will be limited to the locus $\beta = - \frac{1}{8} \alpha^2$.  It would be very interesting to confirm the supersymmetry by analytic computation in gravity, and to understand, from within the CFT, why this locus is special, and why the states of the superstratum are ``lifted'' outside the loci  $\beta = 0$ and $\beta = - \frac{1}{8} \alpha^2$. 
  
One of the obvious limitations of our microstrata is that, by reducing to three dimensions, we have locked  two of the charges to their BPS values:  that is, we have locked in the pure D1-D5 structure.  Our non-BPS deformations all relate to shifting the mass relative to the third charge, $Q_P$ or $J$, in the three-dimensional formulation.  One would obviously like to achieve similar things with the D1 and D5 background charges, and there are certainly suggestions as to how one might achieve such a thing \cite{Mathur:2013nja}.  Maybe one can gain some insight by taking duals of the microstrata constructed here and see how the sub-extremal momentum charge translates into other sub-extremal brane charges. 

This paper  also presents some tantalizing challenges for  the holographic field theory, especially because of our  perturbative analysis.  Indeed, one should be able to replicate some aspects of the gravitational perturbation expansion in $\alpha$ and $\beta$ directly from the CFT.  

Perhaps the most intriguing conceptual challenge  is  the appearance,  for $\omega_0 >0$,   of ``non-normalizable'' terms, like those of (\ref{delta3nu}),  in the scalar, $\nu$, arising at third (or higher orders) in perturbation theory.     The ``normalizable'' terms in  $\nu$ have the standard interpretation of representing states of the system, and this is entirely consistent with the idea that microstrata are duals of states in the D1-D5 CFT.  The appearance of  non-normalizable terms  in modes of a field of conformal dimension, $\Delta =1$, suggests that the holographic field theory of microstrata necessarily involves a relevant, massive deformation of the CFT that drives a flow to a new IR fixed point.  Indeed the holographic dual operator of $\nu$ is well-known,  (see equation (2.6) in \cite{Giusto:2015dfa}\footnote{We are grateful to Rodolfo Russo for pointing us towards this equation.}), and  it is a fermion mass term.

Thus microstrata with $\omega_0 >0$ seem, at higher orders in perturbation theory, to break the conformal invariance in the action of the holographically dual field theory.  One possible interpretation of our results is that the state created by such  a microstratum coincides with a perturbative state of the CFT up to second order in perturbation theory, however, it is only at higher orders that one discovers that the holographic state really belongs to a massive perturbation of the CFT. Hopefully this could be explored within the CFT itself.   It is also instructive to note that all these non-normalizable terms can be cancelled in microstrata with $\omega_0= 0$, and so the breaking of conformal invariance in the field theory action only arises for $\omega_0 >0$. 

It is, of course, possible that this naive application of the standard holographic lore on normalizable and non-normalizable modes is invalid, and that our computations are exposing some subtlety in the holographic dictionary, perhaps similar to the alternative quantization story in four dimensions \cite{Breitenlohner:1982jf,Klebanov:1999tb}. 

On the other hand, physics suggests that one should have anticipated the breaking of conformal invariance in the action of the CFT because non-extremal black holes do not have AdS throats and probably should not ultimately be described by a CFT.  It is doubly gratifying to see that our microstrata not only exhibit this phenomenon but also provide a perturbative route that may well help us understand it more deeply within the field theory: One might be able to leverage this property of microstrata to probe the new holographic field theory that emerges as the black hole is perturbed away from extremality.

For all of these reasons, we think that  this paper will  open the way to a range of exciting new investigations into non-extremal microstate geometries and their CFT duals.  

%%%%%%%%%%%%%%%%%%%%%%%%%%%%%%%%%%%%%
\section*{Note Added}
%%%%%%%%%%%%%%%%%%%%%%%%%%%%%%%%%%%%% 

Since this work was completed and submitted to the archive, there have been two very significant developments

(i) Preliminary computations \cite{GHW1} suggest that  the microstrata with $\omega_0= 0$ are  supersymmetric for all values of $\alpha$ and $\beta$.   They therefore define new, generalized superstrata.  This makes the role of the special locus (\ref{special_locus}) all the more intriguing.

(ii) Stefano Giusto and Rodolfo Russo have informed us that it is possible to ``coiffure'' microstrata with $\omega_0 =2$ so that the log terms no longer appear. This puts the asymptotics of such microstrata firmly in realm of CFT states rather than Lagrangian deformations  \cite{GR}.

%%%%%%%%%%%%%%%%%%%%%%%%%%%%%%%%%%%%%
\section*{Acknowledgments}
%%%%%%%%%%%%%%%%%%%%%%%%%%%%%%%%%%%%% 
\vspace{-2mm}
We would like to thank  Iosif Bena, Stefano Giusto, Emil Martinec, Rodolfo Russo and David Turton  for helpful discussions and Stefano Giusto and Rodolfo Russo for carefully checking many of our computations (and finding several transcription errors).  The work of NW is supported in part by the DOE grant DE-SC0011687. The work of BG, AH and NW is supported in part by the ERC Grant 787320 - QBH Structure. 

%%%%%%%%%%%%%%%%%%%%%%%%%%%%%%%%%%%%%
\section*{Dedication}
%%%%%%%%%%%%%%%%%%%%%%%%%%%%%%%%%%%%% 
\vspace{-2mm}
While we lost Sidney Coleman more than a decade ago, the underpinnings of this work make it entirely appropriate to dedicate this paper to his memory.  Long ago, Sidney was  exceptionally kind and welcoming to a new junior faculty member at the ``technical college down the the road from Harvard.''   Sidney was a generous spirit with a laconic and infectious  sense of humor.  Apart from a being an inspiring colleague, he was a delightful hiking partner in Aspen and a deep well of information about Science Fiction and the authors of many of the Sci-Fi classics.

%%%%%%%%%%%%%%%%%%%%%%%%%%%%%%%%%%%%%%%%%%%%%%%

%%%%%%%%%%%%%%%%%%%%%%%%%%%%%%%%%%%%%%%%%%%%%%%%%%%%%
%\newpage

\begin{adjustwidth}{-1mm}{-1mm} % to adjust the L and R margins

\bibliographystyle{utphys}

\bibliography{microstates}       % calls file "microstates.bib"

\end{adjustwidth}
%%%%%%%%%%%%%%%%%%%%%%%%%%%%%%%%%%%%%%%%%%%%%%%%%%%%%

%%%%%%%%%%%%%%%%%%%%%%%%%%%%%%%%%%%%%%%%%%%%%%%%%%%%%
\end{document}